\documentclass[a4paper,11pt]{article}
\usepackage{axodraw4j}
\usepackage{jcappub} 

\usepackage[normalem]{ulem}
\usepackage[T1]{fontenc} 
\usepackage[utf8]{inputenc}
\usepackage{latexsym}
\usepackage{amssymb}
\usepackage{amsmath}
\usepackage{amsfonts}
\usepackage{graphics}
\usepackage{graphicx}
\usepackage{caption}
\usepackage{subcaption}
\usepackage[many]{tcolorbox}
\usepackage[capitalise]{cleveref}

\usepackage{amsmath,amsfonts,amssymb,amsthm,amstext,amscd,mathtools,eucal,graphicx,xcolor,esint,color,empheq}

\usepackage[all]{xy}
\usepackage{dsfont,epiolmec,comment}

\newcommand{\x}{\boldsymbol{x}}
\newcommand{\q}{\boldsymbol{q}}
\newcommand{\y}{\boldsymbol{y}}
\newcommand{\kk}{\boldsymbol{k}}
\makeatletter
\newcommand{\pushright}[1]{\ifmeasuring@#1\else\omit\hfill$\displaystyle#1$\fi\ignorespaces}
\newcommand{\pushleft}[1]{\ifmeasuring@#1\else\omit$\displaystyle#1$\hfill\fi\ignorespaces}
\makeatother

\usepackage{colortbl}
\definecolor{summersky}{cmyk}{0.71,0.33,0,0.14}

\definecolor{rp}{cmyk}{0.2, 1, 0.6, 0}

\usepackage{tikz}
\newcommand{\Cross}{$\mathbin{\tikz [x=1.4ex,y=1.4ex,line width=.3ex] \draw (0,0) -- (1,1) (0,1) -- (1,0);}$}

\title{\boldmath Derivative Interactions during Inflation: A Systematic Approach
}

\author[1]{Aliakbar Abolhasani}
\affiliation[1]{Sharif University of Technology, Tehran, Iran}

\author[2]{Harry Goodhew}
\affiliation[2]{DAMTP, University of Cambridge, Cambridge, UK}

\emailAdd{aliakbar.abolhasani@gmail.com}
\emailAdd{hfg23@cam.ac.uk}

\abstract{ We present a systematic prescription for calculating cosmological correlation functions for models with derivative interactions through the wavefunction of the universe and compare this result with the "in-in" formalism-- canonical approach. The key step in this procedure is to perform the path integral over conjugate momenta after which a straightforward generalisation of Feynman's Rules can be applied. We show that this integral recovers the classical action plus some additional divergent contributions which are necessary to cancel other divergences that arise due to loop diagrams involving time derivatives. As a side project, for the first time, we introduce the "off-shell" version of the in-in formalism that is sometimes more straightforward, especially for the models with derivative coupling. To examine our prescription, as a specific example, we work out the trispectra of the scalar fluctuation in the model with the $\lambda {\phi'}^3$ derivative coupling.}
    \excludecomment{versionA}

\begin{document}

\maketitle

\section{Introduction}
All late time structures in the Universe are believed to have been seeded by quantum fluctuations stretched to cosmological scales during the exponential expansion of the Universe at very early times-- known as inflation. This period represents a unique window into the Universe's history where we require otherwise unobserved fields to generate such expansion\cite{Guth:1980zm,Linde:1981mu,Albrecht:1982wi} (for a review, see \cite{Baumann:2009ds}). Therefore, understanding exactly which fields were present and how they interacted during this time could significantly advance our knowledge of physics beyond the standard model. Unfortunately, reproducing the conditions present during this period is currently far beyond our technological reach. However, we can still gain insights into this period through observations of the matter and energy distribution across the Universe today, for example through observations of the Cosmic Microwave Background and Large Scale Structure, the statistical properties of which are generated by evolving initial conditions that depend on the physics during inflation \cite{Arkani-Hamed:2015bza,Chen:2009zp,Flauger:2016idt,Wands:2007bd,Chen:2010xka,Baumann:2011nk,Assassi:2012zq}. The quantum nature of these interactions ensures that, even if we knew all the particles and interactions that were present during inflation we could only constrain the probability distribution of the inflationary perturbations, described in terms of the $n$-point functions of the fields at the end of inflation. There are three common ways to calculate these correlation functions: the "in-in" formalism-- sometimes called canonical formulation-- the Schwinger-Keldysh diagrammatic formalism, and the Universe's Wavefunction. 

The "in-in" formalism was developed first by Schwinger and Keldysh \cite{Schwinger:1960qe,Keldysh:1964ud} who presented a systematic approach for calculating the quantum expectation value of quantum operators using the action principle. In this case the approach is to evolve both vacuum states back in time to the infinite past using the time evolution operator and impose a vacuum initial condition. For a neat review of "in-in" formalism-- in its contemporary sense-- see \cite{Weinberg:2005vy}. This approach is commonly used to calculate the non-Gaussianities during inflation  \cite{Maldacena:2002vr,Chen:2006nt,Chen:2009bc,Chen:2010xka}. The Schwinger-Keldysh formalism uses the path-integral approach for this time evolution and permits a diagrammatic representation, \cite{Tsamis:1996qm,Prokopec:2010be,Gong:2016qpq}. Although these diagrammatics provide an excellent way of organizing perturbation theory, this does not simplify the calculations significantly and the Schwinger-Keldysh formalism results in somewhat the same expression as the in-in formalism \cite{Chen:2017ryl}. The final method also uses a path-integral formulation but in this case we calculate the wavefunction of the universe (WFU). Having found the WFU, we can write an effective theory "without-time" on the boundary with simple rules to calculate the quantum expectation value of the cosmological observables \cite{Arkani-Hamed:2017fdk,arkani2020cosmological,baumann2020cosmological,baumann2021cosmological}.
 
 Of course, calculating the same object using any of these methods must give the same result. For non-derivative interactions, the agreement of these approaches is trivial. However, when derivative interactions are present, some nuances must be appreciated when performing the calculations which can sometimes be overlooked. Furthermore, these derivative interactions are not some theoretical quirk but rather naturally occur in the expansion of the effective field theory of inflation \cite{Cheung:2007st,senatore2010non}, for instance, the leading non-linear correction is $\dot\pi^3$. The resulting trispectrum is worked out, at tree level, in \cite{Chen:2009bc, arroja2009full} using the canonical approach, and the results are shown to agree with Schwinger-Keldysh approach  in \cite{Chen:2017ryl}. In this paper, we present a systematic approach to both the in-in and Wavefunction of the Universe and explore some of the complications that derivative couplings introduce, demonstrating the agreement between these two approaches in particular for the ${\phi'}^3$ trispectrum. We do not consider the Schwinger-Keldysh formalism here as it is well documented in \cite{Chen:2017ryl}. It is worth noting that much of the machinery employed here is well understood--  while rather hard to find--- in the context of standard QFT and requires little adaptation to be applicable in cosmology.  However, as far as we know, such a rigorous guideline has not been demonstrated for cosmological calculation yet, so such an account is worth presenting.
\section{Outline}

We first start with a detailed review of the in-in formalism in the presence of derivative couplings in Sec.~\ref{in-in:sec}. We know that in these models, the interaction Hamiltonian is not simply the interaction Lagrangian with a negative sign and therefore some additional interaction terms must be considered. Readers familiar with this calculation can skip this chapter and move to the next directly.

Sec.~\ref{WFU:sec} is devoted to calculating the wavefunction of the universe (WFU). We start with the definition of the wavefunction of the universe and employ the path-integral formulation to calculate the transition amplitude from an initial vacuum in the infinite past to an arbitrary field state $\phi_0$ on the boundary at $\eta_0$. To do this, we rigorously perform the conjugate momentum path integral, and see that we recover divergent corrections to the the standard expression for the wavefunction, $\Psi=\int \mathcal{D} \phi\, e^{iS}$.
 
To show the consistency of the WFU and in-in formalism, as an example, we compare the results of the scalar trispectrum generated by a ${\phi'}^3$ interaction calculated by both methods in Sec.~\ref{sec:WFUTri}. We see that the correction to the effective Hamiltonian in the canonical approach and derivatives of the propagator in the WFU approach give rise to precisely the same contribution to the trispectrum of the perturbations. In Sec.~\ref{cancellation-div:sec} we also examine the divergent corrections perturbatively in this example theory and show that they exactly cancel a particular type of loop divergence that similarly arises due time derivatives. 

In App.~\ref{app:PathIntegral}, we present some additional details on the WFU calculation in the presence of time derivative interaction terms. In App.~\ref{WFU-Feynamn:sec} we reinterpret the results of our calculations as a set of Feynmann Rules that behave analogously to those in flat space. In App.~\ref{prop-deriv:app} we carefully study the time derivatives of the in-in propagators and, for the first time-- as far as we know-- introduce the "off-shell" in-in formalism via the analytical continuation of the Heaviside theta function which greatly simplifies the presentation of the in-in formalism in the presence of derivative couplings.

\section{In-in Calculations: A Review}
\label{in-in:sec}
Within the in-in formalism we calculate correlators by evolving the vacuum states today to the infinite past,
\begin{align}\label{eq:inin}
    \left\langle \Omega(\eta_0) \right\rvert_\text{I} \phi^3_\text{I}(\eta_0)\left\lvert \Omega(\eta_0)\right\rangle_\text{I}&=\left\langle \Omega(\eta_i)\right\rvert_\text{I} U_\text{I}^\dagger(\eta_i,\eta_0) \phi^3_\text{I}(\eta_0) U_\text{I}(\eta_i,\eta_0) \left\lvert \Omega(\eta_i) \right\rangle_\text{I}.
\end{align}
The state $\Omega$ is the vacuum state defined so that it is annihilated by the annihilation operator, $a_k$, at time $\eta_i$. The interaction picture time evolution operator is
\begin{align}\nonumber
    U_\text{I}(\eta_1,\eta_2)&=T\exp\left[-i\int_{\eta_1}^{\eta_2}H_{\text{I}}(\eta)d\eta\right]\\&\approx 1-i\int_{\eta_1}^{\eta_2}H_{\text{I}}(\eta)d\eta-\frac{1}{2}\int_{\eta_1}^{\eta_2}d\eta d\eta' T\left(H_{\text{I}}(\eta)H_{\text{I}}(\eta')\right).
    \label{eq:Dyson}
\end{align}
For models with Hamiltonians that depend on the conjugate momenta $\pi$, only through the free piece,
\begin{align}
    H_{0}=\int d^3x a^4 \left(\frac{\pi^2}{2a^6}+\frac{1}{2a^2}|\nabla \phi|^2 +\frac{1}{2}m^2  \phi^2\right),
\end{align}
the interaction Hamiltonian is equal to the interaction Lagrangian with a negative sign. However, for the models with derivative interactions, matters are not this simple. In the following section, we review how to find the interaction Hamiltonian systematically.    
\subsection{Interaction potential}
To find the interaction potential, we should take the steps below:
\begin{enumerate}
    \item
Start with some theory with action
\begin{align}
    S=\int d^4 x \mathcal{L}[\phi,\phi']
\end{align}
and write down the Hamiltonian as a function of the Heisenberg picture operators $\phi_{\mathrm{H}}(\boldsymbol{x},\eta)$ and $\pi_{\mathrm{H}}(\boldsymbol{x},\eta)$,
\begin{align}
H_{\mathrm{H}}(\eta)=H[\phi_{\mathrm{H}}(\boldsymbol{x},\eta),\pi_{\mathrm{H}}(\boldsymbol{x},\eta)]=\int  d^3 x \mathcal{H}[\phi_{\mathrm{H}}(\x,\eta),\pi_{\mathrm{H}}(\x,\eta)] ,  
\end{align}
where 
\begin{align}
    \pi=\frac{\delta \mathcal{L}}{\delta \phi'}
\end{align}
and
\begin{align}
    \mathcal{H}[\phi,\pi]=\pi\phi'[\phi,\pi]-\mathcal{L}\left[\phi,\phi'[\phi,\pi]\right].
\end{align}
\item The time time dependence of $\phi$ and $\pi$ is then
\begin{align}\label{eq:phiHprime}
    \phi'_{\mathrm{H}}(\x,\eta)&=i\left[H_H(\eta),\phi_{\mathrm{H}}(\x,\eta)\right]\\
    \pi'_{\mathrm{H}}(\x,\eta)&=i\left[H_H(\eta),\pi_{\mathrm{H}}(\x,\eta)\right]
\end{align}
\item These operators evolve in time according to 
\begin{align}
    \phi_{\mathrm{H}}(\eta)= U^{-1}(\eta_i,\eta)\phi_{\mathrm{H}}(\eta_i)U(\eta_i,\eta)
\end{align}
Where we have dropped the $\x$ dependence for brevity. We can differentiate this and use Eq.~\ref{eq:phiHprime} to find the time dependence of $U$,
\end{enumerate}
\begin{align}\nonumber
    iH_{\mathrm{H}}(\eta)\phi_{\mathrm{H}}(\eta)-i\phi_{\mathrm{H}}(\eta)H_{\mathrm{H}}(\eta)&=\frac{dU^{-1}(\eta_i,\eta)}{d\eta}\phi_{\mathrm{H}}(\eta_i) U(\eta_i,\eta)+U^{-1}(\eta_i,\eta)\phi_{\mathrm{H}}(\eta_i)\frac{dU(\eta_i,\eta)}{d\eta}\\&=\frac{dU^{-1}(\eta_i,\eta)}{d\eta}U(\eta_i,\eta)\phi_{\mathrm{H}}(\eta)+\phi_{\mathrm{H}}(\eta)U^{-1}(\eta_i,\eta)\frac{dU(\eta_i,\eta)}{d\eta}.
\end{align}
\begin{enumerate}
    \setcounter{enumi}{3}
    \item[] This is satisfied, for Unitary $U$ and Hermitian $H$, if
\begin{align}
    \frac{d U(\eta_i,\eta)}{d\eta}=-iU(\eta_i,\eta)H_{\mathrm{H}}(\eta)
\end{align}
\item	Split the Hamiltonian into free and interaction pieces by choosing
\begin{align}
    H_0= \int  d^3x\,\frac{1}{2a^2}\left(\pi_{\mathrm{H}}^2 +a^4|\nabla \phi_{\mathrm{H}}|^2 \right)=\int d^3 x \mathcal{H}_0[\phi_{\mathrm{H}}(\x,\eta),\pi_{\mathrm{H}}(\x,\eta)] 
\end{align}
\item
Introduce the interaction picture operators, $\phi_{\mathrm{I}}$ and $\pi_{\mathrm{I}}$ which are defined so that their time dependence is generated by the free Hamiltonian,
\begin{align}
    {\phi_{\mathrm{I}}'}(\eta)&=i[H_0[\phi_{\mathrm{I}}(\eta),\pi_{\mathrm{I}}(\eta)],\phi_{\mathrm{I}}(\eta)],\\ {\pi_{\mathrm{I}}'}(\eta)&=i[H_0[\phi_{\mathrm{I}}(\eta),\pi_{\mathrm{I}}(\eta)],\pi_{\mathrm{I}}(\eta)]
\end{align}
These can also be defined using creation and anihilation operators which annihilate the vacuum of the full theory in the infinite past. This is why it is necessary to introduce the time evolution operators in the in-in formalism, Eq.~\eqref{eq:inin}. The form of $H_0$ and the canonical commutation relations imply that
\begin{equation}\label{eq:piint}
    \pi_I(\eta)=a^2(\eta)\phi_I'(\eta)
\end{equation}
\item	
At $\eta=\eta_i$, the Heisenberg picture operators $\hat{\phi}$ and $\hat{\pi}$ coincide with those in the interaction picture,
\begin{align}
   \phi_{\mathrm{H}}(\eta_i)&=\phi_{\mathrm{I}}(\eta_i)
   \\
   \pi_{\mathrm{H}}(\eta_i)&=\pi_{\mathrm{I}}(\eta_i),
\end{align}
therefore, 
\begin{align}
H_{\mathrm{H}}(\eta_i)=H[\phi_{\mathrm{I}}(\eta_i),\pi_{\mathrm{I}}(\eta_i)]    
\end{align}
\item This will permit analogous solutions to those in the Hamiltonian picture,
\begin{align}
    \phi_{\mathrm{I}}(\eta)=U_{0}^{-1}(\eta_i,\eta)\phi_{\mathrm{I}}(\eta_i)U_{0}(\eta_i,\eta)
\end{align}
with
\begin{align}
    \frac{dU_0}{d\eta}(\eta_i,\eta)&=-iU_0(\eta_i,\eta)H_0[\phi_{\mathrm{I}}(\eta),\pi_{\mathrm{I}}(\eta)]=-iH_0[\phi_{\mathrm{I}}(\eta_i),\pi_{\mathrm{I}}(\eta_i)]U_0(\eta_i,\eta)
\end{align}
\item
Consider the inner product of an operator acting on some state with another state and compare the results in the two pictures to determine how states evolve,
\begin{align}
    \langle \psi_{\mathrm{H}}(\eta)\rvert \mathcal{O}_{\mathrm{H}}(\eta)\lvert \phi_{\mathrm{H}}(\eta)\rangle&=\langle \psi_{\mathrm{H}}(\eta)\rvert U^{-1}(\eta_i,\eta) \mathcal{O}_{\mathrm{H}}(\eta_i)U(\eta_i,\eta)\lvert \phi_{\mathrm{H}}(\eta)\rangle
    \\&=\langle \psi_{\mathrm{H}}(\eta)\rvert U^{-1}(\eta_i,\eta)U_0(\eta_i,\eta)\mathcal{O}_{\mathrm{I}}(\eta)U_0^{-1}(\eta_i,\eta)U(\eta_i,\eta)\lvert \phi_{\mathrm{H}}(\eta)\rangle\\&=\langle \psi_{\mathrm{I}}(\eta)\rvert \mathcal{O}_{\mathrm{I}}(\eta)\lvert \phi_{\mathrm{I}}(\eta)\rangle.
\end{align}
Where we have used that the time evolution of an arbitrary Hermitian operator $\mathcal{O}_{\mathrm{H}}$ is generated in the same way as for the field, $\phi_{\mathrm{H}}$ and its conjugate momentum. Therefore, states in the interaction picture are related to those in the Heisenberg picture by
\begin{equation}
    \lvert \phi_{\mathrm{I}}(\eta)\rangle=U_0^{-1}(\eta_i,\eta)U(\eta_i,\eta)\lvert \phi_{\mathrm{H}}(\eta)\rangle
\end{equation}
\item
Fix this operator by considering its time derivative
\begin{align}
    &\frac{d}{d\eta}\left(U_0^{-1}(\eta_i,\eta)U(\eta_i,\eta)\right)\\&=iH_0[\phi_{\mathrm{I}}(\eta),\pi_{\mathrm{I}}(\eta)]U_0^{-1}(\eta_i,\eta)U(\eta_i,\eta)-iU_0^{-1}(\eta_i,\eta)H[\phi_{\mathrm{H}}(\eta_i),\pi_{\mathrm{H}}(\eta_i)] U(\eta_i,\eta)\\
    &=iH_0[\phi_{\mathrm{I}}(\eta),\pi_{\mathrm{I}}(\eta)]U_0^{-1}(\eta_i,\eta)U(\eta_i,\eta)-iH[\phi_{\mathrm{I}}(\eta),\pi_{\mathrm{I}}(\eta)] U_0^{-1}(\eta_i,\eta) U(\eta_i,\eta)\\
    &=-iH_{\mathrm{I}}[\phi_{\mathrm{I}}(\eta),\pi_{\mathrm{I}}(\eta)] U_0^{-1}(\eta_i,\eta) U(\eta_i,\eta)
\end{align}
This is the third time that we are seeing some form of this equation. We now note that this has a formal solution given by
\begin{align}
    U_{\mathrm{I}}=U_0^{-1}(\eta_i,\eta)U(\eta_i,\eta)=\mathrm{T}\exp\left[-i\int_{\eta_i}^\eta  d\eta' H_{\mathrm{I}}[\phi_{\mathrm{I}}(\eta'),\pi_{\mathrm{I}}(\eta')]\right]
\end{align}
\item Replace $\pi_{\mathrm{I}}$ with its solution in the interaction picture, Eq.~\eqref{eq:piint}. To make the importance of this step clear we will temporarily adopt a slightly cumbersome notation,
\begin{align}
    U_{\mathrm{I}}=\mathrm{T}\exp\left[-i\int_{\eta_i}^\eta d^4x' \mathcal{H}_{\mathrm{I}}\left[\phi_{\mathrm{I}}(\eta'),\pi_{\mathrm{I}}(\eta')=a^2(\eta)\phi_{\mathrm{I}}'(\eta)\right]\right].
\end{align}
Note, in particular, that 
\begin{align}
    \mathcal{H}_{\mathrm{I}}\left[\phi_{\mathrm{I}}(\eta),\pi_{\mathrm{I}}(\eta)=a^2(\eta)\phi_{\mathrm{I}}'(\eta)\right]\neq -\mathcal{L}_{\mathrm{I}}[\phi_{\mathrm{I}}(\eta),\phi_{\mathrm{I}}'(\eta)].
\end{align}
To see this consider the theory with action
\begin{equation}
    S=\int  d^4x a^4\left(\frac{1}{2a^2}{\phi'}^2-\frac{1}{2a^2}\partial_i\phi\partial_i\phi-\frac{\lambda}{a^3} {\phi'}^3\right)
\end{equation}
for which it is straightforward to read off that $\mathcal{L}_{\mathrm{I}}=-a\lambda{\phi'}^3$ but the interaction Hamiltonian is given, to second order in $\lambda$, by
\begin{align}
    \mathcal{H}_{\mathrm{I}}[\phi,\pi]=\lambda \frac{ \pi^3}{a^5}+\frac{9\lambda^2\pi^4}{2a^8}+\mathcal{O}(\lambda^3).
\end{align}
Therefore, the required interaction picture operator is 
\begin{tcolorbox}[ams align, colback=white, colframe=black] \label{interaction-Hamiltonian:eq}
    \mathcal{H}_{\mathrm{I}}\left[\phi_{\mathrm{I}}(\eta),\pi_{\mathrm{I}}(\eta)=a^2(\eta)\phi_{\mathrm{I}}'(\eta)\right]\equiv \mathcal{H}_{\mathrm{I}}(\eta)=\lambda a{\phi'_{\mathrm{I}}}^3+\frac{9\lambda^2}{2}{\phi'_{\mathrm{I}}}^4+\mathcal{O}(\lambda^3).
\end{tcolorbox}
\item Calculate the full Hamiltonian from the Hamiltonian density,
\begin{align}
    H_{\mathrm{I}}(\eta)=\int d^3x \mathcal{H}_{\mathrm{I}}(\eta)
\end{align}
from which we can perform the Dyson series expansion in Eq.~\eqref{eq:Dyson}.
\end{enumerate}
To demonstrate the effect of the increased complexity of the calculations involving time derivative interactions we focus on the specific case of the trispectrum generated by interactions of the form ${\phi'}^3$. We focus on this interaction as it perhaps the simplest case involving derivative interactions but still contains all the interesting behaviour and, as can be seen from Eq.~\eqref{interaction-Hamiltonian:eq} the first non-trivial corrections from this interaction appear at second order in $\lambda$ and so require us to consider at least the tripsectrum.
\subsection{$\lambda {\phi'}^3$ Trispectrum: the Canonical Approach}
In perturbation theory, we get terms like  $\big\langle T(\phi'(\eta)\phi'(\eta'))\big\rangle$. In the interaction picture,
\begin{equation}
    \hat{\phi}'=\int \frac{d^3\kk}{(2\pi)^3}\left[\phi'_k(\eta)a_{\kk}+\phi'^{\ast}_k(\eta) a_{-\kk}^\dagger\right]e^{-i\kk\cdot\x}
\end{equation}
Eq.~\eqref{time-order-derivative:eq} says
\begin{align}\nonumber
    \big\langle T(\phi'(x)\phi'(x')) \big\rangle=\int\frac{d^3\kk}{(2\pi)^3}\phi'_k(\eta)\phi'^{\ast}_k(\eta')\theta(\eta-\eta')
    +\int\frac{d^3\kk}{(2\pi)^3}\phi'_k(\eta')\phi'^{\ast}_k(\eta)\theta(\eta'-\eta)
\end{align}
Therefore, for the interaction Hamiltonian Eq.~\eqref{interaction-Hamiltonian:eq}, the 4-particle correlator is given by
\begin{align}
\nonumber
    T(x_1,x_2,x_3,x_4)=
    &\big\langle \Omega \rvert -\frac{\lambda^2}{2}\iint d^4x\, d^4 x'\,a(\eta) a(\eta')~\bar{T}\left(\hat{\phi}'^3(x)\,\hat{\phi}'^3(x')\right)\hat{\phi}^4(\eta_0)\,\lvert \Omega\big\rangle
    \\
    \nonumber
    +& \big\langle \Omega \rvert -\frac{\lambda^2}{2}\iint d^4x\, d^4 x'\,a(\eta) a(\eta')~\hat{\phi}^4(\eta_0)\,T\left(\hat{\phi}'^3(\eta)\,\hat{\phi}'^3(\eta')\right)\lvert \Omega \big\rangle
    \\
    \nonumber
    +&\big\langle \Omega \rvert ~\lambda^2\,\int d^4 x\,a(\eta)~ \hat{\phi}'^3(x)~\hat{\phi}^4(\eta_0)~\int d^4x'\,a(\eta')~ \hat{\phi}'^3(x')\lvert \Omega\big\rangle
    \\
    +&\big \langle \Omega \rvert ~\dfrac{9i\, \lambda^2}{2}\int d^4 x~ \hat{\phi}'^4(x)~\hat{\phi}^4(\eta_0)\,\lvert \Omega \big \rangle+ \big\langle \Omega \rvert -\dfrac{9i\,\lambda^2}{2}\,\hat{\phi}^4(\eta_0)\,\int d^4x~ \hat{\phi}'^4(x)\lvert \Omega\big\rangle\label{eq:Tinin}
\end{align}
In any term with two three point vertices there are 9 ways to combine one of the momenta from the vertex with the other which leaves 4 momenta that must be paired up with $\phi^4$, there are $4!$  ways to do this. Going to Fourier space, we find
\begin{align}
    T(\kk_1,\kk_2,\kk_3,\kk_4) \equiv T_3+T_4= T^{}_{3\,\mathrm{rr}}+ T^{}_{3\,\mathrm{ll}}+T^{}_{3\,\mathrm{lr}}+T_{4\,\mathrm{r}}+ T_{4\,\mathrm{l}}
\end{align}
From Eq.~\ref{eq:Tinin} it is straightforward to see that \begin{align}
    T_{3\,\mathrm{ll}}&=T_{3\,\mathrm{rr}}^*,& T_{4\,\mathrm{l}}&=T_{4\,\mathrm{r}}^*,& T_{3\,\mathrm{lr}}&=T_{3\,\mathrm{lr}}^*
\end{align}
so we need only write down
\begin{align}
\nonumber
    T^{}_{3\,\mathrm{rr}}&(\kk_1,\kk_2,\kk_3,\kk_4)=-\frac{9\times4}{2}~\lambda^2\,\phi_{k_1}(\eta_0)\phi_{k_2}(\eta_0)\phi_{k_3}(\eta_0)\phi_{k_4}(\eta_0)\times
    \\
    \nonumber
    &\iint_{-\infty}^{0}d\eta d\eta'\,a(\eta)a(\eta')\,\bigg[\phi_{k_1}'^{\ast}(\eta)\phi_{k_2}'^{\ast}(\eta)\phi_{k_3}'^{\ast}(\eta')\phi_{k_4}'^{\ast}(\eta') \,\langle \Omega \rvert T\,\hat{\phi'}_{k_{12}}(\eta)\hat{\phi'}_{-\kk_{12}}(\eta')\lvert \Omega\rangle
    \\
    &~~~~~~~~~~~~~~~~~~~~~~~~~~~~~~~~~~~~~~~~~~~~~~~~~~~~~~~~~~~~~~~~~~~+\kk_{1,2}\leftrightarrow\kk_{3,4}\bigg]+t+u ,
    \label{T-naive-can1:eq}
    \\
    \nonumber
    T^{}_{3\,\mathrm{lr}}&(\kk_1,\kk_2,\kk_3,\kk_4)=~~ 9\times8~\lambda^2\phi_{k_1}(\eta_0)\phi_{k_2}(\eta_0)\phi^{\ast}_{k_3}(\eta_0)\phi^{\ast}_{k_4}(\eta_0)\,
    \\
    &\int_{-\infty}^{0}d\eta\, a(\eta)\,\phi_{k_1}'(\eta)\phi_{k_2}'(\eta)\phi_{k_{12}}'(\eta)\int_{-\infty}^{0}d\eta' a(\eta')\phi_{-\kk_{12}}'^{\ast}(\eta)\phi_{k_3}'^{\ast}(\eta')\phi_{k_4}'^{\ast}(\eta')+t+u.
    \label{T-naive-can3:eq}
    \end{align}

\begin{versionA}
    \begin{align}
    \nonumber
    T^{}_{3\,\mathrm{ll}}&(\kk_1,\kk_2,\kk_3,\kk_4)=-\frac{9\times4}{2}~\lambda^2\,\phi^{\ast}_{k_1}(\eta_0)\phi^{\ast}_{k_2}(\eta_0)\phi^{\ast}_{k_3}(\eta_0)\phi^{\ast}_{k_4}(\eta_0)\times
    \\
    \nonumber
    &\iint_{-\infty}^{0}d\eta d\eta'\,a(\eta)a(\eta')\,\bigg[\phi_{k_1}'(\eta)\phi_{k_2}'(\eta)\phi_{k_3}'(\eta')\phi_{k_4}'(\eta') \,\langle \Omega \rvert \bar{T}\,\hat{\phi'}_{k_{12}}(\eta)\hat{\phi'}_{-\kk_{12}}(\eta')\lvert \Omega\rangle
    \\
    &~~~~~~~~~~~~~~~~~~~~~~~~~~~~~~~~~~~~~~~~~~~~~~~~~~~~~~~~~~~~~~~~~~~+\kk_{1,2}\leftrightarrow\kk_{3,4}\bigg]+t+u
    \label{T-naive-can2:eq}
        \\
    T_{4\,\mathrm{l}}=& +i~4!\times \left(\dfrac{9}{2}\lambda^2 \right)\,\phi^{\ast}_{k_1}(\eta_0)\phi^{\ast}_{k_2}(\eta_0)\phi^{\ast}_{k_3}(\eta_0)\phi^{\ast}_{k_4}(\eta_0)\,   \int_{-\infty}^{0}d\eta \,\phi_{k_1}'(\eta)\phi_{k_2}'(\eta)\phi_{k_3}'(\eta)\phi_{k_4}'(\eta)
\end{align}
    \begin{align}\label{eq:corrections}
    T_{4\,\mathrm{r}}=&-i~4!\times\left(\dfrac{9}{2}\lambda^2\right)\,\phi_{k_1}(\eta_0)\phi_{k_2}(\eta_0)\phi_{k_3}(\eta_0)\phi_{k_4}(\eta_0)\,\int_{-\infty}^{0}d\eta\, \phi_{k_1}'^{\ast}(\eta)\phi_{k_2}'^{\ast}(\eta)\phi_{k_3}'^{\ast}(\eta)\phi_{k_4}'^{\ast}(\eta)
\end{align}
These can be combined to give the total correction to the trispectrum due to this careful treatment of the interaction Hamiltonian as
\end{versionA}
\noindent Note that the factor of 9 comes from the $3\times3$ equivalent ways we can pair up an internal line and the permutations apply to all the terms. The additional $4$-point contact term that arises due to the time derivative interaction likewise generates a contribution to trispectrum at second order in $\lambda$, combining $T_{4\,\mathrm{r}}$ and $T_{4\,\mathrm{l}}$ gives the total correction due to this careful treatment of the interaction Hamiltonian,

\begin{tcolorbox}[ams align, colback=white, colframe=black]
    T_4= \mathrm{Re}\left[-4!\times9i\lambda^2\,\phi_{k_1}(\eta_0)\phi_{k_2}(\eta_0)\phi_{k_3}(\eta_0)\phi_{k_4}(\eta_0)\,\int_{-\infty}^{0}d\eta\, \phi_{k_1}'^{\ast}(\eta)\phi_{k_2}'^{\ast}(\eta)\phi_{k_3}'^{\ast}(\eta)\phi_{k_4}'^{\ast}(\eta)\right]
\label{T-delta-can:eq}
\end{tcolorbox}
\noindent After some algebra, we find
\begin{align}
      T_{3}(\kk_1,\kk_2,\kk_3,\kk_4)&= \dfrac{9H^8 \lambda^2 k_{12}}{8 k_1 k_2 k_3 k_4} \left[ \frac{\left(L^2 M_1^3+3 L M_1^3 M_2+6 M_1^3 M_2^2\right)}{ L^5 M_1^3 M_2^3}+\dfrac{1}{M_1^3 M_2^3} \right]+23~\mathrm{perms},
      \label{Psi-2-a:final:eq}
 \end{align}
 and
 \begin{align}
    T_4&=\dfrac{9H^8\, \lambda^2}{8}\dfrac{-2\times4!}{ k_1 k_2 k_3 k_4 L^5},
\end{align}          
in which, we defined
\begin{align*}
    k_{12} &\equiv |\kk_1+\kk_2|
    \\
    M_1 &\equiv k_1+k_2+k_{12}
    \\
    M_2 &\equiv k_3+k_4+k_{12}
    \\
    L~~ &\equiv k_1+k_2+k_3+k_4.
\end{align*}

\section{Wavefunctional of the universe}
\label{WFU:sec}
It is typical within the cosmology literature \cite{COT,Arkani-Hamed:2017fdk,arkani2020cosmological} to take, as a starting point, the wavefunction of the universe to be
\begin{equation}\label{eq:WFU}
    \Psi[\bar{\phi},\eta]=e^{iS[\phi_{\textrm{cl}}(\bar{\phi},\eta)]}
\end{equation}
Where $S$ is the action and $\phi_{\textrm{cl}}(\eta)$ is the classical solution which extremises the action subject to the boundary condition that $\phi_{\textrm{cl}}(\eta)=\bar{\phi}$ plus whatever initial condition we wish to impose. The expansion of $\phi$ in terms of its background solution and the perturbations introduced by any interactions introduces the bulk-to-bulk and bulk-to-boundary propagators that can be used to define a Feynmanesque diagrammatic calculation. We begin this section by reviewing the derivation of this result for non-derivative interactions as the stationary phase approximation of the integral
\begin{equation}\label{eq:nonderiv}
    \Psi[\bar{\phi},\eta]=\langle \bar{\phi},\eta\rvert \Omega\rangle =\int_{\Omega}^{\bar{\phi}} \mathcal{D}\phi \exp\left[i S[\phi,\eta]\right].
\end{equation}
We then go on to show that, when we introduce derivative interactions, this relationship no longer holds. However it is still possible to perform a stationary phase approximation to express the wavefunction of the universe as a simple exponential,
\begin{equation}
    \Psi[\bar{\phi},\eta]=e^{i\bar{S}[\phi_{\textrm{cl}}(\bar{\phi},\eta)]},
\end{equation}
for some $\bar{S}\neq S $. These additional terms are subleading in Plank's constant and so at tree level (i.e. where this stationary phase approximation is valid) they can be ignored. Furthermore, in the final part of this section we demonstrate that when we do consider loop corrections we find that they are necessary to cancel some divergences that result from taking derivatives of the propagators in a loop that do not arise in the in-in formalism.

\subsection{Canonical to functional formulation}
Without going into too much detail, we write down the well know result \cite{weinberg1995quantum} for the transition amplitude $\langle \bar{\phi},\eta_0| \phi_i,\eta_i \rangle$ as
\begin{align}
    \langle \bar{\phi},\eta_0| \phi_i,\eta_i \rangle = \int_{\phi_i}^{\bar{\phi}} {\cal D}\phi \int {\cal D} \pi ~ \exp  \left[ i \int  d\eta\,d^3 x ({\phi'}(\x,\eta) \pi(\x,t) -{\cal H}(\phi(\x,\eta), \pi(\x,\eta)) \right]
    \label{PI-Hamiltonian:eq}
\end{align}
where we used the fact that two complete sets of eigen-states $|\phi(\x,t)\rangle$ and $|\pi (\x,t)\rangle$
\begin{align}
    \hat{\phi}(\x,\eta)|\phi(\x,\eta)\rangle&= \,\phi(\x,\eta)\, |\phi(\x,\eta)\rangle
    \\
    \hat{\pi}(\x,\eta)|\pi(\x,\eta)\rangle&= \,\pi(\x,\eta)\, |\pi(\x,\eta)\rangle
\end{align}
have the following scalar product
\begin{align}
    \langle \phi(\x,\eta)|\pi(\x',\eta)\rangle = \delta^3(\x-\x')\,\dfrac{1}{\sqrt{2 \pi}} \exp \left(i a^3\int d^3 x\, \phi(\x,\eta) \pi (\x,\eta) \right).
\end{align}
For this, we assume that $\hat{\phi}$ and $\hat{\pi}_{\phi}$ satisfy the canonical commutation relations 
\begin{align}
    [\hat{\phi}(\x,\eta),\hat{\pi}_{\phi}(\x',\eta)]=i\delta^3(\x-\x').
\end{align}
 In the path integral formulation, the field and its conjugate momenta are independent parameters over which one must integrate. For theories which are quadratic in the derivatives of the field,
 \begin{equation}
     S[\phi,{\phi'}]=\int a^4 d^4 x\left( \frac{1}{2a^2}{\phi'}^2+F[\phi,\eta]\right)=\int d\eta L =\int  d\eta d^3x\, \mathcal{L},
 \end{equation}
 the conjugate momentum is
 \begin{equation}
     \pi=\frac{\delta\mathcal{L}}{\delta {\phi'}}=a^2 \phi'.
 \end{equation}
 So, the exponent in this integral is
 \begin{equation}
     i\int d^4 x \left({\phi'}\pi-\frac{\pi^2}{2a^2}+F[\phi,\eta]\right)=-\frac{i}{2}\int d^4 x \left(\frac{\pi}{a}-a \phi'\right)^2+iS[\phi,{\phi}'].
 \end{equation}
Therefore, the integral over $\pi$ is a simple Gaussian which returns a constant, $N$, 
\begin{align}
    \langle \phi_1,\eta_1| \phi_2,\eta_2 \rangle = N \int_{\phi_1}^{\phi_2} {\cal D}\phi ~ \exp  \left[ i S (\phi(\x,\eta), {\phi}'(\x,\eta)) \right].
\end{align}
 However, for more general Hamiltonians, there are some logarithmic corrections to the action. In general, one has
\begin{align}
    \langle \phi_1,\eta_1| \phi_1,\eta_2 \rangle \simeq N \int_{\phi_1}^{\phi_2} {\cal D}\phi ~ \exp  \left[ i\tilde{S}(\phi(\x,\eta), {\phi'}(\x,\eta)) \right]\label{eq:Sbar}
\end{align}
where
\begin{align}
    \tilde{S}= S +i \, \delta^4(0)  \int   d\eta d^3x\, \log f(\phi,\dot{\phi})
    \label{action-correction}
\end{align}
The $S$ in the above relation is the classical action of the theory. To see that this is true first compare Eqs.  \eqref{PI-Hamiltonian:eq} and \eqref{eq:Sbar} so that the effective action is defined by
\begin{align}
    \exp(i\tilde{S})\equiv \int {\cal D} \pi \, \exp  \left[ \frac{i}{\hbar}  \int  d\eta\,d^3x ({\phi'}(\x,t) \pi(\x,t) -{\cal H}(\phi(\x,t), \pi(\x,t)) \right].
\end{align}
Where we have reinserted $\hbar$ which is usually set to one because the additional terms will be suppressed by this factor. To make progress we must first replace the integral over space with a sum over a lattice of space-time patches of infinitesimal 4-volume, $\Omega$, defined in comoving coordinates and discretize the path-integral measure,
\begin{align}
    \exp(i\tilde{S})= \prod_{i} \int \,d\pi_i \exp  \left[ i \,\dfrac{\Omega}{\hbar}   \bigg(\phi'_i \,\pi_i -{\cal H}(\phi_i,\pi_i) \bigg) \right].
    \label{p-integral-leading:eq}
\end{align}
Where $\phi_i=\phi(\eta_i,\x_i)$ and $\pi_i=\pi(\eta_i,\x_i)$. Note, that in this expression, $\Omega$ appears only as $\frac{\Omega}{\hbar}$ and so, despite $\Omega$ being small, this fraction, in the classical limit is large. We perform the integral at each point using the stationary phase approximation,
\begin{align}
    \int dx \,e^{i\, \xi f(x)} = \sum_{x_{\ast}\in \Sigma } e^{i \xi f(x_{\ast}) + i\frac{\pi}{4} \mathrm{sgn}(f''(x_{\ast}))}\sqrt{\dfrac{2\pi}{\xi |f''(x_{\ast})|}} + o (\xi^{-1/2}).
\end{align}
Here, $\Sigma$ indicates the set of all stationary points of the function $f$. In our case $f_i$ is extremised when
\begin{align}
    \phi_i'=\frac{\partial \mathcal{H}}{\partial \pi_i}
\end{align}
Which is one of the classical Hamilton's equations, satisfied on the classical solution of the action and so the integral is \begin{align}
    \int d\pi \exp  \left[ i \,\dfrac{\Omega}{\hbar}   \bigg(\phi' \,\pi -{\cal H}(\phi,\pi) \bigg) \right]=\exp\left[i \,\dfrac{\Omega}{\hbar}  \left(\phi'_{\text{cl}}\pi_{\text{cl}}-\mathcal{H}_{\text{cl}}\right)\right]\sqrt{\frac{2\pi\hbar}{\Omega}}+o\left(\sqrt{\frac{\hbar}{\Omega}}\right)
\end{align}
But, on the classical solution we know that 
\begin{equation}
    \left.\phi'\pi-\mathcal{H}[\phi,\pi]\right\rvert_{\pi=\pi_{\text{cl}}}=\mathcal{L}[\phi,\phi']
\end{equation}
And so we recover
\begin{align}\nonumber
    \exp(i\bar{S})&=\prod_{i}\left(\exp\left[i\frac{\Omega}{\hbar} \mathcal{L}_{\text{cl}}[\phi_i',\phi_i]\right]\sqrt{\frac{2\pi \hbar}{\Omega}}+o\left(\sqrt{\frac{\hbar}{\Omega}}\right)\right)\\&\approx\mathcal{N}\exp\left[\frac{i}{\hbar}\int d^4 x \mathcal{L}_{\text{cl}}[\phi',\phi]\right] .
\end{align}
Where $\mathcal{N}$ is some constant from the product of $\sqrt{\frac{2\pi\hbar}{\Omega}}$ terms, and the approximate sign indicates that we have dropped some terms that come at higher-order in $\hbar$. It is these terms that contribute to the logarithmic corrections in Eq.~\eqref{action-correction}. The delta function is generated by the fact that these additional terms enter the exponential without a factor of $\Omega$, and so when we take the product overall spatial points, we pick up a divergence that we represent using this divergent distribution (this should become more apparent in the example below). This concludes our proof of the relationship in Eq.~\eqref{action-correction}. In particular, the above argument says that knowing the classical action is enough to calculate the coefficients of the Wavefunction of the Universe at tree level.
\subsection{${\phi'}^3$ Wavefunction}
Now, let us do an explicit calculation for the case of ${\cal L}_{\mathrm{int.}} =\lambda {\phi'}^3$. The prescription above is only sufficient to find the wavefunction to leading order in $\hbar$. It is therefore necessary to develop a different procedure to find the higher order terms. It is possible to express this in a diagrammatic manner just as in flat space, a procedure that we elaborate on in App.~\ref{app:PathIntegral}, however, these diagrams just represent shorthand for the various integrals that we will eventually have to perform and can't tell us, for example, which vertices to include and so it is still necessary to fully understand each stage of the calculation before defining the Feynman Rules. To begin with we rewrite Eq.~\eqref{p-integral-leading:eq} by identifying the piece that is quadratic in the conjugate momentum. To do this we start by defining
\begin{align}
    \Phi(\phi,\phi',\pi)\equiv \phi'\pi-\mathcal{H}(\phi,\pi).
\end{align}
So that
\begin{align}
    \exp(i\tilde{S})= \prod_{i} \int \,d\pi_i \exp  \left[ i \,\dfrac{\Omega}{\hbar}    \Phi_i \right].
\end{align}
The stationary phase approximation that we employed before demonstrated that the most significant contribution is from the classical solution and so we find the next to leading order solution by taking $\pi$ close to its classical solution,
\begin{align}
    \pi=\pi_{\text{cl}}=\frac{\delta \mathcal{L}}{\delta \phi'},
\end{align}
around which $\Phi$ is
\begin{align}
    \Phi=\Phi(\phi,\phi',\pi_{\text{cl}})+\frac{\partial \Phi}{\partial \pi}(\pi-\pi_{\text{cl}})+\frac{1}{2}\frac{\partial^2\Phi}{\partial \pi^2}(\pi-\pi_{\text{cl}})^2+\sum_{n=3}^\infty \frac{\Phi^{(n)}}{n!}(\pi-\pi_{\text{cl}})^n
\end{align}
The first term is just the Lagrangian and the second is zero from the equations of motion so we can write
\begin{align}
    \Phi=\mathcal{L}_{\text{cl}}+\frac{1}{2}\Phi^{(2)}(\pi-\pi_{\text{cl}})^2+\sum_{n=3}^\infty \frac{\Phi^{(n)}}{n!}(\pi-\pi_{\text{cl}})^n.
\end{align}
We can define $\tilde{\pi}=\pi-\pi_{\text{cl}}$ so that the wavefunction is given by
\begin{align}
    \exp(i\tilde{S})= \prod_{i} \exp\left[i\frac{\Omega}{\hbar}  \mathcal{L}_{\text{cl}}(\phi_i,\phi'_i)\right] \int \,d\tilde{\pi}_i \exp & \left[ i \,\dfrac{\Omega}{2\hbar}   \Phi^{(2)}_i\tilde{\pi}^2_i \right]\exp\left[i \,\dfrac{\Omega}{\hbar}\sum_{n=3}^\infty      \frac{\Phi^{(n)}_i}{n!}\tilde{\pi}^n_i\right].
\end{align}
Expanding this exponential gives
\begin{align}
    \sum_{N=0}^\infty\frac{1}{N!}\frac{i\Omega}{\hbar}^N\int_{-\infty}^{\infty} d\pi \prod_{i=1}^N \left( \frac{\Phi^{(n_i)}}{n_i!}\pi^{n_i} \right)\exp\left[\frac{i\Omega}{2\hbar}  \Phi^{(2)}\pi^2\right],
\end{align}
which is of the form of the generalised Fresnel integrals,
\begin{align}
\int dx x^m e^{-\frac{i}{2} \alpha x^2} =\frac{1}{2} \left((-1)^m+1\right) (-i \alpha )^{-\frac{m}{2}-\frac{1}{2}} \Gamma \left(\frac{m+1}{2}\right),
\end{align}
so we can perform this integral for even $M=\sum n$,
\begin{align}
    \sqrt{\frac{\hbar}{i\Omega  \Phi^{(2)}}}\sum_{N=0}^\infty\frac{1}{N!}\frac{i\Omega}{\hbar}^N \prod_{i=1}^N \left( \frac{\Phi^{(n_i)}}{n_i!}\right)\left(\frac{\hbar}{i\Omega  \Phi^{(2)}}\right)^{\frac{M}{2}}\Gamma\left(\frac{M +1}{2}\right),
\end{align}
whilst it vanishes for odd $M$. Notice that that as $N$ increases we get additional factors of $\frac{\Omega}{\hbar}$ and so it seems that each term in the series expansion of the exponential becomes increasingly important when $\hbar\rightarrow 0$, however we can see that we also get a factor of  $\left(\frac{\hbar}{\Omega}\right)^\frac{M}{2}$, because $n_i\geq 3$ these terms are actually sub leading in $\hbar$. Therefore, to next to leading order in $\hbar$ we find
\begin{align}
    \exp{i\bar{S}}&=\prod_{i} \exp\left[i\frac{\Omega}{\hbar} \mathcal{L}_{\text{cl}}(\phi_i,\phi'_i)\right]\sqrt{\frac{\pi\hbar}{i\Omega \Phi^{(2)}}} \left(1-i\frac{\hbar}{32\Omega} \frac{\Phi^{(4)}}{  {\Phi^{(2)}}^2}-i\frac{5\hbar}{192\Omega}  \frac{{\Phi^{(3)}}^2}{  {\Phi^{(2)}}^3} \right)
\end{align}
For the interaction that we were considering previously we have
\begin{align}
    \Phi^{(2)}&=-\frac{1}{a^2-6a\lambda\phi'}\\
    \Phi^{(3)}&=-\frac{6\lambda }{a^2\left(a-6\lambda  \phi'\right)^3}\\
    \Phi^{(4)}&=-\frac{108\lambda^2 }{a^3\left(a-6\lambda \phi'\right)^5}
\end{align}

\begin{align}
    \exp{i\bar{S}}&=\prod_{i} \exp\left[i\frac{\Omega}{\hbar} \mathcal{L}_{\text{cl}}(\phi_i,\phi'_i)\right]\sqrt{\frac{i\pi\hbar a (a-6\lambda \phi')}{\Omega  }} \left(1+i\frac{69\hbar}{16 a\Omega} \frac{\lambda^2 }{\left(a-6\lambda\phi'\right)^3}  \right)
\end{align}
If we then expand this about small $\lambda$ we find
\begin{align}
    \tilde{{\cal L}}= \dfrac{a^2}{2 }{\phi'}^2-\frac{a^2}{2}\left(\partial_i\phi\right)^2- \lambda a{\phi' }^3+ \frac{ i \hbar}{\Omega} \left(3 \lambda \,\dfrac{{\phi'}}{a}+ 9  \lambda ^2  \dfrac{{\phi'} ^2}{a^2}\right)+{\cal O}\left( \lambda^3,\frac{\hbar}{\Omega}^2\right)
\end{align}
where we have dropped a field independent term that just gets absorbed into the numerical constant $\mathcal{N}$ that we ignore anyway. The factor of $\Omega^{-1}$, the reciprocal of the space-time volume element may be written as the divergent distribution $\delta^4(0)$, so
\begin{align}
    \tilde{{\cal L}}= \dfrac{a^2}{2 }{\phi'}^2-\frac{a^2}{2}\left(\partial_i\phi\right)^2 -\lambda a{\phi' }^3+ i \hbar\delta^4(0) \left(3 \lambda \,\dfrac{{\phi'}}{a}+ 9  \lambda ^2  \dfrac{{\phi'} ^2}{a^2}\right)+{\cal O}\left( \lambda^3\right).
\label{eff-lagrangian:eq}
\end{align}
Note that the new interactions, in general, appear with a divergent factor of $\delta^4(0)$. In this sense, the wavefunction of the universe can be calculated in the so-called path integral formulation by only following the Feynman recipe
\begin{align}
    \langle \phi',\eta'| \phi,\eta \rangle = N \int_{\phi}^{\phi'} {\cal D}\phi ~ \exp  \left[ i \int d\eta\,d\x ~S(\phi(\x,\eta), {\phi'}(\x,\eta)) \right] +\, \delta^4(0)~ \mathrm{terms}.
\end{align}
We will show, in Section \ref{cancellation-div:sec}, that the $\delta^4(0)$ terms cancel with other divergent terms that arise due to the derivatives on the propagator.
\subsection{${\phi'}^3$~interaction: Trispectrum}\label{sec:WFUTri}
As a consistency check, let us compare the result of the wavefunctional approach for the trispectrum of $\lambda {\phi'}^3$ theory with that of the canonical approach. While it is more straightforward to directly use Eq.~\eqref{Psi-phi0:eq} to calculate the trispectrum, for pedagogical purposes, we continue with the diagrammatic approach introduced in Appendix \ref{WFU-Feynamn:sec}. First of all, using Eq.~\eqref{trispectrum-psi-n:eq}, we get
\begin{align}\nonumber
    &T(\kk_1,\kk_2,\kk_3,\kk_4)=2\prod_{a=1}^4 \phi_{k_a}(\eta_0)\phi_{k_a}^{\ast}(\eta_0)\,
    \\
    &\times\left[\mathrm{Re}\, \psi_4(\kk_1,\kk_2,\kk_3,\kk_4)- \phi_s(\eta_0)\phi^{\ast}_s(\eta_0)\,\mathrm{Re}\, \psi_3(\kk_1,\kk_2,\kk_{12})\,\mathrm{Re} \,\psi_3(\kk_3,\kk_4,\kk_{12})-t-u\right]
\end{align}
Now, using Feynman rules introduced there, we find
\begin{align}
\nonumber
    &\psi_4(\kk_1,\kk_2,\kk_3,\kk_4)
    \\
    \nonumber
    &=-\frac{9\,\lambda^2}{2}\iint_{-\infty}^0 d\eta d\eta' a(\eta)a(\eta')K_{k_1}'(\eta)K_{k_2}'(\eta)K_{k_3}'(\eta')K_{k_4}'(\eta')~\partial_{\eta\,\eta'}G(s;\eta,\eta')+23~\mathrm{perms.}
    \\
    &=-\frac{9\,\lambda^{2}}{2}\iint_{-\infty}^0 d\eta d\eta' a(\eta)a(\eta')\dfrac{\phi_{k_1}'(\eta)}{\phi_{k_1}(\eta_0)}\dfrac{\phi_{k_2}'(\eta)}{\phi_{k_2}(\eta_0)}\dfrac{\phi_{k_3}'(\eta')}{\phi_{k_3}(\eta_0)}\dfrac{\phi_{k_4}'(\eta')}{\phi_{k_4}(\eta_0)}~\partial_{\eta\,\eta'}G(s,\eta,\eta')+23~\mathrm{perms.}
\end{align}
where $s=|\kk_{12}|$. Similarly the $3$-point wavefunction coefficient is given by
\begin{align}
    \psi_3(\kk_1,\kk_2,\kk_3)&=-3!~i\,\lambda \int_{-\infty}^0 d\eta  a(\eta)~K'_{k_1}(\eta)K'_{k_2}(\eta)K'_{k_3}(\eta)
    \\
    &=-3!~i\lambda\int_{-\infty}^0 d\eta  a(\eta)\dfrac{\phi'_{k_1}(\eta)}{\phi_{k_1}(\eta_0)}\dfrac{\phi'_{k_2}(\eta)}{\phi_{k_2}(\eta_0)}\dfrac{\phi'_{k_3}(\eta)}{\phi_{k_3}(\eta_0)}
\end{align}
In this case the combinatorial factors appear somewhat differently, the factor $9\lambda^2$ in the 4-point function comes from the derivative of the Lagrangian. So $T$ is 
\begin{align}
\nonumber
    T&(\kk_1,\kk_2,\kk_3,\kk_4)= \mathcal{T}_4(\kk_1,\kk_2,\kk_3,\kk_4)+ \mathcal{T}_3(\kk_1,\kk_2,\kk_3,\kk_4)
\end{align}
where
\begin{align}
    \nonumber
    \mathcal{T}_4&(\kk_1,\kk_2,\kk_3,\kk_4)=
    -9\lambda^2\mathrm{Re}\bigg[\phi_{k_1}(\eta_0)\phi_{k_2}(\eta_0)\phi_{k_3}(\eta_0)\phi_{k_4}(\eta_0)\\&\qquad\times\iint_{-\infty}^{0}d\eta d\eta'\,a(\eta)a(\eta')\,\phi_{k_1}'^{\ast}(\eta)\phi_{k_2}'^{\ast}(\eta)\phi_{k_3}'^{\ast}(\eta')\phi_{k_4}'^{\ast}(\eta') ~\partial_{\eta\,\eta'}G(s,\eta,\eta') \bigg]+23~\mathrm{perms.}
\end{align}
and noting that $\mathrm{Re}(x)\,\mathrm{Re}(x)=x^2 +{x^{\ast}}^2+2 x x^{\ast}=2\mathrm{Re}(x^2+xx^*)$, we find
\begin{align}
    \nonumber
    &\mathcal{T}_3(\kk_1,\kk_2,\kk_3,\kk_4)= \dfrac{-9\lambda^2}{2} 2\mathrm{Re}
   \left[\frac{\phi^{\ast}_s(\eta_0)}{\phi_s(\eta_0)}\phi_{k_1}^{\ast}(\eta_0)\phi_{k_2}^{\ast}(\eta_0)\phi_{k_3}^{\ast}(\eta_0)\phi_{k_4}^{\ast}(\eta_0)\right.\\\nonumber&\qquad\times\int_{-\infty}^0 d\eta  a(\eta)\phi_{k_1}'(\eta)\phi_{k_2}'(\eta)\phi_{s}'(\eta)~\int_{-\infty}^0  d\eta'a(\eta')\phi_{k_3}'(\eta')\phi_{k_4}'(\eta')\phi_{s}'(\eta')
    \\\nonumber
    &-\phi_{k_1}(\eta_0)\phi_{k_2}(\eta_0)\phi^{\ast}_{k_3}(\eta_0)\phi^{\ast}_{k_4}(\eta_0)\\&\left.\qquad\times\int_{-\infty}^0 d\eta  a(\eta)\phi'^{\ast}_{k_1}(\eta)\phi'^{\ast}_{k_2}(\eta)\phi'^{\ast}_{s}(\eta)~\int_{-\infty}^0  d\eta'a(\eta')\phi'_{k_3}(\eta')\phi'_{k_4}(\eta')\phi'_{s}(\eta')
    \right]+23~\mathrm{perms.}
\end{align}
The reason for the extra factor $1/2$ in the above equation is a bit nuanced. Note that every $\psi_3$ comes in $6$ permutations. Combining two $\psi_3$ in $s,t,u$ channels equals to all permutations of $\kk_1,\kk_2,\kk_3$ and $\kk_4$ times a factor of $9/2$. Note that $3!\times 3!\times 3 = 9/2 \times 4!$. In this theory 
\begin{equation}
    G(s,\eta,\eta')=\Delta_F-\phi^*(\eta)\phi^*(\eta')\frac{\phi(\eta_0)}{\phi^{\ast}(\eta_0)}
\end{equation}
there is some cancellation between terms with $\phi_s/\phi^{\ast}_s$. Therefore,
\begin{align}
    \nonumber
    T&(\kk_1,\kk_2,\kk_3,\kk_4)=
    -9\lambda^2\mathrm{Re} \left[\phi_{k_1}(\eta_0)\phi_{k_2}(\eta_0)\phi_{k_3}(\eta_0)\phi_{k_4}(\eta_0)\right.
    \\\nonumber
    &\qquad\qquad\times\iint_{-\infty}^{0}d\eta d\eta'\,a(\eta)a(\eta')\,\phi_{k_1}'^{\ast}(\eta)\phi_{k_2}'^{\ast}(\eta)\phi_{k_3}'^{\ast}(\eta')\phi_{k_4}'^{\ast}(\eta') ~\partial_{\eta\,\eta'}\Delta_F(s,\eta,\eta') 
    \\
    \nonumber
    &-\phi_{k_1}(\eta_0)\phi_{k_2}(\eta_0)\phi^{\ast}_{k_3}(\eta_0)\phi^{\ast}_{k_4}(\eta_0) \\
    &\left.\qquad\times\int_{-\infty}^0 d\eta  a(\eta)\phi'^{\ast}_{k_1}(\eta)\phi'^{\ast}_{k_2}(\eta)\phi'^{\ast}_{s}(\eta)~\int_{-\infty}^0  d\eta'a(\eta')\phi'_{k_3}(\eta')\phi'_{k_4}(\eta')\phi'_{s}(\eta')
    \right]+23~\mathrm{perms.}
    \label{trispectrum:eq}
\end{align}
The propagator, $\Delta_F$, in Eq.~\eqref{trispectrum:eq} is defined in the same way as the standard Feynman propagator (except with different mode-functions)
\begin{equation}
    \Delta_F=\phi(\eta)\phi^*(\eta')\theta(\eta-\eta')+\theta(\eta'-\eta)\phi(\eta')\phi^*(\eta)
\end{equation}
Notice that here we encounter derivatives of the propagator whereas in the in-in formalism, Section \ref{in-in:sec}, we encountered time ordered products of derivatives of the fields. These two are related to each other according to
\begin{align}
    \partial_{\eta} \partial_{\eta'} \Delta_F(\eta,\eta')&= \langle0\rvert T (\phi'(\eta)\phi'(\eta'))\lvert 0 \rangle+ia^{-2}(\eta)\,\delta(\eta-\eta'),
    \label{total-prop:eq}
\end{align}
details in Appendix \ref{app:Propagators}. This delta function generates what looks like a contact interaction as it removes one of the time integrals so we can define
\begin{align}
    T(\kk_1,\kk_2,\kk_3,\kk_4)=T_{\mathrm{exch.}}(\kk_1,\kk_2,\kk_3,\kk_4)+T_{\mathrm{cont.}}(\kk_1,\kk_2,\kk_3,\kk_4)
\end{align}
where
\begin{align} 
    \nonumber
    T_{\mathrm{exch.}}&(\kk_1,\kk_2,\kk_3,\kk_4)=
    -9\lambda^2\mathrm{Re}\left[\vphantom{\int_\infty^0}\phi_{k_1}(\eta_0)\phi_{k_2}(\eta_0)\phi_{k_3}(\eta_0)\phi_{k_4}(\eta_0)\right.\\ \nonumber&~~~~\times\iint_{-\infty}^{0}d\eta d\eta'\,a(\eta)a(\eta')\,\phi_{k_1}'^{\ast}(\eta)\phi_{k_2}'^{\ast}(\eta)\phi_{k_3}'^{\ast}(\eta')\phi_{k_4}'^{\ast}(\eta') ~\langle 0\rvert T (\phi'(\eta)\phi'(\eta'))\lvert 0 \rangle
   \\\nonumber&-\phi_{k_1}(\eta_0)\phi_{k_2}(\eta_0)\phi^{\ast}_{k_3}(\eta_0)\phi^{\ast}_{k_4}(\eta_0)\,\\ &\left.~~~~\times\int_{-\infty}^0 d\eta a(\eta)\phi'^{\ast}_{k_1}(\eta)\phi'^{\ast}_{k_2}(\eta)\phi'^{\ast}_{s}(\eta) \int_{-\infty}^0 d\eta' d\eta'a(\eta')\phi'_{k_3}(\eta')\phi'_{k_4}(\eta')\phi'_{s}(\eta')
    \right]+23\textrm{ perms.}
    \end{align}
and
\begin{align}
\nonumber
    &T_{\mathrm{cont.}}(\kk_1,\kk_2,\kk_3,\kk_4)=
    \\
    &\mathrm{Re}\left[-9i\lambda^2\phi_{k_1}(\eta_0)\phi_{k_2}(\eta_0)\phi_{k_3}(\eta_0)\phi_{k_4}(\eta_0)\int_{-\infty}^0 d\eta a^2(\eta) \phi_{k_1}'^{\ast}(\eta)\phi_{k_2}'^{\ast}(\eta)\phi_{k_3}'^{\ast}(\eta)\phi_{k_4}'^{\ast}(\eta) +23 ~\mathrm{perms}\right]
\end{align}
By direct comparison with the results in Section \ref{in-in:sec} (\Cref{T-naive-can1:eq,T-naive-can3:eq,T-delta-can:eq}) it is straightforward to see that
\begin{tcolorbox}[ams align, colback=white, colframe=black]
    T_{\mathrm{exch.}}&=T_3,& T_{\mathrm{cont.}}&=T_4.
\end{tcolorbox}
\noindent Therefore, we can see that contributions that were generated by the three point interactions in the in-in formalism are the result of a combination of the three and four point wavefunction coefficients whilst the additional contact piece that is generated by the proper treatment of the Hamiltonian corresponds to a contribution that arises in the wavefunction method due to taking derivatives of the propagator and the two methods generate identical results, as expected. 

\subsection{Divergent Corrections: Position Space} \label{cancellation-div:sec}
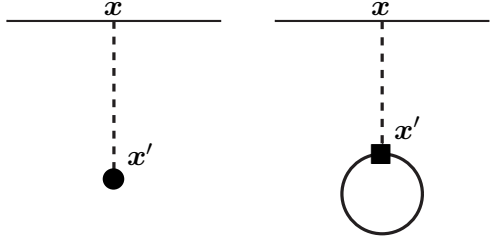
\begin{figure}[ht]
    \begin{picture}(170,90)(-150,-20)
\SetWidth{1.2}
    \Line[dash=True ](40,4)(40,60)
    %\Line[dash=True,arrow,arrowlength=7,arrowwidth=2](20,20)(20,60)
    %\Line[dash=True,arrow,arrowlength=7,arrowwidth=2](60,20)(60,60)
    %\Line[arrow,arrowlength=7,arrowwidth=2](40,0)(100,0)
    \Line[dash=True ](140,14)(140,60)
    \Arc[](140,-5)(15,0,360)
    %\Arc[dash=True,clock=False](40,20)(20,-90,0)
    \SetWidth{0.8}
    \Line[](0,60)(80,60)
    \Line[](100,60)(180,60)
    \SetColor{Black}
    \Text(40,0)[]{\huge $\bullet$}
    \Text(140,10)[]{$\blacksquare$}
    %\Text(10,25)[]{$\q_1$}
    \Text(40,65)[]{$\x$}
    \Text(150,20)[]{$\x'$}
    \Text(50,10)[]{$\x'$}
    \Text(140,65)[]{$\x$}
    %\Text(120,0)[]{$=i\lambda\, \q.(\q'-\q) $}
   \end{picture}
 \caption{Feynman diagrams associated with $\psi_1(\x)$}
\label{psi-1:diagram}
\end{figure}

Up to now, we omitted the divergent corrections to the effective interaction Lagrangian. Keeping them, we find, using the methodology explained in Appendix~\ref{app:PathIntegral} and \ref{WFU-Feynamn:sec},
\begin{align}
\nonumber
    &\Psi [\phi_0(\x)] \propto \\\nonumber&\exp\,\Bigg( -i\lambda\int d^4x\, \bigg[a\bar{\phi}'^3- 3ia\bar{\phi}'^2\left(\partial_{\eta}\dfrac{\delta}{\delta J}\right) -3a\bar{\phi}'  \left(\partial_{\eta}\dfrac{\delta}{\delta J}\right)^2 +ia\left( \partial_{\eta}\dfrac{\delta}{\delta J}\right)^3 
    +3i \lambda\delta^4(0) \dfrac{\bar{\phi}'}{a}\\
    \nonumber
    &+3\lambda\delta^4(0)\frac{1}{a}\left(\partial_{\eta}\frac{\delta}{\delta J}\right)+ 9i \lambda^2\delta^4(0) \dfrac{\bar{\phi}'^2}{a^2}+ 18 \lambda^2\delta^4(0) \dfrac{\bar{\phi}'}{a^2}\left(\partial_{\eta}\dfrac{\delta}{\delta J}\right)
    -9i\lambda^2\delta^4(0)\frac{1}{a^2}\left(\partial_{\eta}\frac{\delta}{\delta J}\right)^2\bigg]\Bigg)
    \\\nonumber
   &\pushright{\exp \left( \dfrac{-1}{2} \iint d^4x \,d^4 y~ G(x,y) J(x)J(y)  \right)\bigg \vert _{J=0}.}\\
    \end{align}
Where $\bar{\phi}$ is the background solution,
\begin{equation}
    \bar{\phi}(\x,\eta)=\int d^3x' K(\x,\x',\eta)\phi_0(\x').
\end{equation}
First we look at the terms that are linear in $\bar{\phi}$ within the exponential,
\begin{align}
\nonumber
    \psi_1(\x') &=3i\lambda\int d^4 x K'(\x',\x,\eta)\\\nonumber& \times\left[\dfrac{i \delta^4(0)}{a}
    -\int d\eta' \delta(\eta-\eta') a\, \partial_{\eta\eta'}\dfrac{\delta}{\delta J}\dfrac{\delta}{\delta J'} \right]
    \exp \left( \dfrac{-1}{2} \iint d^4y d^4 z~ G(y,z) J(y)J(z)  \right)\bigg \vert _{J=0}
   \\
  &=3i\lambda \int  d^4x  K'(\x',\x,\eta) \left[\dfrac{i \delta^4(0)}{a}
    +\int d\eta' \delta(\eta-\eta') a\, \partial_{\eta\eta'}G(\x,\eta',\x,\eta) \right]
    \end{align}
Where we introduced the additional integral over $\eta'$ so that the second derivative doesn't act on both $\frac{\delta}{\delta J}$'s. We can straightforwardly perform the integral to remove the delta function which leaves us with
\begin{equation}
    \left.\partial_{\eta\eta'}G(\textbf{x},\eta',\textbf{x},\eta)\right\rvert_{\eta'=\eta}\neq \partial_{\eta\eta} G(\textbf{x},\eta,\textbf{x},\eta).
\end{equation} 
Therefore, if we want to remove the derivatives on the Green's function it is necessary to perform integration by parts before performing the integral over $\eta'$ which will result in derivatives of the delta function. Such derivatives are typically dealt with by integrating by parts, thereby returning the original expression with derivatives on the Green's function and we therefore conclude that removing these derivatives in this way is not possible. Evaluating this derivative of the Green's function gives
\begin{align}\nonumber
    \partial_{\eta\eta'}G(x,x')&=ia^{-2}\delta^4(x-x')+\int d^3y\left(\phi'(\eta,\y) {\phi^*}'(\eta',\x-\x'-\y)\theta(\eta-\eta')\right.\\\label{eq:Gderiv}&\left.+{\phi^*}'(\eta,\y) \phi'(\eta',\x-\x'-\y)\theta(\eta'-\eta)-{\phi^*}'(\eta,\y){\phi^*}'(\eta',\x-\x'-\y)\right).
\end{align}
It is this delta function that will be important in this section so to simplify the discussion we introduce a label for the remaining term,
\begin{equation}\label{eq:Xidef}
    \partial_{\eta\eta'}G(x,x')=\Xi(x,x')+ia^{-2}\delta^4(x-x').
\end{equation}
This delta function cancels with the $\delta^4(0)$ that arises due to the new term in the effective action to give
\begin{align}
    \psi_1(\x')=-3i\lambda\int d^4xK'(\x',\x,\eta)a\,\Xi(x,x).
\end{align}
It is straightforward to see, from \cref{eq:Gderiv}, that $\Xi(x,x)$ does not depend on $\x$ and so the $\x$ integral gives 
\begin{align}\nonumber
    \psi_1(\x')=-3i\lambda\int d\eta a\,\Xi(x,x) \int d^3x K'(\x',\x,\eta)&=-3i\lambda\int d\eta a\,\Xi(x,x) \int d^3kK_k'(\eta)\delta^3(\kk)\\&=-3i\lambda\int d\eta a\,\Xi(x,x) K_0'(\eta)=0.
\end{align}
We can equivalently perform this calculation by organising these terms as Feynman diagrams in perturbation theory by including new vertices with divergent vertex factors.  In particular, up the third order in lambda, we must introduce three new vertices, shown in Fig.~\ref{div-vertices:fig}. For the sake of clarity, here, we distinguish divergent vertices showing them with bullets instead of the squares used for the standard vertices.
\begin{figure}[!ht]
\begin{align}
\nonumber
\begin{picture}(50,30)(0,0)
\SetWidth{1.2}
    \Line[dash=True ](0,4)(40,4)
    \Text(0,4)[]{\huge $\bullet$}
    \end{picture}
    = -3 \lambda\,\delta^4(0) \int a^{-1}(\eta) d^4x  
    \end{align}
\begin{align}
\nonumber
\begin{picture}(50,30)(0,0)
\SetWidth{1.2}
    \Line[ dash=True ](0,4)(40,4)
    \Line[ dash=True ](0,4)(-40,4)
    \Text(0,4)[]{\huge $\bullet$}
    \end{picture}
    = -9 \lambda^2\,\delta^4(0) \int a^{-2}(\eta) d^4x 
    \end{align}
    \begin{align}
\nonumber
\begin{picture}(50,30)(0,0)
\SetWidth{1.2}
    \Line[ dash=True ](0,4)(40,4)
    \Line[ dash=False ](0,4)(-40,4)
    \Text(0,4)[]{\huge $\bullet$}
    \end{picture}
    = -18 \lambda^2\,\delta^4(0) \int a^{-2}(\eta) d^4x  
\end{align}
\caption{Figure showing the divergent vertices in the wavefunction of the universe for the interaction $\lambda {\phi'}^3$ up to second order in $\lambda$}
\label{div-vertices:fig}
\end{figure}
For example, $\psi_1(x)$ gets contributions from two different diagrams depicted in Fig.~\ref{psi-1:diagram}. Using these Feynman rules we find an identical result to before,
\begin{align}
    \psi_1(\x)&=\psi^{\bullet}_1+\psi^{\rule{0.7ex}{0.7ex}}_1\\&=-3\lambda \int d^4 x'  K'(\x,\x',\eta')
    \left(\delta^4(0)a^{-1}(\eta') +i a(\eta') \left.\partial_{\eta\eta'} G(\textbf{x}',\eta,\textbf{x}',\eta')\right\rvert_{\eta=\eta'}  \right) \\&=-3i\lambda\int d^4x'K'(\x,\x',\eta')a\,\Xi(x',x')=0.
\end{align}

\begin{figure}[ht]
    \begin{picture}(170,90)(-150,-20)
\SetWidth{1.2}
    \Line[dash=True,clock=True](60,20)(60,60)
    \Line[dash=True,clock=True](20,20)(20,60)
    %%%%
    \Line[dash=True,clock=True](165,5)(165,60)
    \Line[dash=True,clock=True](115,5)(115,60)
    %%%%
    \Arc[dash=True,clock=True](40,20)(20,0,180)
    %%%%
    %\Arc[dash=True,clock=True](150,20)(20,-90,0)
    \Arc[dash=True,clock=False](125,5)(10,180,270)
    \Arc[dash=True,clock=False](155,5)(10,270,360)
    %%%
    %\Line[dash=True,arrow,arrowlength=7,arrowwidth=2](140,14)(140,60)
    \Arc[](140,-5)(15,0,180)
    \Arc[](140,-5)(15,180,360)
    %\Arc[dash=True,clock=False](40,20)(20,-90,0)
    \SetWidth{0.8}
    \Line[](0,60)(80,60)
    \Line[](100,60)(180,60)
    \SetColor{Black}
    \Text(40,0)[]{\huge $\bullet$}
    \Text(125,-5)[]{$\blacksquare$}
    \Text(155,-5)[]{$\blacksquare$}
    %\Text(10,25)[]{$\q_1$}
    \Text(60,65)[]{$\x_2$}
    \Text(20,65)[]{$\x_1$}
    \Text(165,65)[]{$\x_2$}
    \Text(115,65)[]{$\x_1$}
    \Text(40,10)[]{$\x$}
    \Text(115,-10)[]{$\x$}
    \Text(165,-10)[]{$\x'$}
    %\Text(120,0)[]{$=i\lambda\, \q.(\q'-\q) $}
   \end{picture}
 \caption{Feynman diagrams associated with $\psi_2(\x_1,\x_2)$}
\label{psi-2:diagram}
\end{figure}
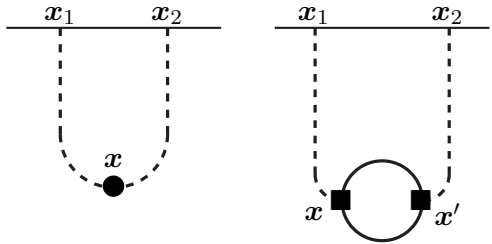

\noindent We can calculate $\psi_2$ in the same fashion. The associated Feynman diagrams appear in Fig.~\ref{psi-2:diagram} and the Feynman rules give
\begin{align}
    \nonumber
    \psi_2(\x_1,\x_2)&=\psi^{\bullet}_2+\psi^{\rule{0.7ex}{0.7ex}}_2= 
    -9 \lambda^2\,\delta^4(0)\int d^4x \,a^{-2}(\eta) K'(\x_1,\x,\eta) K'(\x_2,\x,\eta) 
    \\
    \nonumber
    &+(3i\lambda)^2 \int d^4x d^4x'a(\eta)a(\eta') \,\partial_{\eta\eta'} G(x,x')\,\partial_{\eta\eta'} G(x',x) K'(\x_1,\x,\eta) K'(\x_2,\x',\eta') 
    \\\nonumber
    &=-9\lambda^2\left(\int d^4xd^4x' a(\eta)a(\eta')\Xi(x,x')\Xi(x',x)K'(\x_1,\x,\eta)K'(\x_2,\x',\eta')\right.\\&\pushright{\left.+2i\int d^4x\, \Xi(x,x)K'(\x_1,\x,\eta)K'(\x_2,\x,\eta)\right),\quad\quad}
\end{align}
where once again the divergent vertex contribution has canceled and we are left only with $\Xi(x,x')$ which will, as in the non-loop case, return the same expression as the in-in calculation. 

A brief comment about these loops is in order. Due to the calculation of these results in real as apposed to momentum space the loop in the diagram does not introduce an additional integral, this is exactly the way that loops function in flat space amplitude calculations but may be unfamiliar to readers who have only performed calculations in momentum space. It is non-trivial and exciting that the divergent terms in the effective Hamiltonian density necessary to cancel divergent terms in the propagator arise automatically. The same cancellation is seen but less known in the standard QFT \cite{Gerstein:1971fm,Weinberg:1995mt}.

This cancellation appears identically in momentum space, and also extends to arbitrarily complex theories as long as we restrict ourselves to single loops, we show this in App.~\ref{divergentmomentum:sec} along side a complete expansion of all the terms in the path integral for the theory discussed here as an explicit demonstration of the procedure defined in App.~\ref{app:PathIntegral} in order to illuminate the process and directly connect it to the Feynmann diagram calculation.

\subsection{Other interactions}
Up until this point, we have only considered single derivative interactions of a single field, but two additional cases deserve comment. The first of these are interactions involving more than one field. In flat space, we can integrate them by parts and use the equations of motion to move all derivatives onto one of the fields. However, as was noted in Section \ref{cancellation-div:sec}, when a propagator joins one vertex to itself, it is not possible to integrate the expression by parts to remove the derivatives from Green's function. Similarly, where we have a loop that joins two vertices together, attempting to integrate the expression by parts will move the derivatives over one of the Green's functions to the other, and so the source of the divergence will appear different in this case, but if we are considering the same diagram, then the divergence will be the same. We could equivalently have integrated the action by parts to change the vertices that appear. However, this will generally introduce boundary terms to the action, complicating the calculation, so we do not consider this approach here. The expectation is that, even though the diagrams that contribute to the final result will be different, the final cancellation of these divergences will remain unchanged. 

The second extension that we must consider are interactions with more than one derivative acting on a single field which was identified in \cite{Behbahani_2014} to be important in the effective field theory of inflation\footnote{In this paper, they integrate the interaction terms by parts to show that there are a limited number of shapes that this interaction can take, but one should note that integration by parts in time may introduce non-vanishing boundary terms.}. For such terms, the calculation of the interaction Hamiltonian for the in-in formalism has been well-studied and requires introducing additional momenta for each higher derivative \cite{ostrogradsky1850memoire,woodard2015theorem}. However, this approach is problematic in the wavefunction of the universe method. That is because the contribution from the interactions does not cancel to the lowest order in the perturbations, so it introduces additional path integrals. Therefore, we leave such considerations to future work. An alternative approach to this problem, as considered in \cite{barua1977canonical}, is to eliminate these higher derivative interactions using a field redefinition. To linear order in perturbation theory, we could do this simply by using the equations of motion; however, this is not possible beyond linear order, \cite{criado2019field}, and so it is necessary to find an appropriate field redefinition, for example, to second-order the Lagrangian is
\begin{equation}
    \mathcal{L}\left[\phi+\lambda f\left(\phi^{(n)}\right)\right]=\mathcal{L}[\phi]+\lambda f\sum_i (-1)^i \frac{d}{d\eta^i}\frac{\delta \mathcal{L}}{\delta \phi^{(i)}}+\lambda^2\sum_{i,j}f^{(i)}f^{(j)}\frac{\delta^2 \mathcal{L}}{\delta \phi^{(i)}\delta \phi^{(j)}}
\end{equation}
Where it is necessary to integrate the first term by parts, this is ok as the boundary terms this produces are assumed to vanish when deriving the equations of motion. By fixing the terms in $f$, it should be possible to ensure that this term does not have any derivatives higher than 1. As an illustrative example, we do this for the theory with interaction Lagrangian $\lambda {\phi''}^3$ in Appendix \ref{app:HigherDerivatives}. It appears to be possible for all theories involving higher derivatives, although the authors are not aware of proof of this fact, and such a proof is beyond the scope of this work. Having eliminated these higher derivatives, it is possible to proceed with either the wavefunction of the universe or in-in method to calculate correlators to arbitrary order in perturbation theory.      

\section{Conclusions and Outlook}
In this paper, we presented a systematic prescription for calculating cosmological correlation functions for models with derivative coupling. We elaborated on the calculation of the wavefunction of the universe and compared the resulting correlators with those calculated using the “in-in” formalism. In order to do this we first performed the path integral over conjugate momentum to find the effective Lagrangian necessary for the wavefunction of the universe calculation. We then rigorously showed that the effective Lagrangian is the same as the original Lagrangian apart from some terms that generate divergent vertices which cancel with particular divergences in the loop integrals. 

As a specific example, we calculated the trispectra of the scalar fluctuation in the model with a $\lambda {\phi'}^3$ derivative coupling. We showed that the results found using the path-integral method are entirely consistent with the results from the canonical approach even including the divergent terms because the loop divergences that they cancel similarly don't appear in the canonical approach due to differences in the behaviour of the propagator. These results generalise to any one loop diagrams but extensions to higher loop diagrams (which are expected to cancel with the higher order in $\hbar$ terms in the expansion of the effective action) are left to future work. It is also possible to generalise to theories involving higher derivatives but this relies on being able to eliminate these derivatives to a given order using an appropriate field redefinition. 

In order to perform this comparison it was necessary to develop, for the first time, the "off-shell" version of the in-in formalism which was previously only known on-shell. This development allows for more direct comparisons with off-shell flat space amplitude calculations, a potential avenue for future investigation.

\appendix
\section{Path Integral Formulation: Transition Probability Amplitude}\label{app:PathIntegral}
Of particular interest is to find the transition probability between vacuum state at $\eta=-\infty$ to any arbitrary state at $\eta=\eta_0$. As we have established, this is given, after performing the path integral over the conjugate momentum, by
\begin{align}
    \Psi [\phi_0(\x)] =\langle \phi(\x,\eta_0)|\Omega \rangle=\int^{\phi_0} {\cal D} \phi~ \exp \left( \frac{i}{\hbar} \tilde{S}[\phi] \right) .
    \label{Transition-amp:eq}
\end{align}
Where,
\begin{equation}
    \tilde{S}=S_0+S_{\text{int.}}+S_{\text{div.}}=S_0+\tilde{S}_{\text{int.}},
\end{equation}
and $S_0$ is the quadratic action. As is customary in the path integral formulation of quantum field theory, we  perform this calculation perturbatively in powers of coupling constant. We write 
\begin{align}
    \phi(x)=\bar{\phi}(x)+\varphi(x)
\end{align}
where $\bar{\phi}(x)$ is the solution of the unperturbed equations of motion, 
\begin{align} \label{equation-of-motion:eq}
    \frac{\delta \mathcal{L}_0}{\delta \bar{\phi}(x)}-\frac{d}{d\eta}\frac{\delta \mathcal{L}_0}{\delta \bar{\phi}'(x)}=\mathcal{O}_{\x}\bar{\phi}(x)=0,
\end{align}
subject to the boundary conditions,
\begin{align}
       \bar{\phi}(\eta_0,\x)&=\phi_0(\x),
        \\
        \lim_{\eta\rightarrow -\infty(1-i\epsilon)}\bar{\phi}(\eta ,\x)&= 0.
\end{align}
This solution can be expressed as
\begin{align}
    \bar{\phi}(\eta,\x)=\int d^3x' K(\x,\eta,\x',\eta_0)\phi_0(\x')
\end{align}
where, the so called "bulk-to-boundary" propagator satisfies
\begin{align}
    \mathcal{O}_{\x}
       K(\x,\eta,\x',\eta_0)=0,
\end{align}        
subjected to the following boundary conditions
\begin{align}
        K(\x,\eta_0,\x',\eta_0)&=\delta^3(\x-\x'),
        \\
        \lim_{\eta\rightarrow -\infty(1-i\epsilon)}K(\x,\eta,\x',\eta_0)&= 0.
\end{align}
Equivalently, in Fourier space this convolution becomes a product,
\begin{align}
    \bar{\phi}(\kk,\eta) = K_{k}(\eta,\eta_0)\phi_{\kk}(\eta_0) 
\end{align}
where
\begin{align}
    K_k(\eta_0,\eta_0)&=1,\\
    \lim_{\eta\rightarrow -\infty(1-i\epsilon)}K_k(\eta,\eta_0)&=0.
\end{align}
For a massless scalar field minimally coupled to gravity in dS, this differential equation is
\begin{equation}
    a^2\bar{\phi}_k''+2a'a\bar{\phi}_k'+a^2k^2\bar{\phi}_k=0.
\end{equation}
So, the bulk to boundary propagator is given by
    \begin{align}
    \label{bu-to-bo-prop:eq}
        K_k(\eta,\eta_0) =\dfrac{\phi^{\ast}_{k}(\eta)}{\phi^{\ast}_{k}(\eta_0)}= \dfrac{i+k \eta}{i+k \eta_0} e^{i k (\eta-\eta_0)}.
    \end{align}
where $\phi_{k}\propto (i-k\eta)e^{-ik\eta}$ is the mode function that coincides with the Bunch-Davies vacuum for $\eta \to -\infty$. In terms of this propagator, after integrating the quadratic piece by parts, the effective action is given by
\begin{align}
    \bar{S}=\frac{1}{2}\int \frac{d^3k}{(2\pi)^3} a^2(\eta_0)K'_k(\eta_0){\phi}_{\kk}(\eta_0){\phi}_{-\kk}(\eta_0)-\frac{1}{2}\int  \frac{d^3k}{(2\pi)^3}d\eta\,\varphi_{-\kk}\mathcal{O}_k\varphi_{\kk}+\tilde{S}_{\text{int.}}\left[\bar{\phi}+\varphi\right]
\end{align}
whilst the path integrand is $\mathcal{D}\varphi$. To perform this path integral we follow the standard procedure and define the generating functional,
\begin{align}\label{eq:genfun}
    Z_0[J]=\int \mathcal{D}\varphi \exp\left[-\frac{i}{2\hbar}\int \frac{d^3k}{(2\pi)^3}d\eta  \,\varphi_{-\kk} \mathcal{O}_k\varphi_{\kk}+ 2J_{\kk}\varphi_{-\kk}\right].
\end{align}
To do this path integral we must complete the square, to to this we introduce 
\begin{align}
    \tilde{\varphi}_{\kk}=\varphi_{\kk}+\mathcal{O}_{\kk}^{-1}J_{\kk}
\end{align}
and require that this new term has the same boundary conditions, i.e.
\begin{align}
    \mathcal{O}_k^{-1}J_{\kk}(\eta_0) &=0\\
    \lim_{\eta\rightarrow -\infty}\mathcal{O}_k^{-1}J_{\kk}(\eta)&=0.
\end{align}
These boundary conditions allow us to integrate these terms by parts and so the generating functional can be written as
\begin{align}
    Z_0[J]&=\int \mathcal{D}\tilde{\varphi} \exp\left[-\frac{i}{2\hbar}\int \frac{d^3k}{(2\pi)^3}d\eta  \,\tilde{\varphi}_{-\kk}\mathcal{O}_k \tilde{\varphi}_{\kk} -J_{\kk}\mathcal{O}_k^{-1}J_{-\kk}\right]
\end{align}
The inverse differential operator, $\mathcal{O}_k^{-1}J_{\kk}$ can be defined formally through the Green's Function, $G_k$, which satisfies, 
\begin{align}\label{Green-Function:eq}
    \mathcal{O}_{k}G_k(\eta,\eta')&=-i\delta(\eta-\eta')\\\label{eq:Gbc1}
    G_k(\eta_0,\eta')&=0,\\\label{eq:Gbc2}
    \lim_{\eta\rightarrow -\infty(1-i\epsilon)}G_k(\eta,\eta')&=0.
\end{align}
This Green's Function is usually called the "bulk-to-bulk" propagator due to is role in the construction of Feynman diagrams. The inverse differential operator is then
\begin{equation}
    \mathcal{O}_{k}^{-1}J_{\kk}=i\int d\eta' G_k(\eta,\eta')J_{\kk}(\eta')
\end{equation}
To perform calculations it is necessary to find this propagator. To start with we integrate Eq.~\eqref{Green-Function:eq} over the range $\eta \in [\eta'-\epsilon,\eta'+\epsilon]$, to deduce the following junction conditions,
\begin{align}\nonumber
    &\lim_{\epsilon \rightarrow0}~G_k(\eta'-\epsilon,\eta')- G_k(\eta'+\epsilon,\eta')=0
    \\
   &\lim_{\epsilon \rightarrow0}~ \partial_{\eta}G_k(\eta'+\epsilon,\eta')- \partial_{\eta}G_k(\eta'-\epsilon,\eta')=-i a^{-2}(\eta')
   \label{junction-condition:eq}
\end{align}
In general, the homogeneous equation Eq.~\eqref{equation-of-motion:eq} has two independent solutions, $\phi_{\pm}$ that tend to the usual positive and negative solutions for $\eta \rightarrow - \infty$. In terms of homogeneous solutions, we have
\begin{align}
G_k(\eta,\eta') = \begin{cases}
        A^{-}(\eta')\phi_{+}(\eta)+ B^{-}(\eta')\phi_{-}(\eta), & \text{for } \eta < \eta'\\
        \\
        A^{+}(\eta')\phi_{+}(\eta)+ B^{+}(\eta')\phi_{-}(\eta), & \text{for } \eta > \eta'.
        \end{cases} 
\end{align}
The junction conditions Eq.~\eqref{junction-condition:eq}, imply that
\begin{align}
\Delta A(\eta')=\dfrac{+i\,a^{-2}(\eta')\,\phi_{-}(\eta')}{W(\eta')},\qquad 
\Delta B(\eta')=\dfrac{-i\,a^{-2}(\eta')\,\phi_{+}(\eta')}{W(\eta')}
\end{align}
where $W(\eta')$ denotes the Wronskinan of the homogeneous solutions of Eq.~\eqref{equation-of-motion:eq}, namely $\phi_{\pm}$ and $\Delta A (\eta')=A^{+}(\eta')-A^{-}(\eta')$ and likewise $\Delta B(\eta')=B^{+}(\eta')-B^{-}(\eta')$. Besides, one finds that the Wronskian of Eq.~\eqref{equation-of-motion:eq} is  
\begin{align}
     W(\eta) =\phi_{+}(\eta) \phi'_{-}(\eta)- \phi_{-}(\eta) \phi'_{+}(\eta)= -i H^2 \eta^2 =-i a^{-2}(\eta).
\end{align}
By imposing the boundary conditions in \cref{eq:Gbc1,eq:Gbc2} we get
\begin{align}
A^{-}(\eta')\phi_{+}(-\infty(1-i\epsilon))+B^{-}(\eta')\phi_{-}(-\infty(1-i\epsilon))=0
\\
A^{+}(\eta')\phi_{+}(\eta_0)+B^{+}(\eta')\phi_{-}(\eta_0)=0
\end{align}
We take $\phi_{-}=\phi_{k}(\eta)$ to be the mode function that coincides with Bunch-Davies vacuum at $\eta \to -\infty$ limit. Putting this all together, after some simple algebra, we get
\begin{tcolorbox}[ams align,colback=white, colframe=black]\label{bulk-to-bulk-propagator:eq}
    G_k(\eta,\eta') =  \phi_k(\eta) \phi_k^{\ast}(\eta')\theta(\eta-\eta')+\phi_k^{\ast}(\eta) \phi_k(\eta')\theta(\eta'-\eta)- \phi_k^{\ast}(\eta) \phi_k^{\ast}(\eta')\dfrac{\phi_k(\eta_0)}{\phi_k^{\ast}(\eta_0)} 
\end{tcolorbox}    
\noindent In terms of the bulk-to-bulk propagator, the generating function is given by
\begin{align}\nonumber
    Z_0[J]&=\exp\left[\frac{-1}{2\hbar}\int \frac{d^3k}{(2\pi)^3} d\eta d\eta'  J_{\kk}(\eta)G_k(\eta,\eta') J_{-\kk}(\eta')\right]\int \mathcal{D}\tilde{\varphi} \exp\left[\frac{-i}{2\hbar}\int \frac{d^3k}{(2\pi)^3} d\eta  \tilde{\varphi}_{-\kk}\mathcal{O}_k \tilde{\varphi}_{\kk}\right]\\&=\mathcal{N}\exp\left[\frac{-1}{2\hbar}\int \frac{d^3k}{(2\pi)^3} d\eta d\eta'  J_{\kk}(\eta)G_k(\eta,\eta') J_{-\kk}(\eta')\right].
    \label{eq:gengauss}
\end{align}
Where $\mathcal{N}$ is the result of the integral over $\bar{\varphi}$ and is just some constant. Now consider that, from Eq.~\eqref{eq:genfun}, we have 
\begin{align}\label{eq:dZ1}
    i(2\pi)^3\hbar \frac{d}{dJ_{\textbf{q}}(\eta')}Z_0[J]&= \int \mathcal{D}\varphi \varphi_{-\textbf{q}}(\eta')\exp\left[-\frac{i}{2\hbar}\int \frac{d^3k}{(2\pi)^3} d\eta  \varphi_{-\kk} \mathcal{O}_k\varphi_{\kk}+ 2J_{\kk}\varphi_{-\kk}\right]\\\label{eq:dZ2}
    i(2\pi)^3\hbar \frac{d}{d\eta'}\frac{d}{dJ_{\textbf{q}}(\eta')}Z_0[J]&= \int \mathcal{D}\varphi \varphi_{-\textbf{q}}'(\eta')\exp\left[-\frac{i}{2\hbar}\int \frac{d^3k}{(2\pi)^3} d\eta  \varphi_{-\kk} \mathcal{O}_k\varphi_{\kk}+ 2J_{\kk}\varphi_{-\kk}\right]
\end{align}
We now return to the wavefunction, Eq.~\eqref{Transition-amp:eq}, which can be written as
\begin{align}
    \Psi\left[\phi_{\kk}(\eta_0)\right]=e^{\frac{i}{\hbar}S_B}\int  \mathcal{D}\varphi \,\exp\left[\frac{i}{\hbar}\tilde{S}_{\text{int.}}[\phi_0+\varphi]\right]\exp\left[-\frac{i}{2\hbar}\int \frac{d^3k}{(2\pi)^3} d\eta  \varphi_{-\kk}\mathcal{O}_k\varphi_{\kk}\right],
\end{align}
where
\begin{align}
    S_B=\frac{1}{2}\int \frac{d^3k}{(2\pi)^3}a^2(\eta_0)K_k'(\eta_0)\phi_{\kk}(\eta_0)\phi_{-\kk}(\eta_0).
\end{align}
From \cref{eq:dZ1,eq:dZ2} we can express this in terms of $Z_0[J]$,
\begin{align}
    \Psi\left[\phi_{\kk}(\eta_0)\right]=\left.e^{\frac{i}{\hbar}S_B} \exp\left[\frac{i}{\hbar}\tilde{S}_{\text{int.}}\left[\phi_0+i(2\pi)^3\hbar\frac{d}{dJ}\right]\right]Z_0[J]\right\rvert_{J=0}.
\end{align}
Just as in flat space we can work with $\log Z[J]$ instead of $Z[J]$ to remove disconnected diagrams from the expression so we have
\begin{align}\label{Psi-phi0:eq}
    \Psi^{\text{con.}}\left[\phi_{\kk}(\eta_0)\right]=\left.e^{\frac{i}{\hbar}S_B} \exp\left[\frac{i}{\hbar}\tilde{S}_{\text{int.}}\left[\phi_0+i(2\pi)^3\hbar\frac{d}{dJ}\right]\right]\log Z_0[J]\right\rvert_{J=0}.
\end{align}
\section{${\phi'}^3$ interaction: Feynman Rules}
\label{WFU-Feynamn:sec}
It is almost straightforward to apply the machinery developed in the previous section to a theory with kinetic coupling. In particular, we assume the following interaction 
\begin{align}
\tilde{S} \equiv S_{\mathrm{free.}}+S_{\mathrm{int.}}=&- \int d \eta\, d^3x \,a^2(\eta)  \left(\dfrac{1}{2}(\partial \phi )^2 +\dfrac{m^2\,\phi^2}{2 \eta^2 H^2}\right) - \lambda \int d \eta\, d^3x\,a(\eta)  \,{\phi'}^3
\label{full-action2:eq}
\end{align}
The interaction action $S_{\mathrm{int.}}$ can be expanded as
\begin{align}
\nonumber
   &{S}_{\mathrm{int.}}\big[\bar{\phi}-i\frac{\delta}{\delta J}\big] =
   \\
   &-\lambda \int d\eta a(\eta)\, d\x \left[\bar{\phi}'^3+3\, \bar{\phi}'^2  \left(-i\,\partial_{\eta}\dfrac{\delta}{\delta J}\right) +3\,\bar{\phi}' \left(-i\, \partial_{\eta}\dfrac{\delta}{\delta J}\right)^2 +\left(-i\, \partial_{\eta}\dfrac{\delta}{\delta J}\right)^3 \right]
   \label{S-total:eq}
\end{align}
Hence, the wavefunction of the universe can be calculated perturbatively using Eq.~\eqref{Psi-phi0:eq} as
\begin{align}
\nonumber
    &\Psi [\phi_0(\x)] \propto \\\nonumber&\exp\,\left(-i\lambda \int d\eta a(\eta)\, d\x \left[\bar{\phi}'^3+3\, \bar{\phi}'^2  \left(-i\,\partial_{\eta}\dfrac{\delta}{\delta J}\right) +3\,\bar{\phi}' \left(-i\, \partial_{\eta}\dfrac{\delta}{\delta J}\right)^2 +\left(-i\, \partial_{\eta}\dfrac{\delta}{\delta J}\right)^3 \right] \right)
    \\\nonumber
   &\pushright{\exp \left( \dfrac{-1}{2} \iint d^4x \,d^4 y~ G(x,y) J(x)J(y)  \right)\bigg \vert _{J=0}.}\\
    \end{align}
The wavefunction of the universe in the momentum space can be written as the following expansion
\begin{align}
    \Psi[\phi_0(\boldsymbol{k})]=\exp\left( -\sum_{n=2}\dfrac{1}{n!} \int \prod_{i=1}^n \dfrac{d\q_i}{(2\pi)^3}\, \psi_n(\q_1,..\q_n) \phi_{\q_1}(\eta_0)..\phi_{\q_n}(\eta_0)\right).
\end{align}
In a similar fashion as in flat-space QFT, using the above definition and Eq.~\eqref{Psi-phi0:eq}, the Feynman rules, in Fourier space, are
\begin{itemize}
    \item[$\bullet$]
    \emph{General Structure:} For calculating $\psi_n$ draw a horizontal line-- which corresponds to $\eta_0$ \emph{boundary} hyper-surface. Point $n$ sites on it. Attach a dashed line that denotes a bulk-to-boundary propagator. Now try to glue these lines by appropriate vertices. Perhaps you may need to insert some bulk-to-bulk propagators to do this. Finally, you must take into account all possible permutations of the dashed lines to the $n$-sites.  
    
    \item[$\bullet$]
    \emph{Vertices:}
There are four different kinds of vertices depending on what kind of outgoing lines are attached to them
\begin{align}
\nonumber
\begin{picture}(170,100)(0,-60)
\SetWidth{1.2}
    \Line[arrow,arrowlength=7,arrowwidth=2](40,4)(40,40)
    \Line[arrow,arrowlength=7,arrowwidth=2](40,0)(0,0)
    \Line[arrow,arrowlength=7,arrowwidth=2](40,0)(80,0)
    \SetColor{Black}
    \Text(40,0)[]{$\blacksquare$}
    %\Text(100,-10)[]{$\eta'$}
    \Text(125,0)[]{$=-i\lambda\displaystyle \displaystyle\int d \eta\, a(\eta)$}
\end{picture}
%%%%%%%%%%%%
\begin{picture}(170,80)(-20,-60)
\SetWidth{1.2}
    \Line[arrow,arrowlength=7,arrowwidth=2](40,4)(40,40)
    \Line[arrow,arrowlength=7,arrowwidth=2](40,0)(0,0)
    \Line[dash=True,arrow,arrowlength=7,arrowwidth=2](40,0)(80,0)
    \SetColor{Black}
    \Text(40,0)[]{$\blacksquare$}
    \Text(125,0)[]{$=-3i\lambda\displaystyle \displaystyle\int d \eta\, a(\eta)$}
    %\Text(100,-10)[]{$\eta'$}
    %\Text(120,0)[]{$=i\lambda\, \q.(\q'-\q) $}
\end{picture}
%%%%%%%%%%%%%%%%%%%%%%%%
\\
\nonumber
\begin{picture}(170,40)(0,-20)
\SetWidth{1.2}
    \Line[dash=True,arrow,arrowlength=7,arrowwidth=2](40,4)(40,40)
    \Line[dash=True,arrow,arrowlength=7,arrowwidth=2](40,0)(0,0)
    \Line[arrow,arrowlength=7,arrowwidth=2](40,0)(80,0)
    \SetColor{Black}
    \Text(40,0)[]{$\blacksquare$}
    \Text(125,0)[]{$=-3i\lambda\displaystyle \displaystyle\int d \eta\, a(\eta)$}
    %\Text(100,-10)[]{$\eta'$}
    %\Text(120,0)[]{$=i\lambda\, \q.(\q'-\q) $}
\end{picture}
%%%%%%%%%%%%
\begin{picture}(170,40)(-20,-20)
\SetWidth{1.2}
    \Line[dash=True,arrow,arrowlength=7,arrowwidth=2](40,4)(40,40)
    \Line[dash=True,arrow,arrowlength=7,arrowwidth=2](40,0)(0,0)
    \Line[dash=True,arrow,arrowlength=7,arrowwidth=2](40,0)(80,0)
    \SetColor{Black}
    \Text(40,0)[]{$\blacksquare$}
    \Text(125,0)[]{$=-i\lambda\displaystyle \displaystyle\int d \eta\, a(\eta)$}
    %\Text(100,-10)[]{$\eta'$}
   % \Text(120,0)[]{$=i\lambda\, \q.(\q'-\q) $}
\end{picture}
%\caption{Feynman diagram/rules associated with the Lagrangian of Eq. \eqref{full-action:eq}}
\end{align}
The dashed and solid lines correspond to the bulk to boundary propagators and bulk to bulk propagators, respectively. Moreover, for every vertex total momentum must be conserved, so we add a $\delta^3(\kk_{\mathrm{tot.}})$.    
    
    \item[$\bullet$]
    \emph{Propagators:} There is a subtlety here. It must be noted that, for the derivative coupling interaction, for each vertex, we have a time derivative to vertex time acting on each propagator. In this sense,  each bulk-to-boundary propagator appears with a time derivative, and each bulk-to-bulk propagator appears with two time-derivatives
\begin{align}
    \nonumber
    \begin{picture}(170,40)(0,-20)
\SetWidth{1.2}
    %\Line[dash=True,arrow,arrowlength=7,arrowwidth=2](40,4)(40,40)
    %\Line[dash=True,arrow,arrowlength=7,arrowwidth=2](40,0)(0,0)
    \Line[dash=True,arrow,arrowlength=7,arrowwidth=2](40,0)(100,0)
    \SetWidth{0.9}
    \Line[](100,-10)(100,10)
    \SetWidth{1.2}
    \SetColor{Black}
    \Text(40,0)[]{$\blacksquare$}
    \Text(145,0)[]{$=\partial_{\eta} K(\eta,\eta_0) $}
    \end{picture}
    \\
    \nonumber
    \begin{picture}(170,40)(0,-20)
\SetWidth{1.2}
    %\Line[dash=True,arrow,arrowlength=7,arrowwidth=2](40,4)(40,40)
    %\Line[dash=True,arrow,arrowlength=7,arrowwidth=2](40,0)(0,0)
    \Line[arrow,arrowlength=7,arrowwidth=2](40,0)(100,0)
    \SetColor{Black}
    \Text(40,0)[]{$\blacksquare$}
    \Text(100,0)[]{$\blacksquare$}
    \Text(145,0)[]{$=\partial_{\eta} \partial_{\eta'} G(\eta,\eta') $}
    \end{picture}
\end{align}
\item[$\bullet$]
\emph{Symmetry Factors:} These are calculated just as in flat space. A possible set of rules, \cite{dong2010symmetry}, for the symmetry factor is
\begin{enumerate}
    \item When a propagator starts and ends on the same vertex $\times 2$
    \item When a pair of vertices is connected by $k$ identical propagators $\times k!$
    \item When a vertext can be permuted without affecting the diagram $\times \textrm{permutations}$
    \item For each "double bubble" (a figure of eight diagram) $\times 2$
\end{enumerate}
The integral must then be divided by the resulting number.

\end{itemize}
 
\subsection{Wavefunction of the Universe: $\lambda {\phi'}^3$}

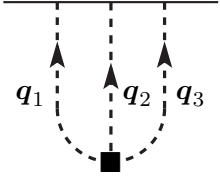
\begin{figure}[ht]
    \begin{picture}(170,90)(-150,-20)
\SetWidth{1.2}
    \Line[dash=True,arrow,arrowlength=7,arrowwidth=2](40,4)(40,60)
    \Line[dash=True,arrow,arrowlength=7,arrowwidth=2](20,20)(20,60)
    \Line[dash=True,arrow,arrowlength=7,arrowwidth=2](60,20)(60,60)
    %\Line[arrow,arrowlength=7,arrowwidth=2](40,0)(100,0)
    \Arc[dash=True,clock=True](40,20)(20,-90,-180)
    \Arc[dash=True,clock=False](40,20)(20,-90,0)
    \SetWidth{0.8}
    \Line[](0,60)(80,60)
    \SetColor{Black}
    \Text(40,0)[]{$\blacksquare$}
    \Text(10,25)[]{$\q_1$}
    \Text(50,25)[]{$\q_2$}
    \Text(70,25)[]{$\q_3$}
    %\Text(120,0)[]{$=i\lambda\, \q.(\q'-\q) $}
   \end{picture}
%\caption{Feynman diagram of $\Psi^{(1)}[\phi_0]$}
 \caption{Feynam diagrams associated with $\psi_3(\q_1,\q_2,\q_3)$}
\label{psi-3:diagram}
\end{figure}

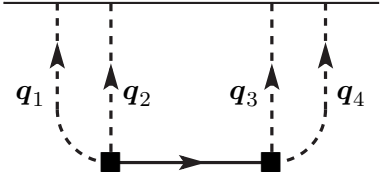
\begin{figure}[ht]
    \begin{picture}(170,90)(-150,-20)
    \SetWidth{1.2}
   \Line[dash=True,arrow,arrowlength=7,arrowwidth=2](40,4)(40,60)
    \Line[dash=True,arrow,arrowlength=7,arrowwidth=2](20,20)(20,60)
    \Line[dash=True,arrow,arrowlength=7,arrowwidth=2](100,4)(100,60)
    \Line[dash=True,arrow,arrowlength=7,arrowwidth=2](120,20)(120,60)
    \Line[arrow,arrowlength=7,arrowwidth=2](40,0)(100,0)
    \Arc[dash=True,clock=True](40,20)(20,-90,-180)
    \Arc[dash=True,clock=False](100,20)(20,-90,0)
    \SetWidth{0.8}
    \Line[](0,60)(140,60)
    \SetColor{Black}
    \Text(40,0)[]{$\blacksquare$}
    \Text(100,0)[]{$\blacksquare$}
    %\Text(33,-2.5)[]{$\bullet$}
    %\Text(107,-2.5)[]{$\bullet$}
    \Text(10,25)[]{$\q_1$}
    \Text(50,25)[]{$\q_2$}
    \Text(90,25)[]{$\q_3$}
    \Text(130,25)[]{$\q_4$}
   % \Text(70,-15)[]{(a)}
    %\Text(120,0)[]{$=i\lambda\, \q.(\q'-\q) $}
   \end{picture}
%\caption{Feynman diagram of $\Psi^{(1)}[\phi_0]$}
 \caption{Feynman diagram for the 4-point exchange diagram from the interaction $\lambda {\phi'}^3$}
\label{psi-4:diagram}
\end{figure}

Having found the Feynman rules, the 3rd order wavefunction coefficients is read from Fig.~\ref{psi-3:diagram},
\begin{align}
    \psi_3(\q_1,\q_2,\q_3)=-i\lambda  \int d \eta a(\eta) \partial_{\eta}K_{q_1}(\eta) \partial_{\eta}K_{q_2}(\eta) \partial_{\eta}K_{q_3}(\eta)+6~\mathrm{perms.}
    \end{align}
$\psi_4$ can, likewise, be read from Fig.~\ref{psi-4:diagram},
\begin{align}
    \nonumber
    &\psi_4(\q_1,\q_2,\q_3,\q_4)=\dfrac{1}{2}\left(-3i \lambda\right)^2 
    \\\nonumber&\iint d\eta_1 \, d\eta_2~a(\eta_1)a(\eta_2) ~ \partial_{\eta_1}K_{q_1}(\eta_1)\,\partial_{\eta_1}K_{q_2}(\eta_1)\partial_{\eta_2} K_{q_3}(\eta_2)\,\partial_{\eta_2} K_{q_4}(\eta_2)~ \partial_{\eta_1}\partial_{\eta_2}G_{|\q_1+\q_2|}(\eta_2,\eta_1)
    \\
    &\hspace*{11 cm} +24~\mathrm{perms.}
    \label{Psi2b:eq}
\end{align}

\subsection{N-point functions}
\begin{figure}[ht!]
\begin{picture}(100,80)(-100,-40)

\SetWidth{1.2}
    \Line[](8.5,-21.5)(51.5,21.5) \Line[](8.5,21.5)(51.5,-21.5) %\Text(30,-36)[]{$K_4(\q_1,\q_2,\q_3,\q_4)$}
    %\Text(30,0)[]{\Cross}
    \Text(30,-30)[]{(a)}
    \end{picture}
    %%%%%%%%%%%%%%%%%%%%%%%%
    %%%%%%%%%%%%%%%%%%%%%%%
    \begin{picture}(100,80)(-100,-40)
\SetWidth{1.2}
    \Line[](8.5,-21.5)(30,0) \Line[](8.5,21.5)(30,0)
    \Line[](60,0)(81.5,21.5) \Line[](60,0)(81.5,-21.5)
    \Line[](30,0)(60,0)
    \Text(45,0)[]{\Cross}
    \Text(45,-30)[]{(b)}
    \end{picture}
    
    \caption{Feynman diagrams for the boundary theory}\label{fig:boundary}
\end{figure}
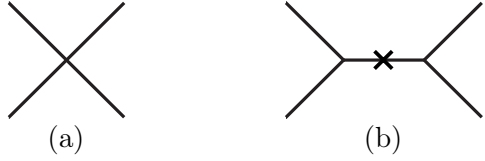

Having found $\psi_n$, quantum expectation values of a product of field fluctuations, $\langle \phi^n\rangle$, can be easily found using the following formula
\begin{align}
    \langle \phi^n\rangle= \dfrac{\displaystyle\int {\cal D}\phi~ \phi_{k_1}\phi_{k_2}...\phi_{k_n} |\Psi[\phi_0]|^2}{\displaystyle\int {\cal D}\phi~ |\Psi[\phi_0]|^2}     
\end{align}
For example, for the first few moments, namely, the power spectrum, bispectrum and trispectrum  are
\begin{align}
 P(\kk_1,\kk_2)\equiv& \langle\Psi[\phi_0]\big| \hat{\phi}_{\kk_1}(\eta_0)\,\hat{\phi}_{\kk_2}(\eta_0)\big|\Psi[\phi_0] \rangle' = \dfrac{1}{2\,\mathrm{Re}\,\psi_2(k_1)}
   \\
\nonumber
    B(\kk_1,\kk_2,\kk_3)\equiv& \langle\Psi[\phi_0]\big| \hat{\phi}_{\kk_1}(\eta_0)\,\hat{\phi}_{\kk_2}(\eta_0)\,\hat{\phi}_{\kk_3}(\eta_0)\,\big|\Psi[\phi_0] \rangle
    \\
    =& -\dfrac{\mathrm{Re} \,\psi_3(\kk_1,\kk_2,\kk_3)}{4\,\mathrm{Re}\,\psi_2(k_1)\,\mathrm{Re}\,\psi_2(k_2)\,\mathrm{Re}\,\psi_2(k_3)}
    \\
\nonumber
    T(\kk_1,\kk_2,\kk_3,\kk_4)\equiv& \langle\Psi[\phi_0]\big| \hat{\phi}_{\kk_1}(\eta_0)\,\hat{\phi}_{\kk_2}(\eta_0)\,\hat{\phi}_{\kk_3}(\eta_0)\,\hat{\phi}_{\kk_4}(\eta_0)\big|\Psi[\phi_0] \rangle
    \\
    \nonumber
    =&- \dfrac{\mathrm{Re} \,\psi_4(\kk_1,\kk_2,\kk_3,\kk_4)}{8\,\mathrm{Re}\,\psi_2(k_1)\,\mathrm{Re}\,\psi_2(k_2)\,\mathrm{Re}\,\psi_2(k_3)\,\mathrm{Re}\,\psi_2(k_4)}
    \\
    &+ \dfrac{\mathrm{Re}\,\psi_3(\kk_1,\kk_2,\kk_{12})\,\mathrm{Re}\,\psi_3(\kk_3,\kk_4,\kk_{12}) }{8\,\mathrm{Re}\,\psi_2(k_{12})\,\mathrm{Re}\,\psi_2(k_1)\,\mathrm{Re}\,\psi_2(k_2)\,\mathrm{Re}\,\psi_2(k_3)\,\mathrm{Re}\,\psi_2(k_4)\,}- \mathrm{perms.}
 \label{trispectrum-psi-n:eq}
   \end{align}
Note that the above result can be recast into Feynman rule of a boundary theory-- with no time. For example, the diagrams contributing to the trispectrum can be seen in Fig.~\ref{fig:boundary} 

\section{Propagator and its derivatives}
\label{prop-deriv:app}
The definition of the time ordered product of a (Bose) field is
\begin{align}
    T (\phi_1(x) \phi_2(y)) = \theta(x^0-y^0)\,\phi_1(x) \phi_2(y)+\theta(y^0-x^0)\,\phi_2(y) \phi_1(x)
\end{align}
By differentiating with respect to the time argument, we find
\begin{align}
    \partial_{x^0}T (\phi_1(x) \phi_2(y)) = T( \partial_{x^0}\,\phi_1(x) \phi_2(y))+\delta(x^0-y^0)\,[\phi_1(x),\phi_2(y)]
\end{align}
In particular, we can relate the second derivative to the time order product of the derivatives,
\begin{align}
    \partial_{x^0}\partial_{y^0}T (\phi(x) \phi(y)) = T \big( \partial_{x^0}\,\phi(x) \partial_{y^0}\phi(y)\big)+W(\phi,\phi^{\ast})\,\delta^4(x-y),
    \label{prop-derivative:eq}
\end{align}
which, by taking quantum expectation value,  gives rise to the following important identity
\begin{tcolorbox}[ams align, colback=white, colframe=black]
     \big\langle T\big( \partial_{x^0}\,\phi(x) \partial_{y^0}\phi(y)\big) \big\rangle = \partial_{x^0}\partial_{y^0} \big\langle T (\phi(x) \phi(y)) \big\rangle - i\delta^4(x-y).
    \label{time-order-derivative:eq}
\end{tcolorbox}
\noindent The expectation value of the time ordered product of two fields in the vacuum is often referred to as the Feynman Propagator,
\begin{align}
\nonumber
\big\langle T (\phi(x) \phi(y)) \big\rangle&=\Delta_{F}(x-y) = \int d^3 k e^{i\boldsymbol{k}.(\boldsymbol{x}-\boldsymbol{y})} \Delta_{F}(\boldsymbol{k};\eta_1,\eta_2)\\& = \int d^3 k e^{i\boldsymbol{k}.(\boldsymbol{x}-\boldsymbol{y})}\bigg[ \theta(\eta_1-\eta_2)\,D_{>}(|\boldsymbol{k}|;\eta_1,\eta_2)+\theta(\eta_2-\eta_1)\,\,D_{<}(|\boldsymbol{k}|;\eta_1,\eta_2)\bigg].
\end{align}
From which we can read off the Fourier transform of the propagator,
\begin{align}
\Delta_{F}(\boldsymbol{k};\eta_1,\eta_2) = \theta(\eta_1-\eta_2)\,D_{>}(|\boldsymbol{k}|;\eta_1,\eta_2)+\theta(\eta_2-\eta_1)\,\,D_{<}(|\boldsymbol{k}|;\eta_1,\eta_2)
\end{align}
\subsection{Time order product \& Equal time commutation relation}
The problem with the above discussion is that the Theta function is not a "proper" function in a strict mathematical sense. For the time ordering of the product of two operators, in the limit of two operators calculated at the same time, we define
\begin{align}
\lim_{t_1-t_2 \rightarrow 0^{+}}T (\varphi_1(x_1) \varphi_2(x_2))=\lim_{t_1-t_2 \rightarrow 0^{-}}T (\varphi_1(x_1) \varphi_2(x_2))
\end{align}
The above definition is consistent as long as long as the two operators commute. However, there will be some complications if the two operator do not commute at equal time. As a result, the equal time commutation relation (ETCR) provides a recipe for the analytical continuation of the propagator as 
\begin{equation}\label{eq:onshellpropagator}
\Delta_{F}(\boldsymbol{k};\eta_1,\eta_2) =
\begin{cases}
 \theta(\eta_1-\eta_2)\,\phi_{|\boldsymbol{k}|}(\eta_1)\phi^{\ast}_{|\boldsymbol{k}|}(\eta_2)+\theta(\eta_2-\eta_1)\,\phi^{\ast}_{|\boldsymbol{k}|}(\eta_1)\phi_{|\boldsymbol{k}|}(\eta_2) \qquad \eta_1\neq \eta_2
\\
|\phi_{|\boldsymbol{k}|}(\eta_1)|^2\qquad \qquad \qquad \qquad \qquad \qquad\qquad\qquad \qquad\qquad~~ \eta_1= \eta_2.
\end{cases}
\end{equation}
However, even using the above definition, one finds that the derivative of the propagator for equal time is not well-defined. So the main question is that how to cure this problem.
 
\subsection{Analytical Continuation}
To consistently define the derivative of the propagator, we propose that  \emph{before differentiating with respect to time}, we must replace the theta function with its analytical continuation
\begin{align}
\theta(\eta)=\dfrac{-1}{2\pi i} \int_{-\infty}^{+\infty} \dfrac{e^{-is \eta}}{s+i\epsilon}\, ds
\label{heaviside-AC:eq}.
\end{align}
This allows us to define an "off-shell" propagator which, as we shall see below, greatly simplifies calculations of models involving derivative interactions as, unlike with the previous definition, it is possible to directly take derivatives of this propagator
\begin{align}
    \theta'(\eta)=\lim_{\epsilon \to 0}\dfrac{1}{2\pi} \int_{-\infty}^{+\infty} \dfrac{s e^{-is \eta}}{s+i\epsilon}\, ds.
\label{dirac_delta-AC:eq}
\end{align}
 This relation is used in \cite{Chen:2017ryl}, which correctly leads to the $\delta$-function corrections to the derivative of the propagator. As emphasized in \cite{Chen:2017ryl} and further demonstrated in Section \ref{sec:WFUTri}, this correction, equivalent to a new contact interaction, is crucial to find agreement between the canonical and path-integral formulations. The above relation has a smooth limit for $\epsilon \to 0$, so
\begin{align}
    \theta'(\eta)= \int_{-\infty}^{+\infty}  \dfrac{ds}{2\pi} e^{-is \eta}\, = \delta (\eta).
\end{align}
In the rest of this section we use this formalism explicitly and show its agreement with conventional (on-shell)  propagators rigorously.
\subsection{Minkowski Space}
As a warm up let us apply our prescription to Minkowski space. Using Eq.~\eqref{heaviside-AC:eq}, we find the (Feynman) propagator of a massless scalar field in Minkowski space as
\begin{align}
\Delta_{F}(x_1-x_2)= \int \dfrac{d^4 k}{(2\pi)^4}\dfrac{i\,e^{-ik.(x_1-x_2)}}{k^2+i \epsilon}.
\end{align}
in which $k^2=|k_0^2-\kk^2|$.  We call the propagator found via analytical continuation before integrating k, the "off-shell" propagator. Once the integral on $k_0$ is taken, we call it an "on-shell" propagator, as we have then enforced the relationship between $\kk\cdot \kk$ and the energy, $k_0$. Taking the derivative of the above propagator, we find
\begin{align}
\partial_{\eta_1}\partial_{\eta_2} \Delta_{F}(x_1-x_2) = \int \dfrac{d^4 k}{(2\pi)^4}\dfrac{i \,k_0 k_0}{k^2+i \epsilon}e^{-ik.(x_1-x_2)}.
\label{off-shell-prop:eq}
\end{align}
This result shows that the Fourier transform of the derivative of the time order product of the field is simply the Fourier transform of that propagator multiplied by $k_0^2$
\begin{align}\label{eq:offshellprop}
\partial_{\eta_1}\partial_{\eta_2} \Delta_{F}(|\kk|;\eta_1,\eta_2) = \int \dfrac{d k_0}{2\pi}\dfrac{i\,k_0^2}{k^2+i \epsilon}e^{-ik_0(\eta_1-\eta_2)}.
\end{align}
Using Cauchy's integral theorem is a bit subtle here. For $\eta_1>\eta_2$ we close the contour below, while for $\eta_2>\eta_1$ we should close the contour above (this particular prescription is what defines the Feynman propagator). It must be noted that the above integral does not vanish on the half-circle with an infinitely large radius. The integral on the half-circle is 
\begin{align}
\int_{C^{-}} \dfrac{d k_0}{2\pi} e^{-ik_0(\eta_1-\eta_2)} = - \delta(\eta_1-\eta_2)
\end{align}
To get to this result, we used the following relation
\begin{align}
\int_{C^{-}} \dfrac{d k_0}{2\pi} e^{-ik_0(\eta_1-\eta_2)}+\int_{\infty}^{+\infty} \dfrac{d k_0}{2\pi} e^{-ik_0(\eta_1-\eta_2)} = 0.
\end{align}
 Putting this all together, we find
\begin{align}
\partial_{\eta_1}\partial_{\eta_2} \Delta_{F}(|\kk|;\eta_1,\eta_2) = \dfrac{1}{2}|\kk| \left[e^{-i|\kk|(\eta_1-\eta_2)} \theta(\eta_1-\eta_2)+e^{+i|\kk|(\eta_1-\eta_2)} \theta(\eta_2-\eta_1) \right] + i \delta(\eta_1-\eta_2).
\end{align}
Noting that for a massless scalar field in the Minkowski space 
\begin{align}
D_{>}(|\kk|;\eta_1,\eta_2) = \dfrac{1}{2|\kk|} e^{-i|\kk|(\eta_1-\eta_2)},
\end{align}
we find
\begin{tcolorbox}[ams align, colback=white, colframe=black]
\nonumber
\partial_{\eta_1}\partial_{\eta_2} &\Delta_{F}(|\kk|;\eta_1,\eta_2)= \\ &\partial_{\eta_1}\partial_{\eta_2}D_{>}(|\kk|;\eta_1,\eta_2) \theta(\eta_1-\eta_2)+\partial_{\eta_1}\partial_{\eta_2}D_{<}(|\kk|;\eta_1,\eta_2) \theta(\eta_2-\eta_1)+i\delta(\eta_1-\eta_2).
\label{on-shell-prop:eq}
\end{tcolorbox}
\noindent This justifies the Eq. (125) of \cite{Chen:2017ryl}. Nevertheless, it is worth emphasizing that the analytical continuation of the Heaviside function provides us with a consistent definition for the derivative of the propagator at $\eta_1=\eta_2$.

 A remark is in order here. We showed that via our definition, the derivative of the off-shell propagator is simple in the sense that its Fourier transform is simply the propagator multiplied by a couple of momenta. While there is no disgusting\footnote{Coleman calls this term "disgusting" in his QFT lecture notes!} term in the derivative of the off-shell propagator, the on-shell propagator gets a divergent correction term. Now, using Eq.~\eqref{prop-derivative:eq}, quite surprisingly, this divergent correction cancels, and we get 
 \begin{tcolorbox}[colback=white, colframe=black]
\begin{equation}
\big\langle \partial_{\eta_1} \varphi(x_1) \partial_{\eta_2} \varphi(x_2)\big\rangle =  \partial_{\eta_1}\partial_{\eta_2}D_{>}(|\kk|;\eta_1,\eta_2) \theta(\eta_1-\eta_2)+\partial_{\eta_1}\partial_{\eta_2}D_{<}(|\kk|;\eta_1,\eta_2) \theta(\eta_2-\eta_1).
\end{equation}
\end{tcolorbox}
\subsection{On-shell vs Off-shell calculations}
In this section we demonstrate the agreement between the standard in-in calculations using the on-shell propagator, Eq.~\ref{eq:onshellpropagator}, and the calculation using the off-shell propagator, Eq.~\ref{eq:offshellprop}. To do this we begin by performing a sample calculation in Minkowski space for a theory involving the following interaction Lagrangian,
\begin{align}
{\cal L}_{\mathrm{int.}} = -\lambda \dot{\phi}^3.
\end{align}
To allow for easier comparison to cosmology, we calculate the quantum expectation value of a product of four scalar fields at some time $\eta_0$, namely $\langle\phi_{k_1}(\eta_0)\,\phi_{k_2}(\eta_0)\,\phi_{k_3}(\eta_0)\,\phi_{k_4}(\eta_0)\rangle$. Using perturbation theory, one finds
\begin{align}\nonumber
&\langle\phi_{k_1}(\eta_0)\,\phi_{k_2}(\eta_0)\,\phi_{k_3}(\eta_0)\,\phi_{k_4}(\eta_0)\rangle\supset  
\\
& \dfrac{-\lambda^2}{2} \phi_{k_1}(\eta_0)\,\phi_{k_2}(\eta_0)\,\phi_{k_3}(\eta_0)\,\phi_{k_4}(\eta_0)\iint d\eta_1  d\eta_2 ~ \dot{\phi}^{\ast}_{k_1}(\eta_1)\,\dot{\phi}^{\ast}_{k_2}(\eta_1)\,\dot{\phi}^{\ast}_{k_3}(\eta_2)\,\dot{\phi}^{\ast}_{k_4}(\eta_2) \langle T \dot{\varphi}_s(\eta_1) \dot{\varphi}_s(\eta_2) \rangle
\end{align}
where $s = |\kk_1+\kk_2|$. Now, using Eq.~\eqref{time-order-derivative:eq}, the integral in the above equation can be written as
\begin{align}
\nonumber
\iint d\eta_1  d\eta_2 ~& \dot{\phi}^{\ast}_{k_1}(\eta_1)\,\dot{\phi}^{\ast}_{k_2}(\eta_1)\,\dot{\phi}^{\ast}_{k_3}(\eta_2)\,\dot{\phi}^{\ast}_{k_4}(\eta_2) \langle T \dot{\varphi}_s(\eta_1) \dot{\varphi}_s(\eta_2) \rangle = 
\\
&\iint d\eta_1  d\eta_2 ~ \dot{\phi}^{\ast}_{k_1}(\eta_1)\,\dot{\phi}^{\ast}_{k_2}(\eta_1)\,\dot{\phi}^{\ast}_{k_3}(\eta_2)\,\dot{\phi}^{\ast}_{k_4}(\eta_2) \,\bigg[ \partial_{\eta_1} \partial_{\eta_2}\langle T \varphi_s(\eta_1) \varphi_s(\eta_2) \rangle -i\delta(\eta_1-\eta_2)\bigg]. 
\end{align}
The integral of the second term in the brackets is straightforward, so we focus on the integral of the former,
\begin{align}
{\cal I}=\iint_{-\infty}^{\eta_0} d\eta_1  d\eta_2 ~ \dot{\phi}^{\ast}_{k_1}(\eta_1)\,\dot{\phi}^{\ast}_{k_2}(\eta_1)\,\dot{\phi}^{\ast}_{k_3}(\eta_2)\,\dot{\phi}^{\ast}_{k_4}(\eta_2)~ \partial_{\eta_1} \partial_{\eta_2}\langle T \varphi_s(\eta_1) \varphi_s(\eta_2) \rangle
\end{align}
It must be noted that the time integrals are calculated --by a Wick rotation-- along a slightly deformed contour $\eta \rightarrow \eta(1-i \epsilon)$. We use the following mode function for the scalar field
\begin{align}
{\phi}_{k}(\eta) = \dfrac{1}{\sqrt{2 k}} e^{-i k \eta}
\end{align}
\subsubsection{Off-shell propagator}
In this section, we use the off-shell propagator defined in Eq.~\eqref{off-shell-prop:eq}. More or less like the standard QFT, we first perform time integral, and then integrated over $k_0$. We therefore have
\begin{align}
{\cal I}=\dfrac{1}{4}\sqrt{k_1k_2k_3k_4}\int \dfrac{d E}{2\pi} \dfrac{iE^2}{E^2-s^2+i\epsilon}\iint_{-\infty}^{\eta_0} d\eta_1  d\eta_2 ~ e^{i E_L\eta_1}\,e^{i E_R\eta_2}\,\, e^{-iE(\eta_1-\eta_2)}
\end{align}
in which $E_L=k_1+k_2$ and $E_R=k_3+k_4$. So 
\begin{align}
{\cal I}=\dfrac{i}{4}\sqrt{k_1k_2k_3k_4}e^{i E_T\eta_0}\int \dfrac{d E}{2\pi}E^2\, \dfrac{1}{(E^2-s^2+i\epsilon)} \dfrac{1}{(E+E_R(1-i\epsilon))} \dfrac{1}{(E-E_L(1-i\epsilon))}.
\end{align}
 The above integral can be calculated using Cauchy's integral theorem. By closing the contour above and noting that the integral on $C_{+}$ is vanishing, we find
\begin{align}
{\cal I}=\dfrac{1}{4}\sqrt{k_1k_2k_3k_4}e^{i E_T\eta_0} \dfrac{s E_T+2 E_L E_R}{2 E_T (E_L+s) (E_R+s)},
\label{I-off-shell:eq}
\end{align}
where $E_T =E_L+E_R$. Note that one gets the same result if uses the contour passing below the real axis.
\subsubsection{On-shell propagator}
Using on-shell propagator, Eq.~\eqref{on-shell-prop:eq}, we have
\begin{align}
\nonumber
{\cal I}=&\dfrac{1}{4}\sqrt{k_1k_2k_3k_4}\times
\\\nonumber
\dfrac{s}{2}&\bigg[\int^{\eta_0}_{-\infty} d\eta_1 \,e^{i (E_L-s)\eta_1}\int_{-\infty}^{\eta_1}  d\eta_2 \,e^{i (E_R+s)\eta_2}\,+\int^{\eta_0}_{-\infty} d\eta_2 \,e^{i (E_R-s)\eta_2} \int_{-\infty}^{\eta_2}  d\eta_1 ~ e^{i (E_L+s)\eta_1} \bigg]
\\&\hspace{10cm}+ i \int^{\eta_0}_{-\infty}  d\eta_1\, e^{i E_T\eta_1}
\end{align}
Again by pushing the integral contour slightly upward to $\eta \rightarrow \eta (1-i \epsilon)$, one gets
\begin{align}
\nonumber
{\cal I}=&\dfrac{1}{4}\sqrt{k_1k_2k_3k_4}\left[\dfrac{s}{2} \left(\dfrac{-i\, e^{iE_T \eta_0}}{E_T} \dfrac{-i}{E_R+s}+\dfrac{-i\, e^{iE_T \eta_0}}{E_T} \dfrac{-i}{E_L+s}\right)+\dfrac{-i\times i\, e^{iE_T \eta_0}}{E_T} \right]
\\
=& \dfrac{1}{4}\sqrt{k_1k_2k_3k_4}e^{i E_T\eta_0} \dfrac{s E_T+2 E_L E_R}{2 E_T (E_L+s) (E_R+s)}
\end{align}
which agrees with the off-shell propagator's result, Eq.~\eqref{I-off-shell:eq}.

\section{Propagators in the inflationary background}\label{app:Propagators}
In the inflationary background, we expand the scalar perturbations as
\begin{align}
    \phi(x) = \int \dfrac{d^3k}{(2\pi)^3} \,e^{i\boldsymbol{k}\cdot\boldsymbol{x}} \left[ \phi_{k}(\eta)\,a_{\kk} + \phi^{\ast}_{k}(\eta) \,a^\dagger_{-\kk} \right]
\end{align}
in which
\begin{align}
\phi_k(\eta) =
\frac{H}{\sqrt{2 k^3}} (-i+ k \eta)e^{-i k\eta}
~,
\end{align}
The corresponding propagator is
\begin{align}\nonumber
    &\langle T\,\varphi(x) \varphi(x')\rangle \\&=\theta(\eta-\eta')\,\int \dfrac{d^3k}{(2\pi)^3}\,e^{i\boldsymbol{k}\cdot(\boldsymbol{x}-\boldsymbol{x}')} \phi_{k}(\eta)\phi^{\ast}_{k}(\eta') + \theta(\eta'-\eta)\,\int \dfrac{d^3k}{(2\pi)^3}\,e^{i\boldsymbol{k}\cdot(\boldsymbol{x}-\boldsymbol{x}')}\phi^{\ast}_{k}(\eta)\phi_{k}(\eta') 
\end{align}
Using Eq.~\eqref{heaviside-AC:eq} and introducing a new integral variable $E=k+s$ in the first integral and $E=-k-s$ in the second, we get
\begin{align}
    \langle T\,\varphi(x) \varphi(x')\rangle =\int \dfrac{d^4k}{(2\pi)^4}\,e^{-ik\cdot(x-x')} \frac{H^2  \left(i(1+k^2\,\eta \eta') - E (\eta-\eta')\right)}{k^2 (k^\mu k_\mu-i\epsilon) } 
\end{align}
where $k^\mu k_\mu=E^2-k^2$. Now, let us investigate the time derivative of this propagator. After some algebra, we find
\begin{align}
    \partial_{\eta} \partial_{\eta'}\langle T\,\varphi(x) \varphi(x')\rangle =\int \dfrac{d^4k}{(2\pi)^4} \,e^{ik\cdot(x'-x)}  \dfrac{H^2 \left( i k^2 (i+E\eta) (-i+E\eta' )- {E}^2 (i+E (\eta-\eta'))\right)}{k^2 (k^\mu k_\mu-i\epsilon) } 
\end{align}
By integrating on $k_0$, using Cauchy's integral theorem, over the closed half circle above for $\eta>\eta'$, and on a closed half circle below for the $\eta<\eta'$, we have therefore
\begin{align}\nonumber
    \partial_{\eta} \partial_{\eta'}\langle T\,&\varphi(x) \varphi(y)\rangle \\&=\int d^3 k\,e^{i\boldsymbol{k}\cdot(\boldsymbol{x}-\boldsymbol{x}')} H^2  \eta\eta' \left[\dfrac{k}{2}\,e^{-i k (\eta-\eta')}\theta(
    \eta-\eta)+\dfrac{k}{2}\,e^{ik (\eta-\eta')}\theta(
    \eta'-\eta) \right] -{\cal I}_{C_{\pm}}
\end{align}
    where
\begin{align}
\nonumber
{\cal I}_{C_{\pm}} &=  \int \dfrac{d^3k}{(2\pi)^3}\,e^{i\boldsymbol{k}\cdot(\boldsymbol{x}-\boldsymbol{x}')}  \int_{C_{\pm}} \dfrac{dE}{2\pi} \dfrac{H^2 \left( ik^2 (i+E\eta) (-i+E\eta' )-i {E}^2 (E (\eta-\eta')+i)\right)}{k^2 (k^\mu k_\mu-i\epsilon) } 
\\
&\simeq\int \dfrac{d^3k}{(2\pi)^3}\,e^{i\boldsymbol{k}\cdot(\boldsymbol{x}-\boldsymbol{x}')}  \int_{C_{\pm}} \dfrac{dE}{2\pi} \dfrac{H^2 e^{-i E (\eta-\eta')} \left(i k^2 \eta \eta'-  E (\eta-\eta')-i\right)}{k^2}
\end{align}
Since the integrand in the above equation is analytic in the entire $E$-plane, its integral over a closed loop is zero. As a result, 
\begin{align}
\nonumber
&{\cal I}_{C_{\pm}} = -\int \dfrac{d^3k}{(2\pi)^3}\,e^{i\boldsymbol{k}\cdot(\boldsymbol{x}-\boldsymbol{x}')} H^2 \times
\\
 &\left[i\int_{-\infty}^{\infty} \dfrac{dE}{2\pi} e^{-i E(\eta-\eta')} \eta \eta' +\dfrac{1}{k^2} \left( - (\eta-\eta') \int_{-\infty}^{\infty} \dfrac{dE}{2\pi} E\,e^{-i E(\eta-\eta')}  -i \int_{-\infty}^{\infty} \dfrac{dE}{2\pi} e^{-i E (\eta-\eta')} \right) \right] .
\end{align}
 Now, by using following identity
\begin{align}
\int_{-\infty}^{+\infty} \dfrac{dk}{2\pi} \,e^{- i k x} = \delta(x),
\end{align}
we get
\begin{align}
{\cal I}_{C_{\pm}} &= -\delta^3(\boldsymbol{x}-\boldsymbol{x}') \times\left[i\,a^{-2}(\eta) \delta (\eta-\eta') +\dfrac{H^2}{k^2} \left( -i (\eta-\eta')\dfrac{\partial}{\partial \eta} \delta (\eta-\eta') - i\delta(\eta-\eta') \right) \right].
\end{align}
Integrating by parts, it can be seen that the term in the parenthesis vanishes, so 
\begin{align}
{\cal I}_{C_{\pm}} &=-i\,a^{-2}(\eta)\,\delta^4(x-x').
\end{align}
Putting this all together, the time derivative of the propagator is found as
 \begin{tcolorbox}[ams align, colback=white, colframe=black]\nonumber
    &\partial_{\eta} \partial_{\eta'}\langle T\,\varphi(x) \varphi(y)\rangle =\int \dfrac{d^3 k}{(2\pi)^3}\,e^{i\boldsymbol{k}\cdot(\boldsymbol{x}-\boldsymbol{y})} \left[\phi'_k(\eta){\phi_k^{\ast}}'(\eta')\theta(
    \eta-\eta)+{\phi_k^{\ast}}'(\eta)\phi'_k(\eta')\theta(
    \eta'-\eta) \right]\\&\hspace{9cm}+ ia^{-2}(\eta) \delta^4(x-x').
    \label{partial-der-prop:eq}
\end{tcolorbox}
\noindent        Quite surprisingly, this result does not change if one treats the Heaviside function naively without performing analytical continuation. Henceforth, like the flat space case, using Eq.~\eqref{time-order-derivative:eq}, we find the propagator-- of field derivatives-- as
 \begin{tcolorbox}[ams align, colback=white, colframe=black]
     \langle T\, \partial_{\eta}\phi(x) \partial_{\eta'}\phi(y)\rangle= 
     \int \dfrac{d^3 k}{(2\pi)^3}\,e^{i\boldsymbol{k}\cdot(\boldsymbol{x}-\boldsymbol{y})} \left[\phi'_k(\eta){\phi_k'}^{\ast}(\eta')\theta(
    \eta-\eta)+{\phi_k'}^{\ast}(\eta)\phi_k'(\eta')\theta(
    \eta'-\eta) \right]
\end{tcolorbox}

\section{Divergent Corrections: Momentum Space}\label{divergentmomentum:sec}
In general we have that
\begin{align}
    \mathcal{L}[\phi,\phi']=\frac{a^2}{2}\left({\phi'}^2-k^2\phi^2\right)+\lambda \mathcal{L}_{\textrm{int}}[\phi,\phi']
\end{align}
from which we find the Hamiltonian to be
\begin{align}
    \mathcal{H}[\phi,\pi]=\frac{\pi^2}{2a^2}+k^2a^2\frac{\phi^2}{2}-\lambda \mathcal{L}_{\textrm{int}}\left[\phi,\frac{\pi}{a^2}\right]+\sum_{n=2} \frac{(-\lambda)^n}{n!}\frac{\delta^{n-2}}{\delta {\phi'}^{n-2}}\left(\frac{\delta \mathcal{L}_{\textrm{int}}}{\delta \phi'}\right)^n.
\end{align}
This therefore gives us the effective action to leading order in $\hbar$ as
\begin{align}
    \tilde{\mathcal{L}}=\mathcal{L}+\frac{i\hbar}{2}\delta^4(0)\sum_{n=1}\frac{(-\lambda)^n}{na^{2n}}\left(\frac{\delta^2\mathcal{L}_{\textrm{int}}}{\delta {\phi'}^2}\right)^n.
\end{align}
Single vertex loop diagrams contribute a divergence of the form
\begin{equation}
    \frac{i\hbar\lambda}{2}\int d\eta \frac{d^3 k}{(2\pi)^3} \frac{\delta^2\mathcal{L}_{\textrm{int}}}{\delta {\phi'}^2}\left.\partial_{\eta\eta'}G_k(\eta,\eta')\right\rvert_{\eta=\eta'}=\frac{i\lambda}{2} \int d\eta \frac{\delta^2\mathcal{L}_{\textrm{int}}}{\delta {\phi'}^2}\left(\frac{i}{a^2}\delta^4(0)+\int\frac{d^3 k}{(2\pi)^3} \Xi_k(\eta,\eta) \right)
\end{equation}
where we have used that the second derivative of the Green's function in momentum space is
\begin{align}
    \partial_{\eta'}\partial_\eta G_k(\eta,\eta')&=\phi_-'(\eta)\phi_+'(\eta')\theta(\eta-\eta')+\phi_+'(\eta)\phi_-'(\eta')\theta(\eta'-\eta)\\&-\phi_+'(\eta)\phi_+'(\eta')\frac{\phi_-(\eta_0)}{\phi_+(\eta_0)}+ia^{-2}\delta(\eta-\eta')=\Xi_k(\eta,\eta')+ia^{-2}(\eta-\eta')
\end{align}
and we have both performed the trivial integrals over delta functions and used that
\begin{align}
    \int \frac{d^3k}{(2\pi)^3}e^{i\kk\cdot\x}=\delta^3(\x)\rightarrow\int \frac{d^3k}{(2\pi)^3}=\delta^3(0).
\end{align}
The two vertex single loop diagrams contribute a divergence like
\begin{align}
    &-\frac{\hbar\lambda^2}{4}\int d\eta d\eta' \frac{d^3 k}{(2\pi)^3} \left(\frac{\delta^2\mathcal{L}_{\textrm{int}}}{\delta {\phi'}^2}\right)^2 \partial_{\eta\eta'}G_k(\eta,\eta')\partial_{\eta\eta'}G_{\lvert \textbf{k}_L-\textbf{k}\rvert}(\eta,\eta')=\\&-\frac{\hbar\lambda^2}{4}\left[\int d\eta d\eta' \frac{d^3 k}{(2\pi)^3} \left(\frac{\delta^2\mathcal{L}_{\textrm{int}}}{\delta {\phi'}^2}\right)^2\Xi_k(\eta,\eta')\Xi_{\lvert \textbf{k}_L-\textbf{k}\rvert}(\eta,\eta')-\int d\eta  \left(\frac{\delta^2\mathcal{L}_{\textrm{int}}}{\delta {\phi'}^2}\right)^2\frac{1}{a^4}\delta^4(0)\right.\\&\left.+i\int d\eta \frac{d^3 k}{(2\pi)^3} \frac{1}{a^2} \left(\frac{\delta^2\mathcal{L}_{\textrm{int}}}{\delta {\phi'}^2}\right)^2\left(\Xi_k(\eta,\eta)+\Xi_{\lvert \textbf{k}_L-\textbf{k}\rvert}(\eta,\eta)\right)\right]
\end{align}
where $\textbf{k}_L$ is the sum of the momenta entering the left hand vertex. It is possible to see that both of these terms (as well as higher order connected diagrams with a single loop) will cancel the divergent terms in the effective action. We similarly expect that diagrams with multiple loops will cancel the divergences that appear at higher order in $\hbar$ in the effective action.

\begin{figure}
    \centering

\tikzset{every picture/.style={line width=0.75pt}} %set default line width to 0.75pt        

\begin{tikzpicture}[x=0.75pt,y=0.75pt,yscale=-1.4,xscale=1.4]
%uncomment if require: \path (0,603); %set diagram left start at 0, and has height of 603

%Shape: Arc [id:dp1625812393570445] 
\draw  [draw opacity=0][dash pattern={on 2.25pt off 1.5pt}] (60,100) .. controls (60,100) and (60,100) .. (60,100) .. controls (60,100) and (60,100) .. (60,100) .. controls (60,105.52) and (55.52,110) .. (50,110) .. controls (44.48,110) and (40,105.52) .. (40,100) -- (50,100) -- cycle ; \draw  [dash pattern={on 2.25pt off 1.5pt}] (60,100) .. controls (60,100) and (60,100) .. (60,100) .. controls (60,100) and (60,100) .. (60,100) .. controls (60,105.52) and (55.52,110) .. (50,110) .. controls (44.48,110) and (40,105.52) .. (40,100) ;
%Straight Lines [id:da3791405559106513] 
\draw  [dash pattern={on 2.25pt off 1.5pt}]  (60,80) -- (60,100) ;
%Straight Lines [id:da6846231109573611] 
\draw  [dash pattern={on 2.25pt off 1.5pt}]  (50,30) -- (50,50) ;
%Shape: Circle [id:dp796850803166562] 
\draw   (40,60) .. controls (40,54.48) and (44.48,50) .. (50,50) .. controls (55.52,50) and (60,54.48) .. (60,60) .. controls (60,65.52) and (55.52,70) .. (50,70) .. controls (44.48,70) and (40,65.52) .. (40,60) -- cycle ;
%Straight Lines [id:da9012307534178678] 
\draw    (120,90) ;
\draw [shift={(120,90)}, rotate = 0] [color={rgb, 255:red, 0; green, 0; blue, 0 }  ][fill={rgb, 255:red, 0; green, 0; blue, 0 }  ][line width=0.75]      (0, 0) circle [x radius= 3.35, y radius= 3.35]   ;
%Straight Lines [id:da9155764315114829] 
\draw    (70,160) -- (100,160) ;
%Shape: Circle [id:dp10145998987257299] 
\draw   (40,350) .. controls (40,344.48) and (44.48,340) .. (50,340) .. controls (55.52,340) and (60,344.48) .. (60,350) .. controls (60,355.52) and (55.52,360) .. (50,360) .. controls (44.48,360) and (40,355.52) .. (40,350) -- cycle ;
%Straight Lines [id:da8698810549294662] 
\draw    (40,350) -- (54,350) -- (60,350) ;
%Straight Lines [id:da3717681879163508] 
\draw  [dash pattern={on 2.25pt off 1.5pt}]  (70,430) -- (70,450) ;
\draw [shift={(70,450)}, rotate = 90] [color={rgb, 255:red, 0; green, 0; blue, 0 }  ][fill={rgb, 255:red, 0; green, 0; blue, 0 }  ][line width=0.75]      (0, 0) circle [x radius= 3.35, y radius= 3.35]   ;
%Shape: Circle [id:dp49106609642777443] 
\draw   (50,500) .. controls (50,494.48) and (54.48,490) .. (60,490) .. controls (65.52,490) and (70,494.48) .. (70,500) .. controls (70,505.52) and (65.52,510) .. (60,510) .. controls (54.48,510) and (50,505.52) .. (50,500) -- cycle ;
%Straight Lines [id:da907649969102039] 
\draw    (20,80) -- (80,80) ;
%Straight Lines [id:da19300480397526987] 
\draw    (20,130) -- (150,130) ;
%Straight Lines [id:da06955251516252647] 
\draw    (20,30) -- (80,30) ;
%Straight Lines [id:da5950365927759831] 
\draw    (20,180) -- (150,180) ;
%Straight Lines [id:da9192694885698158] 
\draw    (20,280) -- (150,280) ;
%Straight Lines [id:da7527024390588892] 
\draw    (20,330) -- (80,330) ;
%Straight Lines [id:da40377215143539735] 
\draw    (20,380) -- (150,380) ;
%Straight Lines [id:da6058198497554941] 
\draw    (20,430) -- (150,430) ;
%Straight Lines [id:da5179101791566301] 
\draw    (20,480) -- (150,480) ;
%Straight Lines [id:da7209792915106599] 
\draw    (20,530) -- (80,530) ;
%Straight Lines [id:da19017160528315125] 
\draw  [dash pattern={on 2.25pt off 1.5pt}]  (60,530) -- (60,560) ;
\draw [shift={(60,560)}, rotate = 90] [color={rgb, 255:red, 0; green, 0; blue, 0 }  ][fill={rgb, 255:red, 0; green, 0; blue, 0 }  ][line width=0.75]      (0, 0) circle [x radius= 3.35, y radius= 3.35]   ;
%Straight Lines [id:da13299163161529792] 
\draw  [dash pattern={on 2.25pt off 1.5pt}]  (40,530) -- (40,560) ;
\draw [shift={(40,560)}, rotate = 90] [color={rgb, 255:red, 0; green, 0; blue, 0 }  ][fill={rgb, 255:red, 0; green, 0; blue, 0 }  ][line width=0.75]      (0, 0) circle [x radius= 3.35, y radius= 3.35]   ;
%Straight Lines [id:da9870256101877983] 
\draw    (100,550) -- (140,550) ;
\draw [shift={(140,550)}, rotate = 0] [color={rgb, 255:red, 0; green, 0; blue, 0 }  ][fill={rgb, 255:red, 0; green, 0; blue, 0 }  ][line width=0.75]      (0, 0) circle [x radius= 3.35, y radius= 3.35]   ;
\draw [shift={(100,550)}, rotate = 0] [color={rgb, 255:red, 0; green, 0; blue, 0 }  ][fill={rgb, 255:red, 0; green, 0; blue, 0 }  ][line width=0.75]      (0, 0) circle [x radius= 3.35, y radius= 3.35]   ;
%Straight Lines [id:da0660568762436029] 
\draw  [dash pattern={on 2.25pt off 1.5pt}]  (70,230) -- (70,250) ;
%Shape: Circle [id:dp5609298382159664] 
\draw   (60,260) .. controls (60,254.48) and (64.48,250) .. (70,250) .. controls (75.52,250) and (80,254.48) .. (80,260) .. controls (80,265.52) and (75.52,270) .. (70,270) .. controls (64.48,270) and (60,265.52) .. (60,260) -- cycle ;
%Straight Lines [id:da8211191060337373] 
\draw    (20,230) -- (150,230) ;
%Shape: Circle [id:dp03749887095987847] 
\draw   (90,350) .. controls (90,344.48) and (94.48,340) .. (100,340) .. controls (105.52,340) and (110,344.48) .. (110,350) .. controls (110,355.52) and (105.52,360) .. (100,360) .. controls (94.48,360) and (90,355.52) .. (90,350) -- cycle ;
%Straight Lines [id:da3313371611288074] 
\draw    (110,350) -- (130,350) ;
%Shape: Circle [id:dp7282545450830809] 
\draw   (130,350) .. controls (130,344.48) and (134.48,340) .. (140,340) .. controls (145.52,340) and (150,344.48) .. (150,350) .. controls (150,355.52) and (145.52,360) .. (140,360) .. controls (134.48,360) and (130,355.52) .. (130,350) -- cycle ;
%Straight Lines [id:da119504921217926] 
\draw  [dash pattern={on 2.25pt off 1.5pt}]  (40,80) -- (40,100) ;
%Shape: Arc [id:dp5845909348548601] 
\draw  [draw opacity=0][dash pattern={on 2.25pt off 1.5pt}] (110,150) .. controls (110,150) and (110,150) .. (110,150) .. controls (110,150) and (110,150) .. (110,150) .. controls (110,155.52) and (105.52,160) .. (100,160) -- (100,150) -- cycle ; \draw  [dash pattern={on 2.25pt off 1.5pt}] (110,150) .. controls (110,150) and (110,150) .. (110,150) .. controls (110,150) and (110,150) .. (110,150) .. controls (110,155.52) and (105.52,160) .. (100,160) ;
%Straight Lines [id:da3629233650150465] 
\draw  [dash pattern={on 2.25pt off 1.5pt}]  (110,130) -- (110,150) ;
%Straight Lines [id:da6664866673853689] 
\draw  [dash pattern={on 2.25pt off 1.5pt}]  (100,130) -- (100,160) ;
%Shape: Square [id:dp02945780101553619] 
\draw  [fill={rgb, 255:red, 0; green, 0; blue, 0 }  ,fill opacity=1 ] (97.08,157.08) -- (102.92,157.08) -- (102.92,162.92) -- (97.08,162.92) -- cycle ;
%Shape: Arc [id:dp14014322262015844] 
\draw  [draw opacity=0][dash pattern={on 2.25pt off 1.5pt}] (70,160) .. controls (70,160) and (70,160) .. (70,160) .. controls (70,160) and (70,160) .. (70,160) .. controls (64.48,160) and (60,155.52) .. (60,150) -- (70,150) -- cycle ; \draw  [dash pattern={on 2.25pt off 1.5pt}] (70,160) .. controls (70,160) and (70,160) .. (70,160) .. controls (70,160) and (70,160) .. (70,160) .. controls (64.48,160) and (60,155.52) .. (60,150) ;
%Straight Lines [id:da4453927681019423] 
\draw  [dash pattern={on 2.25pt off 1.5pt}]  (70,130) -- (70,160) ;
%Shape: Square [id:dp3616542878742681] 
\draw  [fill={rgb, 255:red, 0; green, 0; blue, 0 }  ,fill opacity=1 ] (67.08,157.08) -- (72.92,157.08) -- (72.92,162.92) -- (67.08,162.92) -- cycle ;
%Straight Lines [id:da4334022106546016] 
\draw  [dash pattern={on 2.25pt off 1.5pt}]  (60,130) -- (60,150) ;
%Shape: Arc [id:dp5264421350491577] 
\draw  [draw opacity=0][dash pattern={on 2.25pt off 1.5pt}] (70,210) .. controls (70,210) and (70,210) .. (70,210) .. controls (70,210) and (70,210) .. (70,210) .. controls (64.48,210) and (60,205.52) .. (60,200) -- (70,200) -- cycle ; \draw  [dash pattern={on 2.25pt off 1.5pt}] (70,210) .. controls (70,210) and (70,210) .. (70,210) .. controls (70,210) and (70,210) .. (70,210) .. controls (64.48,210) and (60,205.52) .. (60,200) ;
%Straight Lines [id:da21478865984599627] 
\draw  [dash pattern={on 2.25pt off 1.5pt}]  (70,180) -- (70,210) ;
%Shape: Square [id:dp07499702416648901] 
\draw  [fill={rgb, 255:red, 0; green, 0; blue, 0 }  ,fill opacity=1 ] (67.08,207.08) -- (72.92,207.08) -- (72.92,212.92) -- (67.08,212.92) -- cycle ;
%Straight Lines [id:da835854165354206] 
\draw  [dash pattern={on 2.25pt off 1.5pt}]  (60,180) -- (60,200) ;
%Shape: Square [id:dp375292617294283] 
\draw  [fill={rgb, 255:red, 0; green, 0; blue, 0 }  ,fill opacity=1 ] (67.08,247.08) -- (72.92,247.08) -- (72.92,252.92) -- (67.08,252.92) -- cycle ;
%Straight Lines [id:da7675240568742838] 
\draw  [dash pattern={on 2.25pt off 1.5pt}]  (100,230) -- (100,250) ;
%Shape: Circle [id:dp19903011182879382] 
\draw   (90,260) .. controls (90,254.48) and (94.48,250) .. (100,250) .. controls (105.52,250) and (110,254.48) .. (110,260) .. controls (110,265.52) and (105.52,270) .. (100,270) .. controls (94.48,270) and (90,265.52) .. (90,260) -- cycle ;
%Shape: Square [id:dp2280938387484006] 
\draw  [fill={rgb, 255:red, 0; green, 0; blue, 0 }  ,fill opacity=1 ] (97.08,247.08) -- (102.92,247.08) -- (102.92,252.92) -- (97.08,252.92) -- cycle ;
%Shape: Square [id:dp20406082117307656] 
\draw  [fill={rgb, 255:red, 0; green, 0; blue, 0 }  ,fill opacity=1 ] (47.08,47.08) -- (52.92,47.08) -- (52.92,52.92) -- (47.08,52.92) -- cycle ;
%Shape: Circle [id:dp7587898727961526] 
\draw   (75,310) .. controls (75,304.48) and (79.48,300) .. (85,300) .. controls (90.52,300) and (95,304.48) .. (95,310) .. controls (95,315.52) and (90.52,320) .. (85,320) .. controls (79.48,320) and (75,315.52) .. (75,310) -- cycle ;
%Shape: Arc [id:dp3612588243674375] 
\draw  [draw opacity=0][dash pattern={on 2.25pt off 1.5pt}] (75,310) .. controls (75,310) and (75,310) .. (75,310) .. controls (75,310) and (75,310) .. (75,310) .. controls (69.48,310) and (65,305.52) .. (65,300) -- (75,300) -- cycle ; \draw  [dash pattern={on 2.25pt off 1.5pt}] (75,310) .. controls (75,310) and (75,310) .. (75,310) .. controls (75,310) and (75,310) .. (75,310) .. controls (69.48,310) and (65,305.52) .. (65,300) ;
%Shape: Square [id:dp556609757373379] 
\draw  [fill={rgb, 255:red, 0; green, 0; blue, 0 }  ,fill opacity=1 ] (72.08,307.08) -- (77.92,307.08) -- (77.92,312.92) -- (72.08,312.92) -- cycle ;
%Straight Lines [id:da9735540352925227] 
\draw  [dash pattern={on 2.25pt off 1.5pt}]  (65,280) -- (65,300) ;
%Shape: Square [id:dp45211851828410854] 
\draw  [fill={rgb, 255:red, 0; green, 0; blue, 0 }  ,fill opacity=1 ] (92.08,307.08) -- (97.92,307.08) -- (97.92,312.92) -- (92.08,312.92) -- cycle ;
%Shape: Arc [id:dp4992669770934761] 
\draw  [draw opacity=0][dash pattern={on 2.25pt off 1.5pt}] (105,300) .. controls (105,300) and (105,300) .. (105,300) .. controls (105,300) and (105,300) .. (105,300) .. controls (105,305.52) and (100.52,310) .. (95,310) -- (95,300) -- cycle ; \draw  [dash pattern={on 2.25pt off 1.5pt}] (105,300) .. controls (105,300) and (105,300) .. (105,300) .. controls (105,300) and (105,300) .. (105,300) .. controls (105,305.52) and (100.52,310) .. (95,310) ;
%Straight Lines [id:da723331893499098] 
\draw  [dash pattern={on 2.25pt off 1.5pt}]  (105,280) -- (105,300) ;
%Shape: Circle [id:dp5505941259042209] 
\draw   (100,210) .. controls (100,204.48) and (104.48,200) .. (110,200) .. controls (115.52,200) and (120,204.48) .. (120,210) .. controls (120,215.52) and (115.52,220) .. (110,220) .. controls (104.48,220) and (100,215.52) .. (100,210) -- cycle ;
%Straight Lines [id:da8669897206944546] 
\draw    (70,210) -- (100,210) ;
%Shape: Square [id:dp5315974769588951] 
\draw  [fill={rgb, 255:red, 0; green, 0; blue, 0 }  ,fill opacity=1 ] (97.08,207.08) -- (102.92,207.08) -- (102.92,212.92) -- (97.08,212.92) -- cycle ;
%Shape: Square [id:dp3369618671254435] 
\draw  [fill={rgb, 255:red, 0; green, 0; blue, 0 }  ,fill opacity=1 ] (37.08,347.08) -- (42.92,347.08) -- (42.92,352.92) -- (37.08,352.92) -- cycle ;
%Shape: Square [id:dp5306674659011195] 
\draw  [fill={rgb, 255:red, 0; green, 0; blue, 0 }  ,fill opacity=1 ] (57.08,347.08) -- (62.92,347.08) -- (62.92,352.92) -- (57.08,352.92) -- cycle ;
%Shape: Square [id:dp5517832473182329] 
\draw  [fill={rgb, 255:red, 0; green, 0; blue, 0 }  ,fill opacity=1 ] (107.08,347.08) -- (112.92,347.08) -- (112.92,352.92) -- (107.08,352.92) -- cycle ;
%Shape: Square [id:dp7806430818750767] 
\draw  [fill={rgb, 255:red, 0; green, 0; blue, 0 }  ,fill opacity=1 ] (127.08,347.08) -- (132.92,347.08) -- (132.92,352.92) -- (127.08,352.92) -- cycle ;
%Shape: Arc [id:dp18861490246608992] 
\draw  [draw opacity=0][dash pattern={on 2.25pt off 1.5pt}] (70,410) .. controls (70,410) and (70,410) .. (70,410) .. controls (70,410) and (70,410) .. (70,410) .. controls (64.48,410) and (60,405.52) .. (60,400) -- (70,400) -- cycle ; \draw  [dash pattern={on 2.25pt off 1.5pt}] (70,410) .. controls (70,410) and (70,410) .. (70,410) .. controls (70,410) and (70,410) .. (70,410) .. controls (64.48,410) and (60,405.52) .. (60,400) ;
%Straight Lines [id:da9536395090124945] 
\draw  [dash pattern={on 2.25pt off 1.5pt}]  (70,380) -- (70,410) ;
%Shape: Square [id:dp3119134361354974] 
\draw  [fill={rgb, 255:red, 0; green, 0; blue, 0 }  ,fill opacity=1 ] (67.08,407.08) -- (72.92,407.08) -- (72.92,412.92) -- (67.08,412.92) -- cycle ;
%Straight Lines [id:da4276740357825113] 
\draw  [dash pattern={on 2.25pt off 1.5pt}]  (60,380) -- (60,400) ;
%Straight Lines [id:da2558486926624908] 
\draw    (70,410) -- (100,410) ;
\draw [shift={(100,410)}, rotate = 0] [color={rgb, 255:red, 0; green, 0; blue, 0 }  ][fill={rgb, 255:red, 0; green, 0; blue, 0 }  ][line width=0.75]      (0, 0) circle [x radius= 3.35, y radius= 3.35]   ;
%Straight Lines [id:da03075589791989697] 
\draw  [dash pattern={on 2.25pt off 1.5pt}]  (100,430) -- (100,450) ;
%Shape: Circle [id:dp4966337948637096] 
\draw   (90,460) .. controls (90,454.48) and (94.48,450) .. (100,450) .. controls (105.52,450) and (110,454.48) .. (110,460) .. controls (110,465.52) and (105.52,470) .. (100,470) .. controls (94.48,470) and (90,465.52) .. (90,460) -- cycle ;
%Shape: Square [id:dp940736834034221] 
\draw  [fill={rgb, 255:red, 0; green, 0; blue, 0 }  ,fill opacity=1 ] (97.08,447.08) -- (102.92,447.08) -- (102.92,452.92) -- (97.08,452.92) -- cycle ;
%Straight Lines [id:da13752540050313633] 
\draw  [dash pattern={on 2.25pt off 1.5pt}]  (120,30) -- (120,50) ;
\draw [shift={(120,50)}, rotate = 90] [color={rgb, 255:red, 0; green, 0; blue, 0 }  ][fill={rgb, 255:red, 0; green, 0; blue, 0 }  ][line width=0.75]      (0, 0) circle [x radius= 3.35, y radius= 3.35]   ;
%Shape: Circle [id:dp6287094890495788] 
\draw   (110,100) .. controls (110,94.48) and (114.48,90) .. (120,90) .. controls (125.52,90) and (130,94.48) .. (130,100) .. controls (130,105.52) and (125.52,110) .. (120,110) .. controls (114.48,110) and (110,105.52) .. (110,100) -- cycle ;
%Shape: Square [id:dp786745583026792] 
\draw  [fill={rgb, 255:red, 0; green, 0; blue, 0 }  ,fill opacity=1 ] (67.08,497.08) -- (72.92,497.08) -- (72.92,502.92) -- (67.08,502.92) -- cycle ;
%Straight Lines [id:da9121819139104981] 
\draw    (70,500) -- (100,500) ;
\draw [shift={(100,500)}, rotate = 0] [color={rgb, 255:red, 0; green, 0; blue, 0 }  ][fill={rgb, 255:red, 0; green, 0; blue, 0 }  ][line width=0.75]      (0, 0) circle [x radius= 3.35, y radius= 3.35]   ;
%Straight Lines [id:da2704962966159248] 
\draw    (50,110) ;
\draw [shift={(50,110)}, rotate = 0] [color={rgb, 255:red, 0; green, 0; blue, 0 }  ][fill={rgb, 255:red, 0; green, 0; blue, 0 }  ][line width=0.75]      (0, 0) circle [x radius= 3.35, y radius= 3.35]   ;
%Straight Lines [id:da17486470959606848] 
\draw    (90,80) -- (150,80) ;
%Straight Lines [id:da057633805667051785] 
\draw    (90,30) -- (150,30) ;
%Straight Lines [id:da10506131914670447] 
\draw    (90,330) -- (150,330) ;
%Straight Lines [id:da269165315105653] 
\draw    (90,530) -- (150,530) ;

\end{tikzpicture}

\caption{Feynman Diagrams for each of the terms in Eq.~\eqref{eq:FeynmanDiagrams}, each line in this figure shows the diagrams corresponding to the same line in the equation. The vertices are those defined in Section \ref{cancellation-div:sec} and Appendix \ref{WFU-Feynamn:sec} but the arrows and momentum labels have been dropped for conciseness.} 
    \label{fig:FeynmanDiagrams}
\end{figure}
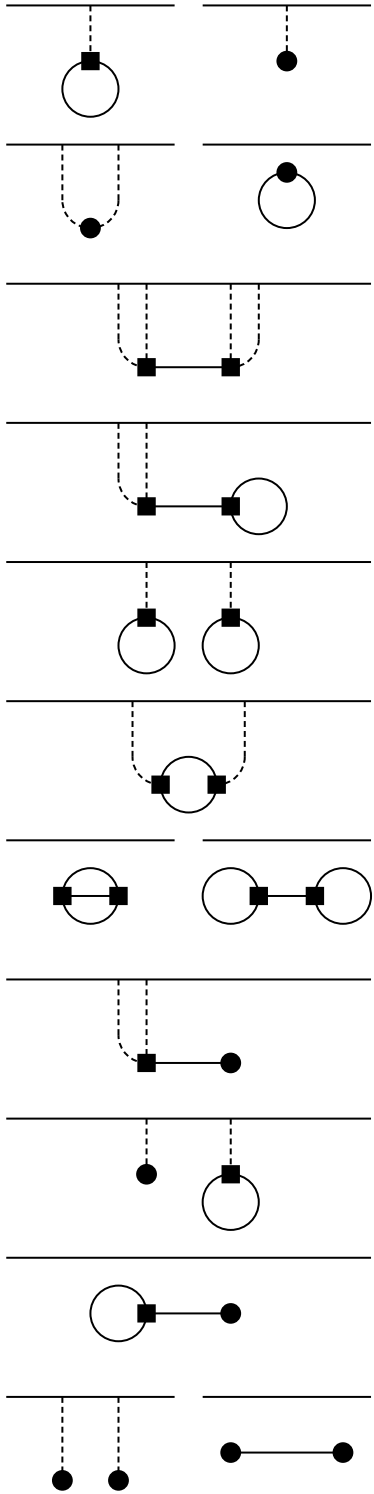

As a concrete example consider the theory with interacting Lagrangian ${\phi'}^3$ including the divergent corrections Eq.~\eqref{eff-lagrangian:eq}, in momentum space we have
\begin{align}\nonumber
    \frac{i}{\hbar}\tilde{S}_{\text{int.}}[\phi]=&-\frac{i}{\hbar}\int \frac{d\kk_1\,d\kk_2\,d\kk_3}{(2\pi)^6}\delta^3(\kk_1+\kk_2+\kk_3)\int d\eta\,a(\eta)\, \lambda\,  \phi'_{\kk_1}\phi'_{\kk_2}\phi'_{\kk_3}\\&-3\lambda\, \delta^4(0)\int d\kk  \delta(\kk)\,\int d\eta  \frac{\phi'_{\kk}}{a} -9\lambda^2 \delta^4(0)\int \dfrac{d\kk}{(2\pi)^3} \int d\eta  \frac{\phi'_{\kk}\phi'_{-\kk}}{a^2}
\end{align}
Expanding the exponential to second order in $\lambda$ and keeping only even derivatives-- as when we set $J=0$, any odd derivative terms will vanish-- we find 
\begin{align}\nonumber
    \exp&\left(\frac{i}{\hbar}\tilde{S}_{\text{int.}}\left[\bar{\phi}+i(2\pi)^3\hbar\frac{d}{d J}\right]\right)=\exp\left(\frac{i}{\hbar}S_{\text{int.}}\left[\bar{\phi}\right]\right)\\\nonumber&\times\left(1+3i(2\pi)^6\lambda\hbar\int_{\kk^3}d\eta_1 d\eta_2 a_1 \delta_{12} {\bar{\phi}_{\kk_1}}'\frac{d}{d\eta_1}\frac{d}{d J_{\kk_2}}\frac{d}{d\eta_2}\frac{d}{d J_{\kk_3}}-3\lambda\delta^4(0)\int_{\kk} \frac{d\eta_1}{a_1}{\bar{\phi}_{\kk_1}}'\right.\\\nonumber&-9\lambda^2\delta^4(0)\int_{\kk^2} \frac{d\eta_1}{a^2_1}\left({\bar{\phi}_{\kk_1}}'{\bar{\phi}_{\kk_2}}'
    -(2\pi)^6\hbar^2\int d\eta_2\delta_{12}\frac{d}{d\eta_1}\frac{d}{d J_{\kk_1}}\frac{d}{d\eta_2}\frac{d}{d J_{\kk_2}}\right)\\\nonumber&-\frac{\lambda^2}{2}(2\pi)^6\int_{\kk^3}\int_{\kk^3}d\eta_1 d\eta_2 a_1 a_2\left( -9{\bar{\phi}_{\kk_1}}'{\bar{\phi}_{\kk_2}}'\frac{d}{d\eta_1}\frac{d}{d J_{\kk_3}}{\bar{\phi}_{\kk_4}}'{\bar{\phi}_{\kk_5}}'\frac{d}{d\eta_2}\frac{d}{d J_{\kk_6}}\right.\\\nonumber&+\hbar^2(2\pi)^6\int d\eta_3d\eta_4\left[6\delta_{23}\delta_{24}{\bar{\phi}_{\kk_1}}'{\bar{\phi}_{\kk_2}}'\frac{d}{d\eta_1}\frac{d}{d J_{\kk_3}}\frac{d}{d\eta_2}\frac{d}{d J_{\kk_4}}\frac{d}{d\eta_3}\frac{d}{d J_{\kk_5}}\frac{d}{d\eta_4}\frac{d}{d J_{\kk_6}}\right.\\ \nonumber&+9\delta_{13}\delta_{24}{\bar{\phi}_{\kk_1}}'\frac{d}{d\eta_1}\frac{d}{d J_{\kk_2}}\frac{d}{d\eta_3}\frac{d}{d J_{\kk_3}}{\bar{\phi}_{\kk_4}}'\frac{d}{d\eta_2}\frac{d}{d J_{\kk_5}}\frac{d}{d\eta_4}\frac{d}{d J_{\kk_6}}\\\nonumber&\left.\left.+\hbar^2(2\pi)^6\int d\eta_5d\eta_6 \delta_{13}\delta_{14}\delta_{25}\delta_{26} \frac{d}{d\eta_1}\frac{d}{d J_{\kk_1}}\frac{d}{d\eta_3}\frac{d}{d J_{\kk_2}}\frac{d}{d\eta_4}\frac{d}{d J_{\kk_3}}\frac{d}{d\eta_2}\frac{d}{d
    J_{\kk_4}}\frac{d}{d\eta_5}\frac{d}{d J_{\kk_5}}\frac{d}{d\eta_6}\frac{d}{d J_{\kk_6}}\right]\right)\\\nonumber&-3i\lambda^2\hbar(2\pi)^6\delta^4(0)\int_{\kk^3}\int_{\kk} \frac{d\eta_2}{a_2}d\eta_1 a_1\left( 3{\bar{\phi}_{\kk_1}}'{\bar{\phi}_{\kk_2}}'\frac{d}{d\eta_1}\frac{d}{d J_{\kk_3}}\frac{d}{d\eta_2}\frac{d}{d J_{\kk_4}}\right.\\\nonumber&\left.+3\int d\eta_3 \delta_{13}\left[  {\bar{\phi}_{\kk_1}}'\frac{d}{d\eta_1}\frac{d}{d J_{\kk_2}}\frac{d}{d\eta_3}\frac{d}{d J_{\kk_3}}{\bar{\phi}_{\kk_4}}'\right.\right.\\\nonumber &\left.\left.-\hbar^2(2\pi)^6\int d\eta_4\delta_{14}\frac{d}{d\eta_1}\frac{d}{d J_{\kk_1}}\frac{d}{d\eta_3}\frac{d}{d J_{\kk_2}}\frac{d}{d\eta_4}\frac{d}{d J_{\kk_3}}\frac{d}{d\eta_2}\frac{d}{d J_{\kk_4}}\right]\right)\\&\left.+\frac{9\lambda^2}{2}\delta^8(0)\int_{\kk}\int_{\kk} \frac{d\eta_1}{a_1} \frac{d\eta_2}{a_2}\left({\bar{\phi}_{\kk_1}}'{\bar{\phi}_{\kk_2}}'-\hbar^2(2\pi)^6\frac{d}{d\eta_1}\frac{d}{d J_{\kk_1}}\frac{d}{d\eta_2}\frac{d}{d J_{\kk_2}}\right) \right),
\end{align}
where $\delta_{ij}=\delta(\eta_i-\eta_j)$, and for the sake of brevity we used the notation
\begin{align}
\displaystyle\int_{\kk^n}\equiv\displaystyle\int \frac{d\kk_1}{(2\pi)^3}..\frac{d\kk_n}{(2\pi)^3} (2\pi)^3\delta^3(\kk_1+...\kk_n).
\end{align}
All the terms with $J$ derivatives depend on 
\begin{align}\nonumber
    &\left.\frac{d}{dJ_{\kk_1}}\frac{d}{dJ_{\kk_2}}\exp\left[-\frac{1}{2\hbar}\int \frac{d^3k}{(2\pi)^3} d\eta d\eta'  J_{\kk}(\eta)G_{k}(\eta,\eta')J_{-\kk}(\eta')\right]\right\rvert_{J=0}=\\&-\frac{1}{\hbar(2\pi)^3}\delta^3(\kk_1+\kk_2)G_{k_1}(\eta_1,\eta_2),\\\nonumber
    &\left.\frac{d}{dJ_{\kk_1}}\frac{d}{dJ_{\kk_2}}\frac{d}{dJ_{\kk_3}}\frac{d}{dJ_{\kk_4}}\exp\left[-\frac{1}{2\hbar}\int \frac{d^3k}{(2\pi)^3} d\eta d\eta'  J_{\kk}(\eta)G_{k}(\eta,\eta')J_{-\kk}(\eta')\right]\right\rvert_{J=0}=\\&\frac{1}{\hbar^2(2\pi)^6}\delta^3(\kk_1+\kk_2)\delta^3(\kk_3+\kk_4)G_{k_1}(\eta_1,\eta_2)G_{k_3}(\eta_3,\eta_4)+2\textrm{ perms},\\\nonumber
    &\left.\frac{d}{dJ_{\kk_1}}\frac{d}{dJ_{\kk_2}}\frac{d}{dJ_{\kk_3}}\frac{d}{dJ_{\kk_4}}\frac{d}{dJ_{\kk_5}}\frac{d}{dJ_{\kk_6}}\exp\left[-\frac{1}{2\hbar}\int \frac{d^3k}{(2\pi)^3} d\eta d\eta'  J_{\kk}(\eta)G_{k}(\eta,\eta')J_{-\kk}(\eta')\right]\right\rvert_{J=0}=\\&-\frac{1}{\hbar^3(2\pi)^9}\delta^3(\kk_1+\kk_2)\delta^3(\kk_3+\kk_4)\delta^3(\kk_5+\kk_6)G_{k_1}(\eta_1,\eta_2)G_{k_3}(\eta_3,\eta_4)G_{k_5}(\eta_5,\eta_6)+14\textrm{ perms}.
\end{align}
So the wavefunction is
\begin{align}\nonumber
    \exp&\left(\frac{i}{\hbar}\tilde{S}_{\text{int.}}\left[\bar{\phi}+i(2\pi)^3\hbar\frac{d}{dJ}\right]\right)= \exp\left(\frac{i}{\hbar}S_{\text{int.}}\left[\bar{\phi}\right]\right)\\\nonumber&\times\left(1-3i(2\pi)^3\lambda\int_{\kk^3}d\eta_1 d\eta_2 a_1 \delta_{12} {\bar{\phi}_{\kk_1}}'\delta^3(\kk_2+\kk_3)G_{k_2}''(\eta_1,\eta_2)-3\lambda\delta^4(0)\int_{\kk} \frac{d\eta_1}{a_1}{\bar{\phi}_{\kk_1}}'\right.\\\nonumber&-9\lambda^2\delta^4(0)\int_{\kk^2} \frac{d\eta_1}{a^2_1}\left({\bar{\phi}_{\kk_1}}'{\bar{\phi}_{\kk_2}}'
    +(2\pi)^3\hbar\int d\eta_2\delta_{12}\delta^3(\kk_1+\kk_2)G_{k_1}''(\eta_1,\eta_2)\right)\\\nonumber&-\frac{\lambda^2}{2\hbar}(2\pi)^3\int_{\kk^3}\int_{\kk^3}d\eta_1 d\eta_2 a_1 a_2\left( 9{\bar{\phi}_{\kk_1}}'{\bar{\phi}_{\kk_2}}'{\bar{\phi}_{\kk_4}}'{\bar{\phi}_{\kk_5}}' \delta^3(\kk_3+\kk_6)G_{k_1}''(\eta_1,\eta_2)\right.\\\nonumber&+\hbar(2\pi)^3\int d\eta_3d\eta_4\left[18\delta_{23}\delta_{24}{\bar{\phi}_{\kk_1}}'{\bar{\phi}_{\kk_2}}'\delta^3(\kk_3+\kk_4)\delta^3(\kk_5+\kk_6)G_{k_3}''(\eta_1,\eta_2)G_{k_5}''(\eta_3,\eta_4)\right.\\ \nonumber&+9\delta_{13}\delta_{24}{\bar{\phi}_{\kk_1}}'{\bar{\phi}_{\kk_4}}'\delta^3(\kk_2+\kk_3)\delta^3(\kk_5+\kk_6)G_{k_2}''(\eta_1,\eta_3)G_{k_6}''(\eta_2,\eta_4)\\\nonumber&+18\delta_{13}\delta_{24}{\bar{\phi}_{\kk_1}}'{\bar{\phi}_{\kk_4}}'\delta^3(\kk_2+\kk_5)\delta^3(\kk_3+\kk_6)G_{k_2}''(\eta_1,\eta_2)G_{k_3}''(\eta_3,\eta_4)-15\hbar(2\pi)^3\delta_{13}\delta_{14}\\\nonumber&\times\left.\left.\int d\eta_5d\eta_6 \delta_{25}\delta_{26}\delta^3(\kk_1+\kk_2)\delta^3(\kk_3+\kk_4)\delta^3(\kk_5+\kk_6)G_{k_1}''(\eta_1,\eta_3)G_{k_3}''(\eta_4,\eta_2)G_{k_5}''(\eta_5,\eta_6)\right]\right)\\\nonumber&+3i\lambda^2(2\pi)^3\delta^4(0)\int_{\kk^3}\int_{\kk} \frac{d\eta_2}{a_2}d\eta_1 a_1\left( 3{\bar{\phi}_{\kk_1}}'{\bar{\phi}_{\kk_2}}'\delta^3(\kk_3+\kk_4)G_{k_3}''(\eta_1,\eta_2)\right.\\\nonumber&\left.+3\int d\eta_3 \delta_{13}\left[  {\bar{\phi}_{\kk_1}}'\delta^3(\kk_2+\kk_3)G_{k_2}''(\eta_1,\eta_3){\bar{\phi}_{\kk_4}}'\right.\right.\\\nonumber &\left.\left.+3\hbar(2\pi)^3\int d\eta_4\delta_{14}\delta^3(\kk_1+\kk_2)\delta^3(\kk_3+\kk_4)G_{k_1}''(\eta_1,\eta_3)G_{k_3}''(\eta_2,\eta_4) \right]\right)\\&\left.+\frac{9\lambda^2}{2}\delta^8(0)\int_{\kk}\int_{\kk} \frac{d\eta_1}{a_1} \frac{d\eta_2}{a_2}\left({\bar{\phi}_{\kk_1}}'{\bar{\phi}_{\kk_2}}'+\hbar(2\pi)^3\delta^3(\kk_1-\kk_2)G_{k_1}''(\eta_1,\eta_2)\right) \right)
    \label{eq:FeynmanDiagrams}
\end{align}
Note that the two-loop term contains several permutations that are slightly too complicated to list and not particularly important at this point. We also present these results in a diagrammatic form in Fig.~\ref{fig:FeynmanDiagrams}, which is usually the more straightforward way to generate these terms but seeing how they are formed explicitly, at least to this order, is potentially enlightening to some readers. The two-loop term is also split into the two diagrams it contains in the diagram as it is easier to represent.

As we have seen previously the new divergent terms that have been introduced due to the derivative interactions cancel with the divergent terms that arise from the $\delta$-function in loop diagrams. To see this, first consider the divergent $\mathcal{O}(\lambda)$ terms,
\begin{align}\nonumber
    -3\lambda \delta^4(0)\int_{\kk}d\eta a^{-1}{\bar{\phi}_{\kk}}'&-3i\lambda (2\pi)^3 \int_{\kk^3}d\eta d\eta' a {\bar{\phi}_{\kk_1}}'\delta(\eta-\eta')\delta^3(\kk_2+\kk_3)\left(ia^{-2}\delta(\eta-\eta')\right)\\&=-3\lambda \delta^4(0)\int_{\kk}d\eta a^{-1}{\bar{\phi}_{\kk}}'+3\lambda\delta(0)\,\int \frac{d\kk' }{(2\pi)^3} \int_{\kk}d\eta a^{-1} {\bar{\phi}_{\kk}}'  =0.
\end{align}
At order $\lambda^2$ we have several types of terms. The first is the exchange diagram,
\begin{align}
    \frac{9\lambda^2}{2\hbar}(2\pi)^3\left(\int d\eta a \int_{\kk^3}{\bar{\phi}_{\kk_2}}'{\bar{\phi}_{\kk_3}}' \right)^2\delta^3(\kk_1+\kk_4)\partial_{\eta}\partial_{\eta'}G_{k_1}(\eta,\eta'),
\end{align}
which contains no divergences. The second are terms that include vertices in addition to the single loop or the new vertex, these will cancel in exactly the same way. Finally, we have the two vertex loop diagram which cancels with the new two point vertex that was introduced,
\begin{align}\nonumber
    &\int_{\kk^3}\int_{\kk^3}d\eta_1d\eta_2d\eta_3d\eta_4a_1^{-1}a_2^{-1}\lambda^2(2\pi)^69\delta_{13}\delta_{24}{\bar{\phi}_{\kk_1}}'{\bar{\phi}_{\kk_4}}'\delta^3(\kk_2+\kk_5)\delta^3(\kk_3+\kk_6)\delta_{12}\delta_{34}\\&-9\lambda^2\delta^4(0)\int_{\kk^2} \frac{d\eta_1}{a^2_1}{\bar{\phi}_{\kk_1}}'{\bar{\phi}_{\kk_2}}'=9\lambda^2\int_{\kk^2}\frac{d\eta_1}{a_1^2}{\bar{\phi}_{\kk_1}}'{\bar{\phi}_{\kk_2}}'\left(\int \frac{d^3k_3}{(2\pi)^3}d\eta_2\delta_{12}^2-\delta^4(0)\right)=0.
\end{align}
This cancellation is the same as what was observed by Weinberg, \cite{Gerstein:1971fm,Weinberg:1995mt}, and we expect it to generalize to more complicated diagrams. In particular, we do not see a cancelation with the two-loop term because we truncated the effective action in $\hbar$ so that the necessary term has already been dropped. It is worth noting that at the level of these diagrams this cancellation is somewhat non-trivial; the exact numerical terms required to generate this cancellation relies on not just the numerical factors of each of the terms in the effective action but also the combinatorics of the two vertex term, specifically not including the disconnected diagram which comes from the same term in action as the loop diagram.
\section{Higher Derivatives}
\label{app:HigherDerivatives}
To see how a field redefinition can be used to eliminate higher-order derivatives, consider the theory with interaction Lagrangian $\lambda {\phi''}^3$. By trialing the field redefinition ansatz 
\begin{align}
    \phi&\rightarrow \phi+f(\phi,\phi',\phi'',\phi^{(3)},\phi^{(4)})\\&=\phi +\lambda \sum_{i=0}^2\sum_{j=i}^2A_{ij}\phi^{(i)}\phi^{(j)}+\lambda^2\sum_{i=0}^4\sum_{j=i}^4\sum_{k=j}^4A_{ijk}\phi^{(i)}\phi^{(j)}\phi^{(k)}+\mathcal{O}(\lambda^3)
\end{align}
It is possible to solve for the coefficients $A$. Note that these can be solved purely algebraically as they first enter through linear perturbations to the action that we can necessarily integrate by parts for the equations of motion to be valid. The resulting field redefinition is
\begin{align}
     &f=-\frac{\lambda}{a^4} \left(\left(k^2 \phi +2 \frac{a'}{a} \phi'\right)^2- \phi'' \left( k^2 \phi +2 \frac{a'}{a} \phi'\right)+ {\phi''}^2\right)\\&-\frac{\lambda^2}{2 a^{14} \left(\phi''+2 \frac{a'}{a} \phi'+k^2 \phi \right)} \left(-8 a^6 \phi^4 k^{10}+416 a^4 \phi^4 a'^2 k^8+5 a^6 \phi^2 {\phi'}^2 k^8\right.\\&\left.-156 a^5 \phi^3 a' {\phi'} k^8-36 a^5 \phi^4 a'' k^8-2 a^6 \phi^3 {\phi''} k^8+20 a^5 \phi a' {\phi'}^3 k^6-764 a^4 \phi^2 a'^2 {\phi'}^2 k^6\right.\\&\left.+3 a^6 \phi^2 {\phi''}^2 k^6+3512 a^3 \phi^3 a'^3 {\phi'} k^6+20 a^5 \phi^2 {\phi'}^2 a'' k^6-472 a^4 \phi^3 a' {\phi'} a'' k^6\right.\\&\left.+64 a^4 \phi^3 a'^2 {\phi''} k^6+4 a^6 \phi {\phi'}^2 {\phi''} k^6-52 a^5 \phi^2 a' {\phi'} {\phi''} k^6-8 a^5 \phi^3 a' {\phi^{(3)}} k^6+4 a^6 \phi^2 {\phi'} {\phi^{(3)}} k^6\right.\\&\left.+20 a^4 a'^2 {\phi'}^4 k^4-1440 a^3 \phi a'^3 {\phi'}^3 k^4-2 a^6 \phi {\phi''}^3 k^4+11108 a^2 \phi^2 a'^4 {\phi'}^2 k^4\right.\\&\left.+20 a^4 \phi^2 {\phi'}^2 a''^2 k^4-148 a^4 \phi^2 a'^2 {\phi''}^2 k^4-a^6 {\phi'}^2 {\phi''}^2 k^4+52 a^5 \phi a' {\phi'} {\phi''}^2 k^4\right.\\&\left.+12 a^5 \phi^2 a'' {\phi''}^2 k^4-13 a^6 \phi^2 {\phi^{(3)}}^2 k^4+80 a^4 \phi a' {\phi'}^3 a'' k^4-2008 a^3 \phi^2 a'^2 {\phi'}^2 a'' k^4\right.\\&\left.+8 a^5 a' {\phi'}^3 {\phi''} k^4-200 a^4 \phi a'^2 {\phi'}^2 {\phi''} k^4+464 a^3 \phi^2 a'^3 {\phi'} {\phi''} k^4+16 a^5 \phi {\phi'}^2 a'' {\phi''} k^4\right.\\&\left.-80 a^4 \phi^2 a' {\phi'} a'' {\phi''} k^4+16 a^5 \phi a' {\phi'}^2 {\phi^{(3)}} k^4-56 a^4 \phi^2 a'^2 {\phi'} {\phi^{(3)}} k^4+8 a^5 \phi^2 {\phi'} a'' {\phi^{(3)}} k^4\right.\\&\left.+80 a^5 \phi^2 a' {\phi''} {\phi^{(3)}} k^4-10 a^6 \phi {\phi'} {\phi''} {\phi^{(3)}} k^4-12 a^6 \phi^2 {\phi''} {\phi^{(4)}} k^4-944 a^2 a'^4 {\phi'}^4 k^2\right.\\&\left.+a^6 {\phi''}^4 k^2+15600 a \phi a'^5 {\phi'}^3 k^2+132 a^4 \phi a'^2 {\phi''}^3 k^2-16 a^5 a' {\phi'} {\phi''}^3 k^2-12 a^5 \phi a'' {\phi''}^3 k^2\right.\\&\left.+80 a^3 \phi a' {\phi'}^3 a''^2 k^2+96 a^4 a'^2 {\phi'}^2 {\phi''}^2 k^2-672 a^3 \phi a'^3 {\phi'} {\phi''}^2 k^2-4 a^5 {\phi'}^2 a'' {\phi''}^2 k^2\right.\\&\left.+128 a^4 \phi a' {\phi'} a'' {\phi''}^2 k^2-52 a^5 \phi a' {\phi'} {\phi^{(3)}}^2 k^2+16 a^6 \phi {\phi''} {\phi^{(3)}}^2 k^2+80 a^3 a'^2 {\phi'}^4 a'' k^2\right.\\&\left.-3520 a^2 \phi a'^3 {\phi'}^3 a'' k^2-208 a^3 a'^3 {\phi'}^3 {\phi''} k^2+1104 a^2 \phi a'^4 {\phi'}^2 {\phi''} k^2+16 a^4 \phi {\phi'}^2 a''^2 {\phi''} k^2\right.\\&\left. +32 a^4 a' {\phi'}^3 a'' {\phi''} k^2-352 a^3 \phi a'^2 {\phi'}^2 a'' {\phi''} k^2+16 a^4 a'^2 {\phi'}^3 {\phi^{(3)}} k^2-128 a^3 \phi a'^3 {\phi'}^2 {\phi^{(3)}} k^2\right.\\&\left.-92 a^5 \phi a' {\phi''}^2 {\phi^{(3)}} k^2+4 a^6 {\phi'} {\phi''}^2 {\phi^{(3)}} k^2+32 a^4 \phi a' {\phi'}^2 a'' {\phi^{(3)}} k^2-20 a^5 a' {\phi'}^2 {\phi''} {\phi^{(3)}} k^2\right.\\&\left.+340 a^4 \phi a'^2 {\phi'} {\phi''} {\phi^{(3)}} k^2-20 a^5 \phi {\phi'} a'' {\phi''} {\phi^{(3)}} k^2+12 a^6 \phi {\phi''}^2 {\phi^{(4)}} k^2-48 a^5 \phi a' {\phi'} {\phi''} {\phi^{(4)}} k^2\right.\\&\left.+8208 a'^6 {\phi'}^4-72 a^4 a'^2 {\phi''}^4+12 a^5 a'' {\phi''}^4+288 a^3 a'^3 {\phi'} {\phi''}^3-48 a^4 a' {\phi'} a'' {\phi''}^3\right.\\&\left.+80 a^2 a'^2 {\phi'}^4 a''^2-756 a^2 a'^4 {\phi'}^2 {\phi''}^2-4 a^4 {\phi'}^2 a''^2 {\phi''}^2+216 a^3 a'^2 {\phi'}^2 a'' {\phi''}^2\right.\\&\left.-52 a^4 a'^2 {\phi'}^2 {\phi^{(3)}}^2-16 a^6 {\phi''}^2 {\phi^{(3)}}^2+32 a^5 a' {\phi'} {\phi''} {\phi^{(3)}}^2-2208 a a'^4 {\phi'}^4 a''+864 a a'^5 {\phi'}^3 {\phi''}\right.\\&\left.+32 a^3 a' {\phi'}^3 a''^2 {\phi''}-384 a^2 a'^3 {\phi'}^3 a'' {\phi''}-96 a^2 a'^4 {\phi'}^3 {\phi^{(3)}}+72 a^5 a' {\phi''}^3 {\phi^{(3)}}\right.\\&\left.-192 a^4 a'^2 {\phi'} {\phi''}^2 {\phi^{(3)}}+8 a^5 {\phi'} a'' {\phi''}^2 {\phi^{(3)}}+32 a^3 a'^2 {\phi'}^3 a'' {\phi^{(3)}}+360 a^3 a'^3 {\phi'}^2 {\phi''} {\phi^{(3)}}\right.\\&\left.-40 a^4 a' {\phi'}^2 a'' {\phi''} {\phi^{(3)}}-12 a^6 {\phi''}^3 {\phi^{(4)}}+24 a^5 a' {\phi'} {\phi''}^2 {\phi^{(4)}}-48 a^4 a'^2 {\phi'}^2 {\phi''} {\phi^{(4)}}\right)
\end{align}
 When we apply this field redefinition the Lagrangian becomes
\begin{align}
    \mathcal{L}&=\frac{1}{2}a^2({\phi'}^2-k^2\phi^2)+\lambda \frac{1}{a^2}\left(k^2\phi+2\frac{a'}{a}\phi'\right)^3-\frac{9\lambda^2}{2a^{6}}\left(k^2\phi+2\frac{a'}{a}\phi'\right)^2\\&\times\left(k^4\left(k^2-48\frac{a'}{a}^2+4\frac{a''}{a}\right)\phi^2+k^2\left(16k^2\frac{a'}{a}-216\frac{a'}{a}^3+40\frac{a'a''}{a^2}\right)\phi\phi'\right.\\&\left.+\left(-k^4+32k^2\frac{a'}{a}^2-244\frac{a'}{a}^4-4k^2\frac{a''}{a}+72\frac{{a'}^2a''}{a^3}-4\frac{a''}{a}^2\right){\phi'}^2\right)\\&=\mathcal{L}^{(0)}+\lambda\mathcal{L}^{(1)}+\lambda^2\mathcal{L}^{(2)}.
\end{align}
The Hamiltonian for this theory is 
\begin{align}
    \mathcal{H}&=\frac{1}{2}a^2\left(k^2\phi^2+\frac{\pi^2}{a^4}\right)-\frac{\lambda}{a^2}\left(k^2\phi+2\frac{a'}{a}\frac{\pi}{a^2}\right)^3+\frac{9\lambda^2}{2a^6}\left(k^2\phi+2\frac{a'}{a}\frac{\pi}{a^2}\right)^2\\&\times\left( k^4\left(k^2-\frac{44 {a'}^2}{a^2}+4\frac{a''}{a}\right)\phi^2+k^2\left(16k^2\frac{a'}{a}-200\frac{a'}{a}^3+40\frac{a'a''}{a^2}\right)\phi\frac{\pi}{a^2}\right.\\&\left.+\left(-k^4+32k^2\frac{a'}{a}^2-228\frac{a'}{a}^4-4k^2\frac{a''}{a}+72\frac{a''{a'}^2}{a^3}-4\frac{a''}{a}^2\right)\frac{\pi^2}{a^4}\right).
\end{align}
Performing the integral over $\pi$ for the time integral gives the effective Lagrangian in the wavefunction of the universe as
\begin{align}
    \tilde{\mathcal{L}}&=\mathcal{L}+i\delta^4(0)\left(-12\lambda\frac{a'}{a}^2\frac{1}{a^4}\left(k^2\phi+2\frac{a'}{a}\phi'\right)\right.\\&\left.-\frac{9\lambda^2}{2a^8}\left(k^4\left(k^4-100k^2\frac{a'}{a}^2+4k^2\frac{a''}{a}+1268\frac{a'}{a}^4-248\frac{{a'}^2a''}{a^3}+4\frac{a''}{a}^2\right)\phi^2\right.\right.\\&\left.+4k^2\frac{a'}{a}\left(3k^4-144k^2\frac{a'}{a}^2+12k^2\frac{a''}{a}+1348\frac{a'}{a}^4-336\frac{a''{a'}^2}{a^3}+12\frac{a''}{a}^2\right)\phi\phi'\right.\\&\left.\left.+\frac8{a'}{a}^2\left(3k^4-96k^2\frac{a'}{a}^2+12k^2\frac{a''}{a}+716\frac{a'}{a}^4-216\frac{{a'}^2a''}{a^3}+12\frac{a''}{a}^2\right){\phi'}^2\right)\right)\\&=\mathcal{L}+i\delta^4(0)\left(\lambda\tilde{\mathcal{L}}^{(1)}+\lambda^2\tilde{\mathcal{L}}^{(2)}\right).
\end{align}
These divergent terms cancel in exactly the same way as the ${\phi'}^3$ theory:
\begin{align}
    \psi_1(\textbf{k})&=-\lambda \int d\eta\left( \tilde{\mathcal{L}}^{(1)}+\frac{1}{2a^2}\frac{\delta^2\mathcal{L}^{(1)}}{\delta {\phi'}^2}\right)\delta^4(0)=0\\
    \psi_2(\textbf{k}_1,\textbf{k}_2)&=-\lambda^2\int d\eta \left(\tilde{\mathcal{L}}^{(2)}+\frac{1}{2a^2}\frac{\delta^2 \mathcal{L}^{(2)}}{\delta{\phi'}^2}+\frac{1}{4a^4}\frac{\delta^2\mathcal{L}^{(1)}}{\delta {\phi'}^2}^2\right)\delta^4(0)=0.
\end{align}
Interestingly, at second-order, both a single vertex and double vertex loop diagram are needed to cancel the divergence that arises in the effective Lagrangian. This behavior is more complicated than we saw in the theory with a ${\phi'}^3$ interaction due to the presence of contact terms in the Lagrangian at second order in the perturbation.

\acknowledgments
We are grateful to Sadra Jazayeri for the extensive discussions that lead to this work. Our special thanks to Xingang Chen for the critical discussions that helped us to clear up the results.  We also would like to thank Mehrdad Mirbabayi, Paolo Creminelli, Aaron Wall and Enrico Pajer for their valuable comments and discussion. HG is supported by jointly by the Science and Technology Facilities Council through a postgraduate studentship and the Cambridge Trust Vice Chancellor's Award.

\bibliographystyle{unsrt}
\bibliography{JCAP_Draft.bib}

\begin{thebibliography}{10}

\bibitem{Guth:1980zm}
Alan~H. Guth.
\newblock {The Inflationary Universe: A Possible Solution to the Horizon and
  Flatness Problems}.
\newblock {\em Phys. Rev. D}, 23:347--356, 1981.

\bibitem{Linde:1981mu}
Andrei~D. Linde.
\newblock {A New Inflationary Universe Scenario: A Possible Solution of the
  Horizon, Flatness, Homogeneity, Isotropy and Primordial Monopole Problems}.
\newblock {\em Phys. Lett. B}, 108:389--393, 1982.

\bibitem{Albrecht:1982wi}
Andreas Albrecht and Paul~J. Steinhardt.
\newblock {Cosmology for Grand Unified Theories with Radiatively Induced
  Symmetry Breaking}.
\newblock {\em Phys. Rev. Lett.}, 48:1220--1223, 1982.

\bibitem{Baumann:2009ds}
Daniel Baumann.
\newblock {Inflation}.
\newblock In {\em {Theoretical Advanced Study Institute in Elementary Particle
  Physics}: {Physics of the Large and the Small}}, pages 523--686, 2011.

\bibitem{Arkani-Hamed:2015bza}
Nima Arkani-Hamed and Juan Maldacena.
\newblock Cosmological collider physics, 2015.

\bibitem{Chen:2009zp}
Xingang Chen and Yi~Wang.
\newblock {Quasi-Single Field Inflation and Non-Gaussianities}.
\newblock {\em JCAP}, 04:027, 2010.

\bibitem{Flauger:2016idt}
Raphael Flauger, Mehrdad Mirbabayi, Leonardo Senatore, and Eva Silverstein.
\newblock {Productive Interactions: heavy particles and non-Gaussianity}.
\newblock {\em JCAP}, 10:058, 2017.

\bibitem{Wands:2007bd}
David Wands.
\newblock {Multiple field inflation}.
\newblock {\em Lect. Notes Phys.}, 738:275--304, 2008.

\bibitem{Chen:2010xka}
Xingang Chen.
\newblock {Primordial Non-Gaussianities from Inflation Models}.
\newblock {\em Adv. Astron.}, 2010:638979, 2010.

\bibitem{Baumann:2011nk}
Daniel Baumann and Daniel Green.
\newblock {Signatures of Supersymmetry from the Early Universe}.
\newblock {\em Phys. Rev. D}, 85:103520, 2012.

\bibitem{Assassi:2012zq}
Valentin Assassi, Daniel Baumann, and Daniel Green.
\newblock {On Soft Limits of Inflationary Correlation Functions}.
\newblock {\em JCAP}, 11:047, 2012.

\bibitem{Schwinger:1960qe}
Julian~S. Schwinger.
\newblock {Brownian motion of a quantum oscillator}.
\newblock {\em J. Math. Phys.}, 2:407--432, 1961.

\bibitem{Keldysh:1964ud}
L.~V. Keldysh.
\newblock {Diagram technique for nonequilibrium processes}.
\newblock {\em Zh. Eksp. Teor. Fiz.}, 47:1515--1527, 1964.

\bibitem{Weinberg:2005vy}
Steven Weinberg.
\newblock {Quantum contributions to cosmological correlations}.
\newblock {\em Phys. Rev. D}, 72:043514, 2005.

\bibitem{Maldacena:2002vr}
Juan~Martin Maldacena.
\newblock {Non-Gaussian features of primordial fluctuations in single field
  inflationary models}.
\newblock {\em JHEP}, 05:013, 2003.

\bibitem{Chen:2006nt}
Xingang Chen, Min-xin Huang, Shamit Kachru, and Gary Shiu.
\newblock {Observational signatures and non-Gaussianities of general single
  field inflation}.
\newblock {\em JCAP}, 01:002, 2007.

\bibitem{Chen:2009bc}
Xingang Chen, Bin Hu, Min-xin Huang, Gary Shiu, and Yi~Wang.
\newblock Large primordial trispectra in general single field inflation.
\newblock {\em Journal of Cosmology and Astroparticle Physics}, 2009(08):008,
  2009.

\bibitem{Tsamis:1996qm}
N.~C. Tsamis and R.~P. Woodard.
\newblock {The Quantum gravitational back reaction on inflation}.
\newblock {\em Annals Phys.}, 253:1--54, 1997.

\bibitem{Prokopec:2010be}
Tomislav Prokopec and Gerasimos Rigopoulos.
\newblock {Path Integral for Inflationary Perturbations}.
\newblock {\em Phys. Rev. D}, 82:023529, 2010.

\bibitem{Gong:2016qpq}
Jinn-Ouk Gong, Min-Seok Seo, and Gary Shiu.
\newblock {Path integral for multi-field inflation}.
\newblock {\em JHEP}, 07:099, 2016.

\bibitem{Chen:2017ryl}
Xingang Chen, Yi~Wang, and Zhong-Zhi Xianyu.
\newblock Schwinger-keldysh diagrammatics for primordial perturbations.
\newblock {\em Journal of Cosmology and Astroparticle Physics}, 2017(12):006,
  2017.

\bibitem{Arkani-Hamed:2017fdk}
Nima Arkani-Hamed, Paolo Benincasa, and Alexander Postnikov.
\newblock Cosmological polytopes and the wavefunction of the universe, 2017.

\bibitem{arkani2020cosmological}
Nima Arkani-Hamed, Daniel Baumann, Hayden Lee, and Guilherme~L Pimentel.
\newblock The cosmological bootstrap: inflationary correlators from symmetries
  and singularities.
\newblock {\em Journal of High Energy Physics}, 2020(4):1--107, 2020.

\bibitem{baumann2020cosmological}
Daniel Baumann, Carlos~Duaso Pueyo, Austin Joyce, Hayden Lee, and Guilherme~L
  Pimentel.
\newblock The cosmological bootstrap: weight-shifting operators and scalar
  seeds.
\newblock {\em Journal of High Energy Physics}, 2020(12):1--51, 2020.

\bibitem{baumann2021cosmological}
Daniel Baumann, Carlos Duaso~Pueyo, Austin Joyce, Hayden Lee, and Guilherme
  L~Pimentel.
\newblock The cosmological bootstrap: spinning correlators from symmetries and
  factorization.
\newblock {\em SciPost Physics}, 11(3):071, 2021.

\bibitem{Cheung:2007st}
Clifford Cheung, Paolo Creminelli, A.~Liam Fitzpatrick, Jared Kaplan, and
  Leonardo Senatore.
\newblock {The Effective Field Theory of Inflation}.
\newblock {\em JHEP}, 03:014, 2008.

\bibitem{senatore2010non}
Leonardo Senatore, Kendrick~M Smith, and Matias Zaldarriaga.
\newblock Non-gaussianities in single field inflation and their optimal limits
  from the wmap 5-year data.
\newblock {\em Journal of Cosmology and Astroparticle Physics}, 2010(01):028,
  2010.

\bibitem{arroja2009full}
Frederico Arroja, Shuntaro Mizuno, Kazuya Koyama, and Takahiro Tanaka.
\newblock Full trispectrum in single field dbi inflation.
\newblock {\em Physical Review D}, 80(4):043527, 2009.

\bibitem{COT}
Harry Goodhew, Sadra Jazayeri, and Enrico Pajer.
\newblock The cosmological optical theorem.
\newblock {\em Journal of Cosmology and Astroparticle Physics}, 2021(04):021,
  2021.

\bibitem{weinberg1995quantum}
Steven Weinberg.
\newblock {\em The quantum theory of fields}, volume~2.
\newblock Cambridge university press, 1995.

\bibitem{Gerstein:1971fm}
Ira~S Gerstein, Roman Jackiw, Benjamin~W Lee, and Steven Weinberg.
\newblock Chiral loops.
\newblock {\em Physical Review D}, 3(10):2486, 1971.

\bibitem{Weinberg:1995mt}
Steven Weinberg.
\newblock {\em {The Quantum theory of fields. Vol. 1: Foundations}}.
\newblock Cambridge University Press, 6 2005.

\bibitem{Behbahani_2014}
Siavosh~R. Behbahani, Mehrdad Mirbabayi, Leonardo Senatore, and Kendrick~M.
  Smith.
\newblock New natural shapes of non-gaussianity from high-derivative
  interactions and their optimal limits from {WMAP} 9-year data.
\newblock {\em Journal of Cosmology and Astroparticle Physics},
  2014(11):035--035, nov 2014.

\bibitem{ostrogradsky1850memoire}
Michael Ostrogradsky.
\newblock {\em M{\'e}moire sur les {\'e}quations diff{\'e}rentielles relatives
  an probl{\'e}me des isop{\'e}rim{\'e}tres}.
\newblock 1850.

\bibitem{woodard2015theorem}
Richard~P Woodard.
\newblock The theorem of ostrogradsky.
\newblock {\em arXiv preprint arXiv:1506.02210}, 2015.

\bibitem{barua1977canonical}
Debojit Barua and Suraj~N Gupta.
\newblock Canonical quantization of fields with higher-derivative couplings.
\newblock {\em Physical Review D}, 16(2):413, 1977.

\bibitem{criado2019field}
Juan~Carlos Criado and Manuel Perez-Victoria.
\newblock Field redefinitions in effective theories at higher orders.
\newblock {\em Journal of High Energy Physics}, 2019(3):1--41, 2019.

\bibitem{dong2010symmetry}
Phung~Van Dong, Le~Tho Hue, HT~Hung, Hoang~Ngoc Long, and NH~Thao.
\newblock Symmetry factors of feynman diagrams for scalar fields.
\newblock {\em Theoretical and Mathematical Physics}, 165(2):1500--1511, 2010.

\end{thebibliography}

\end{document}